%% file: TEXSpap.tex
\documentclass[aip,rsi,amsmath,amssymb,reprint,floatfix]{revtex4-1}
\pdfoutput=1
\usepackage[T1]{fontenc}
\usepackage[utf8]{inputenc}
\usepackage[english]{babel}
\usepackage{times}
\usepackage{graphicx}
\usepackage{colortbl}
\usepackage[usenames,dvipsnames,svgnames,table,rgb,xcdraw]{xcolor}
\convertcolorsUfalse
\usepackage{soul}
\usepackage{color}
\usepackage{booktabs}
\usepackage{longtable}
\usepackage{multirow}
\usepackage[hypertexnames=false]{hyperref} 
\hypersetup{%
  colorlinks=true,
  linkcolor=blue,
  urlcolor=blue,
  citecolor=blue,
  pdftitle={TEXS: tender X-ray emission spectrometer},%
  pdfauthor={M. Rovezzi et al.},%
  pdfsubject={X-ray spectrometers},%
  pdfkeywords={}%
}
\begin{document} 
\setlength{\parskip}{0pt} 

\title[TEXS spectrometer]{TEXS: in-vacuum tender X-ray emission spectrometer with 11 Johansson crystal analyzers}
\author{Mauro Rovezzi}
\email{mauro.rovezzi@esrf.fr}
\affiliation{Univ. Grenoble Alpes, CNRS, IRD, Irstea, Météo France, OSUG, FAME, 71 avenue des Martyrs, CS 40220, 38043, Grenoble, France}
\affiliation{European Synchrotron Radiation Facility, 71 avenue des Martyrs, CS 40220, 38043, Grenoble, France}
\author{Alistair Harris}
\affiliation{Design et Mécanique, Les Coings, 38210, Montaud, France}
\author{Blanka Detlefs}
\affiliation{European Synchrotron Radiation Facility, 71 avenue des Martyrs, CS 40220, 38043, Grenoble, France}
\author{Timothy Bohdan}
\affiliation{European Synchrotron Radiation Facility, 71 avenue des Martyrs, CS 40220, 38043, Grenoble, France}
\author{Artem Svyazhin}
\affiliation{European Synchrotron Radiation Facility, 71 avenue des Martyrs, CS 40220, 38043, Grenoble, France}
\affiliation{M. N. Miheev Institute of Metal Physics, Ural Branch of the Russian Academy of Science, 620990, Ekaterinburg, Russia}
\author{Alessandro Santambrogio}
\affiliation{European Synchrotron Radiation Facility, 71 avenue des Martyrs, CS 40220, 38043, Grenoble, France}
\author{David Degler}
\affiliation{European Synchrotron Radiation Facility, 71 avenue des Martyrs, CS 40220, 38043, Grenoble, France}
\author{Rafal Baran}
\affiliation{European Synchrotron Radiation Facility, 71 avenue des Martyrs, CS 40220, 38043, Grenoble, France}
\author{Benjamin Reynier}
\affiliation{European Synchrotron Radiation Facility, 71 avenue des Martyrs, CS 40220, 38043, Grenoble, France}
\author{Pedro Noguera Crespo}
\affiliation{Added Value Solutions (AVS), Pol. Ind. Sigma Xixilion Kalea 2, Bajo Pabellón 10, 20870, Elgoibar, Spain}
\author{Catherine Heyman}
\affiliation{Design et Mécanique, Les Coings, 38210, Montaud, France}
\author{Hans-Peter Van Der Kleij}
\affiliation{European Synchrotron Radiation Facility, 71 avenue des Martyrs, CS 40220, 38043, Grenoble, France}
\author{Pierre Van Vaerenbergh}
\affiliation{European Synchrotron Radiation Facility, 71 avenue des Martyrs, CS 40220, 38043, Grenoble, France}
\author{Philippe Marion}
\affiliation{European Synchrotron Radiation Facility, 71 avenue des Martyrs, CS 40220, 38043, Grenoble, France}
\author{Hugo Vitoux}
\affiliation{European Synchrotron Radiation Facility, 71 avenue des Martyrs, CS 40220, 38043, Grenoble, France}
\author{Christophe Lapras}
\affiliation{European Synchrotron Radiation Facility, 71 avenue des Martyrs, CS 40220, 38043, Grenoble, France}
\author{Roberto Verbeni}
\affiliation{European Synchrotron Radiation Facility, 71 avenue des Martyrs, CS 40220, 38043, Grenoble, France}
\author{Menhard Menyhert Kocsis}
\affiliation{European Synchrotron Radiation Facility, 71 avenue des Martyrs, CS 40220, 38043, Grenoble, France}
\author{Alain Manceau}
\affiliation{ISTerre, Universit{\'e} Grenoble Alpes, CNRS, CS 40700, 38058, Grenoble, France}
\author{Pieter Glatzel}
\email{glatzel@esrf.fr}
\affiliation{European Synchrotron Radiation Facility, 71 avenue des Martyrs, CS 40220, 38043, Grenoble, France}
\date{\today}


\begin{abstract}
We describe the design and show first results of a large solid angle X-ray emission spectrometer that is optimized for energies between 1.5~keV and 5.5~keV. The spe\-ctro\-me\-ter is based on an array of 11 cylindrically bent Johansson crystal analyzers arranged in a non-dispersive Rowland circle geometry. The smallest achievable energy bandwidth is smaller than the core hole lifetime broadening of the absorption edges in this energy range. Energy scanning is achieved using an innovative design, maintaining the Rowland circle conditions for all crystals with only four motor motions. The entire spectrometer is encased in a high-vacuum chamber that allocates a liquid helium cryostat and provides sufficient space for \emph{in situ} cells and \emph{operando} catalysis reactors.
\end{abstract}
\keywords{X-ray instrumentation, wavelength dispersive spectrometer, X-ray optics, tender x-rays, Johansson crystal analyzers}
\maketitle
%
%
\section{Introduction}\label{sec:introduction}
X-ray emission spectroscopy is attracting growing interest as a tool to characterize local electronic and atomic structure. We define emission spectroscopy as measuring the emitted X-rays from a fluorescence source with an energy bandwidth that is on the order of the core hole lifetime broadening. The spectroscopic techniques with such an instrument include non-resonant X-ray emission spectroscopy (XES), that does not require a monochromatic incoming beam to excite the analyte atom, as well as resonant inelastic X-ray scattering (RIXS) and high energy resolution fluorescence detected (HERFD) X-ray absorption near edge structure (XANES) spectroscopy. A series of books and review papers describe the applications of XES, RIXS and HERFD-XANES in solid state physics, materials science, coordination chemistry and biology~\cite{Meisel:1989_book,Glatzel:2005_CCR,Rueff:2013_JESRP,Hayashi:2013_book,Rovezzi:2014_SST,Bauer:2014_PCCP,Sa:2014_book,DeBeer:2016_book,Proux:2017_JEQ}. High performance and user-friendly X-ray emission spectrometers at synchrotron radiation facilities and in the laboratory, combined with tools for theoretical interpretation of the data, have greatly helped the adoption of the techniques by a large community.

The X-ray energy range between 1.5~keV and 5.5~keV, the so-called ``tender'' X-ray range, covers the K absorption edges of light elements (\emph{e.g.} Al, S, Cl), the L-edges of 4\(d\) transition metals and the M-edges of 5\(d\) transition metals and actinides. The chemical sensitivity of fluorescence lines measured in XES mode can be used to characterize the electronic structure and local coordination of a target element~\cite{AlonsoMori:2009_AC,AlonsoMori:2010_IC}. It is very attractive for heavy elements to use absorption edges of shallow instead of deep core holes (\emph{e.g.} L- instead of K-edge for 4\(d\) transition metals), because of the greatly reduced spectral broadening arising from the core hole lifetime. This has recently been exploited in high pressure and magnetism research~\cite{Wilhelm:2016_HPR} and in chemistry and catalysis~\cite{Thompson:2015_JSR}. The spectral broadening can be further reduced by using an X-ray emission spectrometer for recording the fluorescence lines instead of a conventional solid state detector. To date, this approach has been applied in the field of catalysis~\cite{Thomas:2015_JPCC}, Li--S batteries~\cite{Kavcic:2016_JPCC}, and actinides chemistry~\cite{Kvashnina:2013_PRL,Kvashnina:2019_Ang}.

Despite its relevance, the tender X-ray range is rarely exploited under high ener\-gy resolution because of technical challenges. First of all, the low energies require vacuum conditions in order to minimize absorption by air. Furthermore, the high energy resolution can be achieved either with wavelength dispersive optics or detectors that use materials in their superconducting state~\cite{Doriese:2017_RSI}. Wavelength dispersive optics can be realized using either gratings or perfect crystals. As the efficiency of gratings decreases rapidly towards higher energies, perfect crystals are generally used in tender X-ray spectrometers. Soft crystals with large \emph{d}-spacing such as lithium fluoride (LiF), pentaerythritol (PET), thallium acid phthalate (KAP), Pb stearate and micas were employed for this energy range in some XES instruments. These crystals are not stable with time and when testing them we were not able to achieve the required energy resolution. Thus, standard high-quality hard crystals such as silicon (Si), germanium (Ge), and quartz (\(\beta\)-SiO\(_2\)) are preferred.

X-ray emission spectrometers can generally be grouped into non-dispersive (also referred to as ``point-to-point focusing'') and dispersive geometries. The main difference is in the way the captured solid angle (\emph{i.e.} the total crystal surface) is partitioned. A non-dispersive geometry minimizes the angular difference (\(\Delta\theta = |\theta_i - \theta_B|\)) between the Bragg angle at any point of the crystal (\(\theta_i\)) with respect to the Bragg angle at the centre of the crystal (\(\theta_B\)). This condition is satisfied with a bent optics along the diffraction direction. As a consequence, an emission spectrum is recorded by scanning the optics. In a dispersive set-up, the crystal surface is usually flat along the diffraction direction. The energy of the diffracted X-rays is directly related to the position on the detector. A prominent example is the von Hamos geometry~\cite{Hoszowska:1996_NIMA, Dousse:2014_book}. In this case, an emission spectrum is recorded in a stationary configuration with a position sensitive detector. We note that there are also hybrid instruments working with the crystal off of the Rowland circle and a position sensitive detector~\cite{Welter:2005_JSR,Huotari:2006_RSI,Kavcic:2012_RSI,Holden:2017_RSI}.

An important application for the instrument presented here is HERFD-XANES in samples with low analyte concentration. In this case, the captured solid angle must be used in the smallest possible energy window. Consequently, a non-dispersive geometry was chosen. The interested reader may refer to the seminal study by Wittry and Li~\cite{Wittry:1993_RSI} for a quantitative comparison of scanning spectrometers \emph{versus} dispersive set-ups.

Reviewing all existing instruments is well beyond the scope of this manuscript. We simply note that a multi-analyzer configuration is usually adopted to maximize the effective solid angle of detection. Whereas this strategy is well established for hard X-rays, it has not been realized yet for the tender X-ray energy range. In fact, existing instruments in this energy range use a single crystal analyzer~\cite{Kavcic:2012_RSI,Abraham:2019_JSR}.

The manuscript is organized as follows. In Section~\ref{sec:concept} the concept design is described. The mechanical design and description of the manufactured instrument are reported in Section~\ref{sec:mechanics}. Finally, the results and performances obtained from the commissioning phase are reported and discussed in Section~\ref{sec:results}.

\section{Concept design}\label{sec:concept}
The concept design of the spectrometer is based on the following specifications: 1) energy range between 1.5~keV and 5.5~keV, continuously covered with standard high quality crystals, \emph{e.g.} Si or Ge and \(\beta\)-SiO\(_2\); 2) energy bandwidth close or below the core-hole lifetime broadening of the emission lines measured; 3) optimized solid angle per energy bandwidth, specifically a Rowland circle diameter \(\le\)1~m and a multi-analyzer set-up; 4) windowless from the ultra-high vacuum of the host beamline up to the detector, in order to minimize X-ray attenuation.

\begin{figure}
  \centering
  \includegraphics[width=0.48\textwidth]{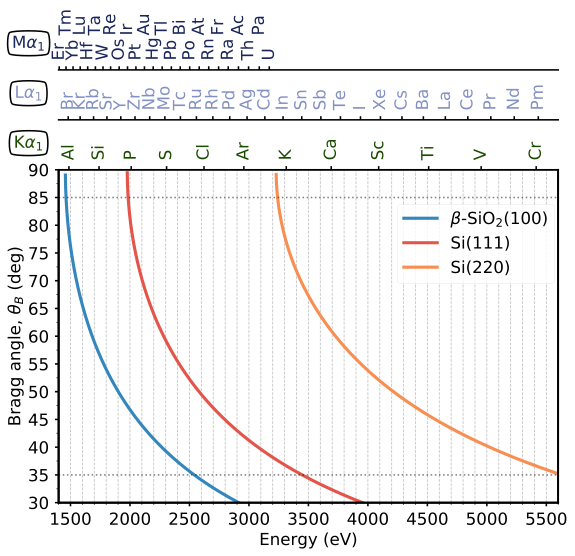}
  \caption{\label{fig1-range}Bragg angular range required to continuously cover the tender X-ray range with $\beta$-SiO$_2$(100), Si(111) and Si(220) reflections. The elements and corresponding energies of three selected families of emission lines, namely, K\(\alpha_1\), L\(\alpha_1\) and M\(\alpha_1\), are indicated in the top axes. The horizontal lines indicate the angular range covered by the spectrometer.}
\end{figure}

Three sets of crystal cuts, $\beta$-SiO$_2$(100), Si(111) and Si(110), are chosen to continuously cover the 1.5--5.5~keV energy range. The Bragg angle ($\theta_B$) ranges from 85$^\circ$ to 35$^\circ$, as shown in Figure~\ref{fig1-range}. The elements and emission lines accessible in this range are also reported in the figure, where for simplicity only the K$\alpha_1$, L$\alpha_1$ and M$\alpha_1$ lines are shown.

\begin{figure*}
  \centering
  \includegraphics[width=0.6\textwidth]{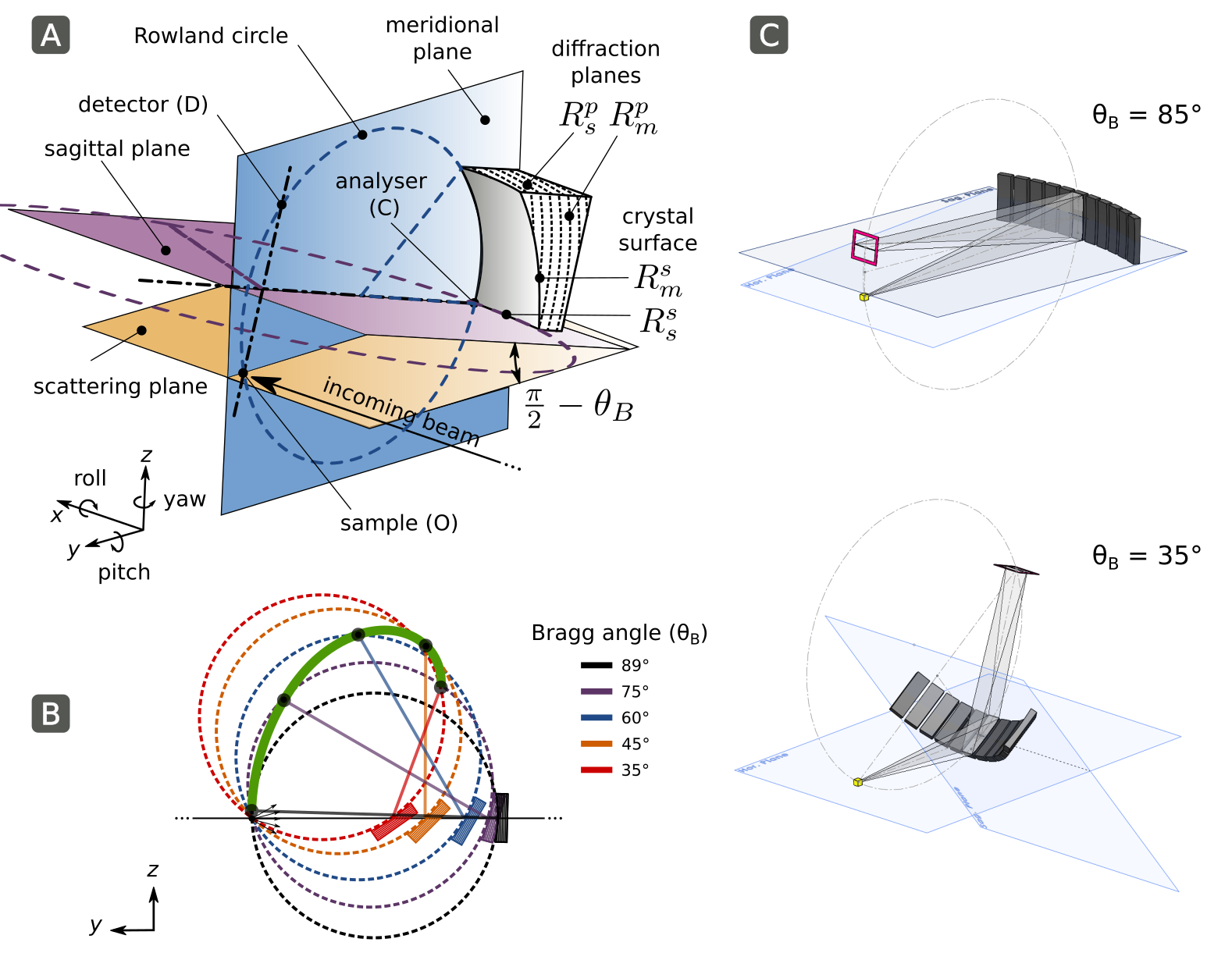}
  \caption{\label{fig2-geometry}Rowland circle geometry employed for the instrument design (A and B, dimensions not to scale): a quarter section of a generic doubly-bent diffractor; how the crystal analyzer and detector move on the Rowland circle when changing the Bragg angle (bottom). Cylindrical segmentation of the analyzer in the sagittal direction for a multi-analyzer configuration at the two extreme Bragg angles (C). The horizontal and sagittal planes are shown. For simplicity, the Rowland circle and rays volume are displayed only for the central analyzer.}
\end{figure*}

The Rowland geo\-me\-try~\cite{Rowland:1882_PhilMag} that was adopted for the design of the spectrometer is shown in Figure~\ref{fig2-geometry}A. The points representing the sample (O), centre of the crystal analyzer (C) and detector (D) define the Rowland circle with radius R. The plane containing the Rowland circle defines the meridional or dispersive direction. In symmetric Bragg reflections, the source-to-analyzer (OC) and the analyzer-to-detector (CD) distances are equal. The source volume is given by the point where the incoming X-rays impinge on the sample surface. The scattering geometry is important for anisotropic radiation emitted from the sample. This is the case for elastic, Compton and X-ray Raman scattering, which give rise to background in the energy range of the fluorescence lines. When light is polarized linearly in the horizontal plane (scattering plane, XY), it is thus advantageous to align the spectrometer normal to the incoming beam (X) and on the scattering plane. This explains why we chose to have the centre of the crystal analyzer at Z=0 for all Bragg angles, as shown in Figure~\ref{fig2-geometry}B. In this configuration, OC stays on an axis (Y) and the detector follows a trajectory resulting from pivoting the Rowland circle around the sample during the Bragg scan (green line in Figure~\ref{fig2-geometry}B). As a consequence, it minimizes the window opening required for sample environment set-ups, such as \emph{in situ} and \emph{operando} cells or cryostats.

The sagittal plane is perpendicular to the sample-detector axis (OD). It is inclined by $\frac{\pi}{2}-\theta_B$ with respect to the scattering plane. A rotation of the Rowland circle around the sample-detector axis describes a circle on the sagittal plane with radius 2R~$\sin(\theta_B)^2$, which we call the sagittal circle. Each point on this circle thus defines a Rowland circle containing points O and D and O\^CD angle $\pi-2\theta_B$. This is the condition for dynamical sagittal focusing as it depends on the Bragg angle. This could be realized by dynamically bending a crystal wafer. However, this would lead to anticlastic bending in the meridional direction. The solution is a segmentation of the crystals in the sagittal direction (Figure~\ref{fig2-geometry}C). The smaller the segments, the better is the approximation to the correct sagittal bending radius. Our choice was to use an array of 11 cylindrical Johansson crystal analyzers with fixed shape and dynamically placed on the sagittal circle. This choice was governed by the availability of Johansson crystals and ultimately costs. The concept described in the following can be extended to more segments and/or smaller crystal bending radii.

\begin{figure}
  \centering
  \includegraphics[width=0.48\textwidth]{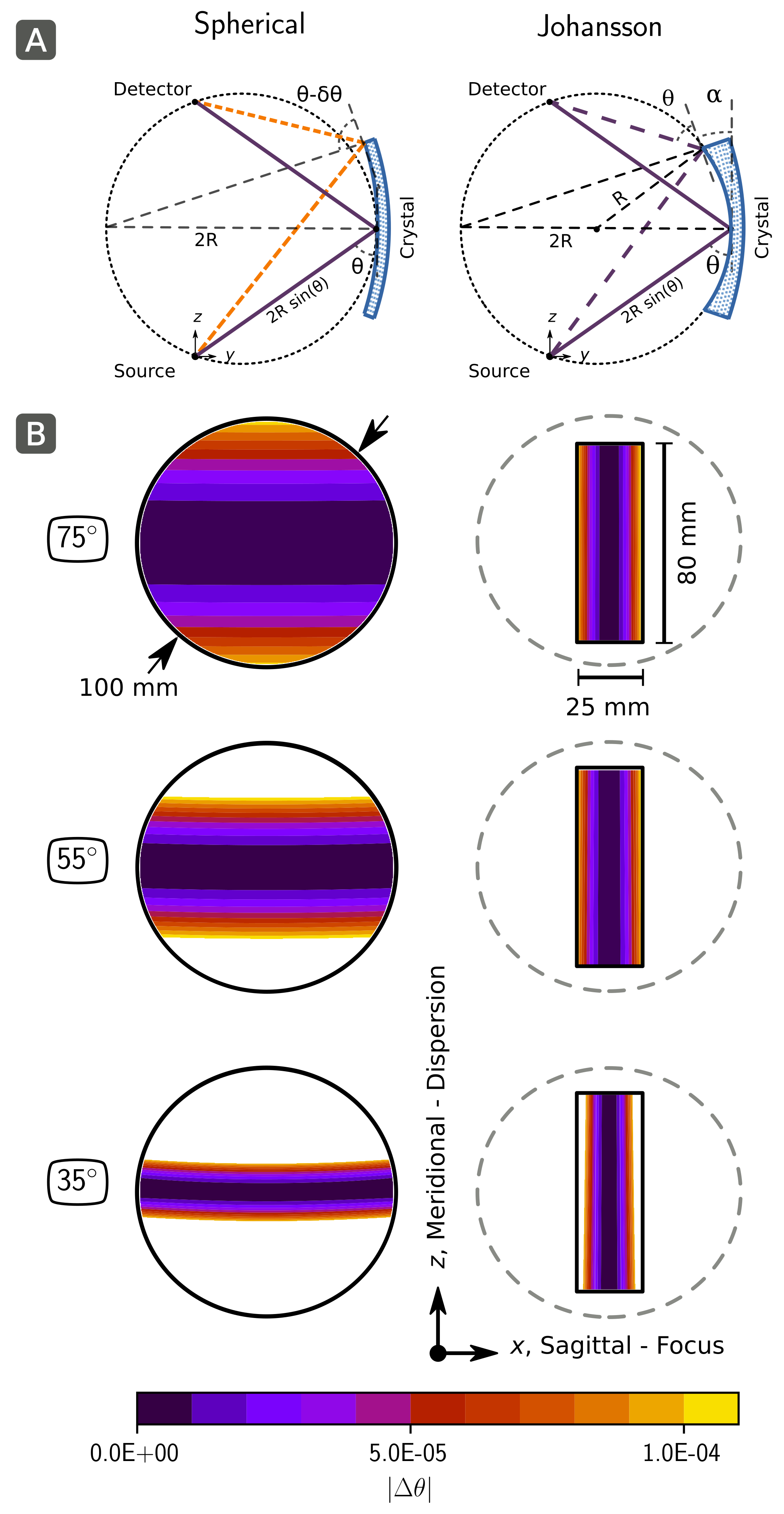}
  \caption{\label{fig3-effscatt}Spherical (left column) \emph{versus} Johansson analyzer (right column). Summary of the geometry projected on the Bragg diffraction plane, YZ (A, not to scale). Effective scattering area plots (B) for Si(111) reflection with a bending radius of 1~m. The region of the crystal that reflects the 0.5~eV bandwidth at three Bragg angles, namely 75\(^\circ\), 55\(^\circ\) and 35\(^\circ\), is shown with a colour scale representing the deviation from the Bragg angle at the centre of the crystal (\(\Delta \theta\)).}
\end{figure}

The optimization of the performances for a generic doubly bent crystal analyzer is based on an analytical study~\cite{Wittry:1990_JAP,Wittry:1990_JAP_a,Wittry:1991_JAP,Wittry:1992_JAP}. It consists in calculating analytically the $\Delta\theta(x,z)$ distribution over the area of the crystal analyzer as a function of the Bragg angle $\theta_B$ at the centre of the crystal. To achieve this purpose, the crystal analyzer is parametrized with four bending radii: R$_{m}^{s}$, R$_{m}^{p}$, R$_{s}^{s}$, R$_{s}^{p}$ which are the bending radius of the surface (R$^{s}$) and crystal planes (R$^{p}$) in the meridional direction (R$_{m}$) or sagittal direction (R$_{s}$), as shown in Figure~\ref{fig2-geometry}A. We included in the calculation only two types of crystal analyzers, based on well established production technologies (Figure~\ref{fig3-effscatt}A): 1) spherically bent crystal analyzers, SBCAs~\cite{Rovezzi:2017_RSI} and references therein); and 2) cylindrically bent Johansson crystal analyzers, CBJCAs~\cite{Johansson:1933_ZP}. SBCAs are widely used on hard X-ray instruments where it is possible to find a reflection to work in back-scattering conditions (\(75^\circ < \theta_B < 90^\circ\)). CBJCAs are mainly developed as X-ray optics for laboratory diffractometers with the goal of monochromatizing the X-ray tube source, \emph{e.g.} selecting only the K$\alpha_1$ emission line, and usually working in grazing-incidence asymmetric Bragg reflections. Furthermore, to take into account the limited bandwidth of the fluorescence line of interest, $\Delta$E, a $\theta_B$-dependent threshold in $\Delta\theta$ is established using the differential form of the Bragg equation, $|\Delta E/E| = \Delta \theta / \tan \theta$. For example, sulfur K$\alpha_1$ has an intrinsic line width $\Delta$E~=~0.61~eV at E~=~2307.8~eV, as tabulated in {\sc xraylib}~\cite{Schoonjans:2011_SAB}. With Si(111) this energy corresponds to $\theta$~=~58.95$^\circ$. The angular width of this line is then $\Delta \theta = 4.3 \times 10^{-4}$~rad. We chose a $\Delta$E threshold of 0.5~eV to determine the area of the crystal diffracting such bandwidth. The results are shown in Figure~\ref{fig3-effscatt}B. An ``effective solid angle per energy bandwidth'' as a function of the Bragg angle and for a given type of crystal analyzer was calculated.

The main differences between SBCAs and CBJCAs are shown in Figure~\ref{fig3-effscatt}. We consider first the meridional/dispersive direction. SBCAs are built by bending a thin crystalline wafer (typically Si or Ge of $\approx$100~$\mu$m) on a spherical substrate of circular shape (typically 100~mm diameter) and radius 2R, the diameter of the Rowland circle. As a consequence, all four radii are equal, R$_{m,s}^{s,p}$~=~2R. A chromatic aberration ($\delta \theta$) is introduced when moving out of the centre of the crystal. This aberration, also known as Johann error, scales with $\cot^2(\theta_B)$ and becomes prohibitive for small Bragg angles. Assuming a threshold of 0.5~eV, Figure~\ref{fig3-effscatt} shows that a smaller area of the analyzer contributes to the scattered amplitude for $\theta_B<75^\circ$. This aberration is removed in CBJCAs. In fact, the Johansson analyzer is manufactured via a bending plus grinding process (Section~\ref{sec:comm_analyzers}). It results in R$_{m}^{p}$~=~2R and R$_{m}^{s}$~=~R. This condition ensures identical Bragg angle over the crystal surface in the meridional direction because the crystal planes have a variable miscut angle $\alpha = \frac{1}{2} \arcsin(\frac{l}{\mathrm{R}})$, where $l$ is the distance from the centre of the analyzer. Without chromatic aberration, the CBJCAs efficiently diffract the X-rays for Bragg angles below 75$^\circ$.

Regarding the sagittal direction, the ideal condition would require R$_{s}^{s}$~=~R$_{s}^{p}$~=~2R~$\sin(\theta_B)^2$. SBCAs are superior to CBJCAs because of the double bending. In fact, CBJCAs are flat in this direction (R$_{s}^{s}$~=~R$_{s}^{p}$~=~$\infty$) and for this reason suffer a chromatic aberration that has a second order dependency on the Bragg angle. For a point source, $\Delta\theta=\tan\theta\left(1\bigg/\sqrt{1+(\frac{w}{4 \mathrm {R}\sin\theta})^2}-1\right)$, where $w$ is the width of the flat side. This disadvantage can be corrected relatively easily by reducing the size of each analyzer in the sagittal direction and increasing the number of analyzers in the spectrometer. With the given threshold of 0.5~eV, it is found that a width of 25~mm keeps the full efficiency of the analyzer in the whole angular range. For those applications requiring better energy resolution, it may be beneficial to further reduce the width of the flat side with a rectangular mask, especially at low Bragg angles. Considering the current limitations in the production of CBJCAs from wafers of 100~mm diameter, the size in the meridional direction is reduced to 80~mm, allowing two analyzers per wafer.

\begin{figure}
  \centering
  \includegraphics[width=0.48\textwidth]{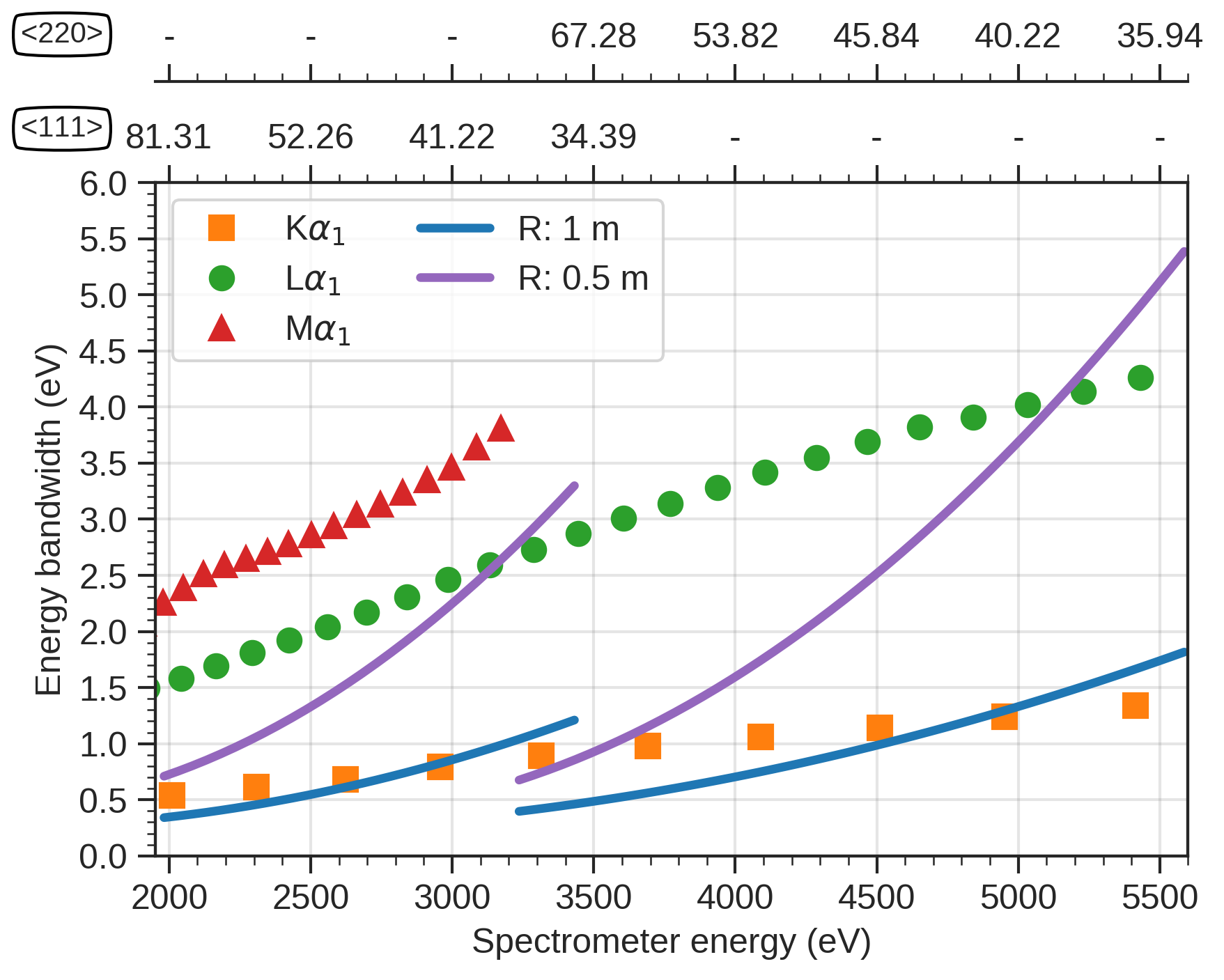}
  \caption{\label{fig4-rt}Results of the ray tracing study for a Si CBJCA of 25\(\times\)80~mm\(^2\) size plus an elliptical fluorescence source size of 0.1\(\times\)0.5~mm\(^2\). The tabulated width of the emission lines for all the elements accessible in the given energy range are shown for comparison (symbols). The top scales give the corresponding Bragg angle for Si(111) and Si(220) reflections.}
\end{figure}

Two additional non-negligible factors that contribute to the energy broadening are: 1) the finite source size, given by the footprint of the exciting radiation (X-rays or particles) on the sample; 2) the strain and defects introduced in the analyzer crystal lattice by the bending and grinding. To correctly take into account these two effects, the previous analytical study was complemented by ray tracing calculations. The simulations were carried out using the \textsc{shadow3}~\cite{SanchezdelRio:2011_JSR} code for a cylindrically-bent Johansson-type crystal analyzer with rectangular shape of 25~mm (sagittal) and 80~mm (meridional). Two bending radii, 1~m and 0.5~m, were considered for each of Si(111) and Si(110) analyzers. The simulations assumed an elliptical source of 0.1~mm (vertical, Z) and 0.5~mm (horizontal, X) and also considered the distortion of the crystal planes by the strain resulting from the bending using the multilamellar model~\cite{SanchezDelRio:2015_JAC}. The tabulated energy widths of the K\(\alpha_1\) (K-L\(_3\)), L\(\alpha_1\) (L\(_3\)-M\(_5\)) and M\(\alpha_1\) (M\(_5\)-N\(_6\)) lines for all elements accessible in the covered energy range are plotted in Figure~\ref{fig4-rt} and compared with the expected performances from ray tracing. 
When studying L- or M-edges, both 0.5~m and 1~m configurations offer an experimental energy broadening below the width of the emission lines. For the K-edges of light elements, or when the highest energy resolution is required, a configuration with 1~m bending radius is preferred. In fact, to collect HERFD-XANES spectra with sharper spectral features it may be necessary to reduce the experimental bandwidth as low as $\approx$0.5~eV, even for L- and M-edges. Nonetheless, for applications aiming to perform XANES spectroscopy on trace elements with dilution levels below 10 parts-per-millions and complex matrices, the gain in collected solid angle provided by the 0.5~m bending radius configuration would be an asset.

A schematic view of the final concept design with 11 analyzers is shown in Figure~\ref{fig2-geometry}C for the two extreme angular positions of the spectrometer, namely, 85\(^\circ\) and 35\(^\circ\). The Rowland circle geometry of Figure~\ref{fig2-geometry}A is used: the sample and the incoming beam are laying on the horizontal plane, whereas the analyzers reside on the sagittal plane, their inclination depending on the Bragg angle. The 11 analyzers at 2R\(=\)0.5~m cover a solid angle ranging from 50~msr (\(\theta_B=35^\circ\)) up to 87~msr (\(\theta_B=85^\circ\)). The solid angle is about four times smaller when 2R\(=\)1~m.

\section{Mechanical design and production}\label{sec:mechanics}
This section presents the mechanical design of the spectrometer as built and installed in the first experimental hutch of beamline ID26 at the ESRF. The instrument was manufactured, assembled and tested by the company Added Value Solutions (Spain).

\begin{figure*}
  \centering
  \includegraphics[width=0.9\textwidth]{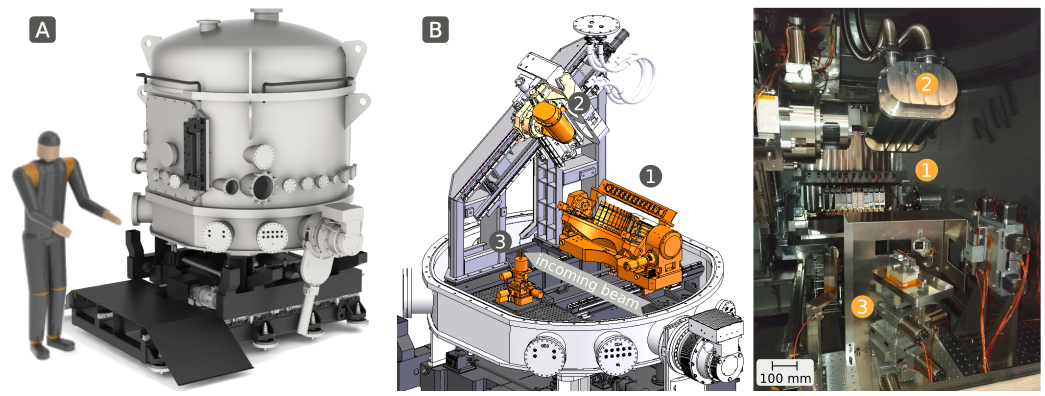}
  \caption{\label{fig5-layout}Mechanical layout of the spectrometer. Three dimensional (3D) rendering of the external view (A). Internal view (B) as 3D drawing (left, upper parts of the vacuum vessel are removed for clarity) and picture (right), showing the spectrometer frame with the main sub-assemblies: analyzer table (1), detector (2) and sample environment (3). The incoming beam direction (X) is represented with a grey arrow.}
\end{figure*}

X-ray attenuation by air and windows is strong in the tender energy range. 1~cm of air or 8~\(\mu\)m of Kapton absorb, respectively, 50\% and 40\% of X-rays at 2~keV. For this reason, a fully in-vacuum solution was adopted. The whole spectrometer mechanics and sample environment are encased in a vacuum vessel of \(\approx\)4~m\(^3\), roughly, a cylinder of 1.74~m dia\-me\-ter and 1.67~m height (Figure~\ref{fig5-layout}A). The dimension of the vessel is optimized for a complete movement on the scanning trajectories of the crystal analyzers with a bending radius comprised between 1020~mm and 480~mm and to accommodate the displacement of the X-ray detector. The vacuum level is high, with a pressure below 1\(\times\)10\(^{-5}\)~mbar. It allows connecting the spectrometer directly to the ultra-high vacuum of the host beamline - via a differential pumping system - and is sufficient for thermal insulation, allowing the use of a windowless liquid helium cryostat (section~\ref{sec:sampenv}). The required vacuum level is obtained by introducing only high vacuum compatible components and avoiding trapped volumes. The obtained out-gassing rate is 3\(\times\)10\(^{-10}\)~mbar\(\cdot\)L/s/cm\(^2\), comparable to unbaked stainless steel after 24~h of pumping. Primary and turbo pumps are mounted on a CF~250 port. The pumping capacity is 80 m\(^3\)/h (Leybold Leyvac 80 dry compressing pump) and 2100~l/s (Turbovac MagW 2200iP turbomolecular magnetically levitated pump). Via this system, the pumping time is less then 90~mi\-nu\-tes to reach the low 10\(^{-5}\)~mbar range. The spectrometer is operated in vacuum during the experiment and the change of sample is performed via a load lock system (section~\ref{sec:sampenv}). Installation of an \emph{in situ} cell is possible by inverse vacuum set-ups and feedthroughs in the available chamber ports. Gas, liquid and electrical connections are possible.

The spectrometer counts a total of fifty-two motorized axes controlled by the BLISS software~\cite{Guijarro:20180_conf} via IcePap stepper motor drivers. The exact Rowland circle tracking is achieved via a device server implemented in Python~\cite{_fn:sloth}. Critical motor axes are equipped with absolute encoders. The vacuum vessel is mounted on motorized long rails. It allows aligning the origin of the spectrometer along the Y direction, perpendicular to the incoming X-ray beam in the horizontal plane, and moving out the whole spectrometer in a ``maintenance mode'', when disconnected from the beamline. The whole spectrometer is isolated from the vacuum vessel via an externally motorized four-point ball-screw lifting system plus three guiding columns connected to the vacuum chamber only via edge-welded bellows. This system allows aligning the spectrometer along the Z direction and avoids transmitting vibrations from the vacuum pumps to the mechanics. The vibrations from the ground are minimized by the internal damping factor of the cast iron Y carriage. The structural analysis performed via finite elements calculations confirms that the vibration modes do not act in the direction of interest, that is, along the pathway of the diffracted beam. The first main mode consists of a global lateral oscillation along X with a maximum displacement of 0.4~mm and a frequency of 16.3~Hz. This mode does not affect the energy resolution as the whole spectrometer oscillates around the incoming beam focal point.

The spectrometer frame consists of a welded stainless steel structure divided in three main sub-assemblies (Figure~\ref{fig5-layout}B): the crystal analyzer table (1); the detector arm (2); and the sample tower (3). These components are described in the following sub-sections.

\subsection{Analyzer table}
The 11 segments supporting the analyzer crystals must follow the motion of the sagittal circle as described previously. This can be achieved by two translations (\emph{e.g.} along Y and Z) and two rotations (\emph{e.g.} $\theta$ and $\chi$ with rotation axes in and perpendicular to the Rowland circle). In previous multi-analyzer spectrometers, this was achieved by piling-up multi-stages with each crystal~\cite{MorettiSala:2018_JSR}. We dismissed this solution because of space constraints and costs. Furthermore, the acceptable angular error of \(<\)10~\(\mu\)rad in the meridional direction over the scan range is beyond the achievable mechanical specifications of the components, as the angular errors for each of the four components add up.

\begin{figure}
  \centering
  \includegraphics[width=0.48\textwidth]{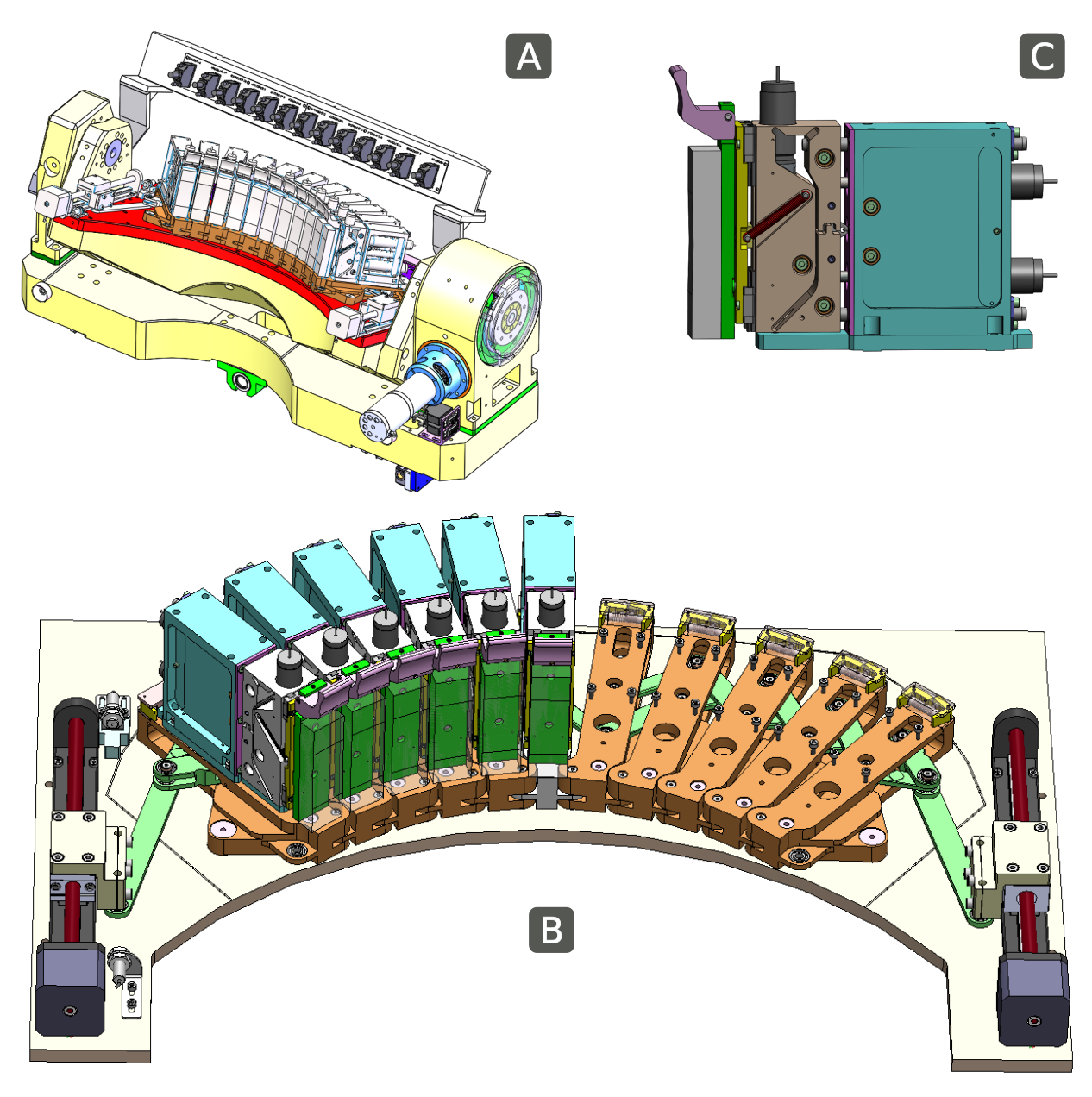}
  \caption{\label{fig6-panto}Components of the analyzer table (A): pantograph (B) and analyzer module (C).}
\end{figure}

We conceived a new design for the displacement of the crystal analyzers that minimizes the angular errors and is, at the same time, less expensive than the above described four-axes solution. All crystal analyzers are mounted on a custom-designed table (Figure~\ref{fig6-panto}A). The table translates along Y (TY) and rotates by $\theta_B$ around an axis that passes through the centre of the central analyzer. Ten out of the 11 analyzer crystals are mounted on each side of the central crystal on a chained mechanical system (bronze pieces shown in Figure~\ref{fig6-panto}B), which is kept in contact with an underlying plate via three points that are held in place by magnets. The chained mechanics is actioned by two linear actuators via a ball-bearing accordion-like system. We call this mechanical system a ``pantograph'' as it resembles a pantograph drawing tool. The underlying plate is made out of carbon steel (magnetic). It is rectified, polished and hardened via a surface nitridation process, ensuring minimal angular errors during the motion of the pantograph and resistance at friction with bronze. The measured flatness is below 6~$\mu$m peak-to-valley over the whole surface of 0.15~m$^2$. The plate is parallel to the sagittal plane and thus ensures correct positioning of all crystals. This friction design mechanics allows obtaining the required dynamical sagittal focusing for all analyzers with only two motors. An important property of this solution is that it can fit a larger number of crystals of different bending radii.

The analyzer table provides combined movements for the Rowland tracking (energy scan). Additional degrees of freedom are implemented per analyzer via a three axis motorized module (Figure~\ref{fig6-panto}C). The module carrying the analyzer allows fine tuning the alignment of each analyzer with respect to the central one, taken as reference. It compensates for the defects introduced during the analyzers production and the inevitable mis-alignment of the mechanics in the whole scanning range. The motorized axis is driven by three compact high-precision linear actuators, the ESRF fixed-head micro-jacks: 1) vertical adjustment of $\pm$2~mm stroke, 2+3) horizontal adjustment of $\pm$5~mm, and $\theta$ adjustment. The last two axes are obtained by driving the actuators either in the same or the opposite directions. We chose to have three motorized degrees of freedom for each analyzer for commissioning and validation of the concept. A simplified design would only require one angular adjustment in the meridional plane.

Each analyzer module with its motorized degrees of freedom is mounted on the pantograph with an additional degree of freedom that allows the fine angular alignment in the sagittal direction. This alignment is performed manually with a laser (Section~\ref{sec:comm-align}). The crystal analyzers are mounted separately with wax or glue on holders, that are held in place by three magnets.

\subsection{Multi-wire gas detector}
\begin{figure*}
  \centering
  \includegraphics[width=0.9\textwidth]{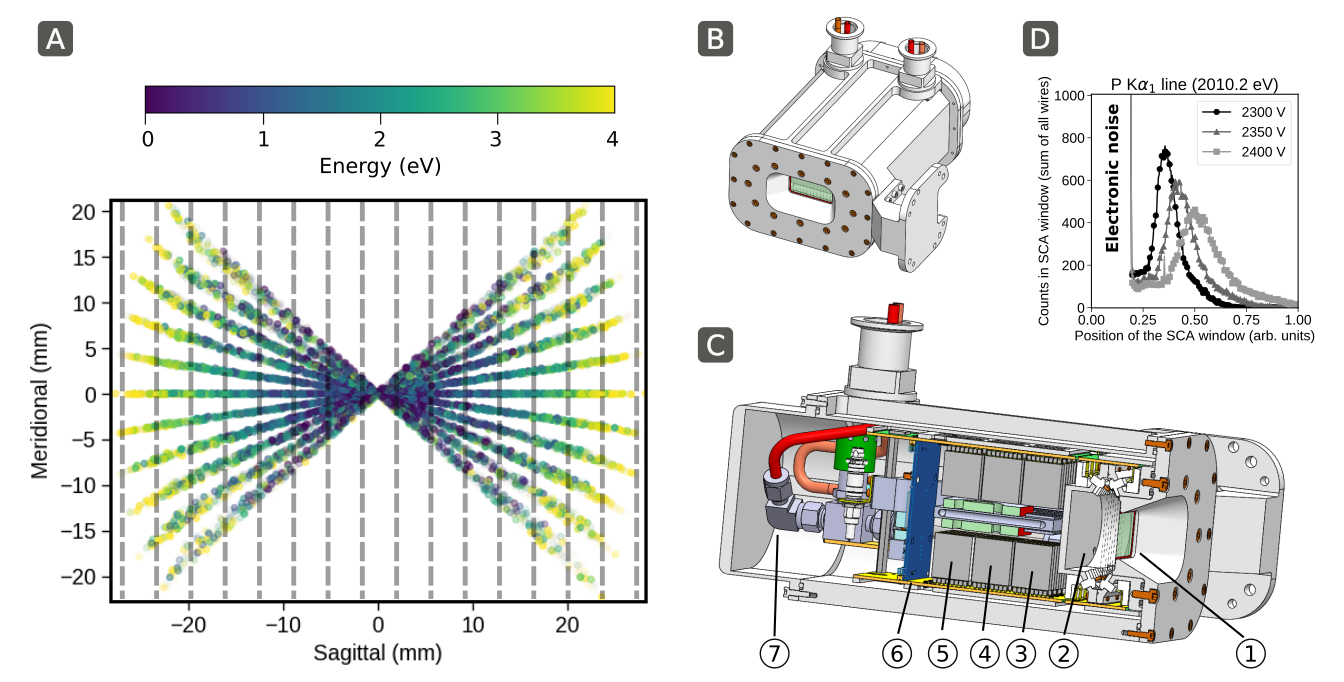}
  \caption{\label{fig7-detector}Multi-wire gas detector. Ray tracing simulation (A) of the focal image from the 11 analyzers at 500~mm bending radius and 35\(^\circ\) Bragg angle. The overlay grey dashed lines represent the chosen wires distribution. Inverse vacuum chamber (B). Main components (C) of the detector: entrance X-ray window (1); gas chamber with sixteen wires (2); charge-sensitive pre-amplifier (3); baseline restore amplifier (4); Gaussian shaper (5); single channel analyzer board (6); gas and cooling pipes (7). Spectral response (energy distribution) as a function of the applied potential (D) for a P K\(\alpha_1\) emission line.}
\end{figure*}

The second core component of the spectrometer is the detector. The first step in defining the required specifications was to simulate, via ray tracing, the focal image of the 11 cylindrical Johansson analyzers. The detector is centred on the YZ plane facing the centre of the analyzer table. The configuration with 2R~=~0.5~m and at \(\theta_B\)~=~35\(^\circ\) is shown in Figure~\ref{fig7-detector}A. This is the configuration where the focus spreads the most. The focal image has a butterfly-like shape because each analyzer contributes with a line focus 50~mm long (twice the size of the flat side) and approximately 0.1~mm thick (vertical size of the source). The focal image of the central analyzer is a horizontal line whereas the focal images of the side analyzers are tilted. The tilting angle depends on the $\chi$ angle. For the highest tilt configuration (Figure~\ref{fig7-detector}A) the total height of the combined focal image is approximately 40~mm. Furthermore, and as expected from Figure~\ref{fig3-effscatt} and shown in Figure~\ref{fig7-detector}A, the focal image has an energy dispersion in the horizontal direction, starting from the centre. This implies that a detector with spatial resolution along the horizontal direction will enable us to perform an in-focus energy correction. The ray tracing simulations with a source size of 0.1~$\times$~0.5~mm$^2$ show that the energy spread along the horizontal direction does not significantly change within 5~mm. Thus, an in-focus energy correction can be applied even when using a detector with a coarse position sensitivity of the order of a few millimeters.

The main specifications required for the detector are the following: 1) active area of 50~\(\times\)~40~mm\(^2\). This is obtained from the analysis of the simulated focal images. 2) overall quantum efficiency (sensor absorption and electronic threshold) \(\ge\)80\% from 1.5 keV to 5.5~keV. 3) electronic noise without X-rays \(\le\)1~event in 10~s over the whole area at minimum photon energy discriminator's threshold. This requirement is dictated by the fact that we would like to be able to detect rare events with count rates of only a few counts per second and therefore we want low electronic noise signal within the counting interval. 4) $\le$10\% non-linearity at 1~MHz counting rate over the whole area. In fact, assuming a negligible background signal, our target statistics is 10\(^{-3}\). 5) continuous data read-out with \(\le\)2~ms delay. The read-out dead-time during continuous scans of the monochromator should be kept as low as possible. In fact, for radiation sensitive samples, the XANES scans ideally last less than 5~s with acquisition times of a few ms per data point. 6) vacuum compatibility and 7) weight not to exceed 8~kg. The detector must be mounted on moving stages covering almost 1~m$^2$ area.

According to these specifications, we surveyed the commercial solutions during the design phase. The conclusion was that the detector fulfilling all specifications was not available ``off the shelf'' at the time or within our budget (supplementary Section~S1). For these reasons, we designed and built a multi-wire gas flow proportional counter at the ESRF. The reader interested in more details about gas detectors may refer to the book by Knoll~\cite{Knoll:2000_book} or the didactic work of Winkler \emph{et al.}~\cite{Winkler:2015_AJP}. Proportional counters are frequently used in the detection and spectroscopy of low-energy X-ray radiation. This is because of their low noise and an energy resolution below 20\% of the measured energy. Historically, flow proportional counters were mounted in wavelength-dispersive spectrometers for micro-analysis at electron microscopy stations (\emph{e.g.} \cite{Wuhrer:2018_conf} for a recent overview).

We developed the sixteen-wires flow proportional counter shown in Figure~\ref{fig7-detector}B-C. The wires are gold-coated tungsten of 50~$\mu$m diameter mounted vertically (along the meridional direction, Y) with a horizontal spacing of 4 mm (along the sagittal direction, X). The choice of a multi-wire configuration has two advantages: 1) to increase the detector linearity; 2) to obtain a spatial resolution in the sagittal direction. A 90\% linearity per wire is measured at approximately 50~kHz (counts/s), as shown in supplementary Figure~S5. The sixteen wires provide an overall linearity up to 1~MHz count-rate as the cylindrical crystals distribute the intensity homogeneously in the sagittal direction. The spatial sampling of the horizontal direction gives access to the energy-dispersion correction within the focal image, improving the energy resolution of the instrument. Furthermore, it allows controlling the spectrometer alignment. In fact, in case of mis-alignment not all line foci will cross the central wires, showing a signal drifting to the side wires. The data analysis presented in Section~\ref{sec:comm-eres} make use of such information. The naming convention used for the commissioning and the distribution of the wires within the geometry of the spectrometer is given in supplementary Figure~S1.

The gas chamber plus the electronics are encased in an ``inverse vacuum'' chamber, where the vacuum is outside and the atmospheric pressure inside. The chamber is mounted on the detector arm, consisting of one rotation and two translation stages. This allows positioning the detector on all required points in the Rowland tracking space, as shown in supplementary Figure~S2, while always facing the central analyzer crystal. The detector chamber is connected to the outside of the vacuum vessel via two highly flexible, annular corrugated metallic hoses (Witzenmann GmbH, model RS 321). The choice of these specific hoses was made after building a dedicated prototype for measuring the forces acting on the detector during the movement (supplementary Figure~S3). The hoses allow carrying the gas and cooling pipes plus the power, Ethernet and signal cables. The dimensions of the gas chamber are 113~$\times$~50~$\times$~40~mm$^3$, (width~$\times$~height~$\times$~thickness). It is filled with a gas mixture composed of 15\% carbon dioxide (CO$_2$) in Argon (Ar). The 40~mm path results in an efficiency of \(\ge\)90\%, decreasing to 60\% before the Ar K-edge (3205.9~eV) (supplementary Figure~S4). An X-ray window separates the gas chamber from the spectrometer vacuum. The window (MOXTEK ProLINE 20) is composed of an ultra-thin polymer of 0.6~$\mu$m thickness coated with 45~nm of Al and deposited on a metal support grid. It can hold 1.2~atm differential pressure sustaining hundreds of cycles, and has a total (including the metal support grid) X-ray transmission $>$90\% above 1~keV.

The wires are connected directly to an application-specific integrated circuit (ASIC) composed of a primary analog stage and a secondary digital part based on a field-programmable gate array (FPGA). The analog electronics implements an RC reset scheme of 100~$\mu$s damping time. The signal is amplified with a charge sensitive pre-amplifier, then transformed to Gaussian pulse via a shaper amplifier of 500~ns and baseline restore. The ASIC allocates commercially available amplifiers that can be easily replaced in case of failure or performance loss. Currently, the charge sensitive amplifiers are provided by Cremat~Inc. The digital electronics implements a single channel analyzer for the signal of each wire and outputs sixteen NIM-type signals transferred to the control computer via a differential line. Having the analogical signal processed close to the wires allows reducing the electronic noise. Furthermore, the whole electronics is water-cooled. All detector settings are remotely controlled via an Ethernet connection to a local embedded ARM computer (Qseven).

The spectral response of the detector and separation from the electronic noise is visible in Figure~\ref{fig7-detector}D, obtained by scanning the level of a small window in the single channel analyzer (SCA). The fluorescence peak of the P K$\alpha_1$ line at 2010.2~eV is well separated from the electronic noise peak at a voltage of 2300~V. Increasing the voltage up to 2400~eV shifts the fluorescence signal peak to higher threshold values. At this voltage, the Si K\(\alpha\) lines (\(\approx\)1739~eV) can be clearly separated from the electronic noise peak (not shown). The low energy threshold of the SCA is set to cut the electronic noise below the acceptable level of 1 count per 10 seconds over the entire detector area. The high energy threshold allows reducing the signal arising from cosmic rays and possible background from fluorescence lines that are excited by higher harmonics of the incident beam energy. Events arising from cosmic rays are very short and can thus easily be identified as spikes in single data points. The affected data points are removed during the data processing. 

\subsection{Sample environment}\label{sec:sampenv}
The vacuum vessel allocates a versatile sample environment volume of approximately 500~mm wide (supplementary Figure~S6). At the base of this volume, a breadboard allows mounting all required equipment. The standard configuration is composed of a four-axis goniometer, which provides X, Y, Z translations plus a rotation around the Z axis. The goniometer allows aligning the sample on the beam with micro-metric precision. A stroke of 40~mm is available on each translation axis, whereas the rotation could in principle perform 360$^\circ$ unless the sample environment constrains the motion.

A thick aluminum shielding positioned around the sample prevents the direct fluorescence from reaching the detector or the X-ray scattering background from reaching the crystal analyzers or the vacuum vessel walls. A set of photo-diodes are employed to measure the total fluorescence yield. An incoming X-ray beam monitor with horizontal and vertical slits is mounted upstream of the sample. The beam monitor consists of a four-quadrant diode with a hole, reading the back-scattering signal from a thin calcium-free Kapton foil ($<$8~$\mu$m thickness).

The vacuum vessel has two large doors of square shape with a side of 700~mm in front of the sample area (supplementary Figure~S6A) and in the back of the analyzer table. During maintenance mode, the sample environment volume and the detector are accessible via the front door. Four flanges mounted on the front door are dedicated to: 1) a load-lock system for quick sample change; 2) a solid state detector for partial fluorescence yield analysis; 3) a pressure gauge; 4) a glass window for visual inspection.

The core components of the sample environment routinely available for users are the load lock set-up and the liquid He cryostat. In addition, two experimental cells were built: one for studying \emph{operando} gas sensors (supplementary Section~S7) and another for \emph{in situ} catalytic reactions.

The load lock set-up was initially built at the ESRF with a simple rod design. However, it turned out to be not suited for user operation. The current design is a commercial solution (Ferrovac Gmbh) featuring a fast entry load lock with a quick access door and a CF40 single shaft sample transport rod of 750~mm linear travel. The transfer rod head is customized for inserting a multi-sample holder of 45~mm length and 30~mm height into the sample receptacle. The sample transfer chamber of $\approx$0.7~l volume is pumped from ambient pressure down to \(10^{-5}\)~mbar in less than a minute thanks to a combination of pre-pumping by a small primary pump and a turbo pump unit, which are connected by a three-way valve ensuring continuous operation of the system. This quick sample load is crucial to transfer frozen samples (\emph{e.g.} frozen solutions) from a liquid nitrogen tank into the cold sample receptacle.

The standard sample holders are made of copper (supplementary Figure~S7) and designed to receive either three 13~mm pellets or eight 5~mm pellets (two rows of four) with a base plate having a locking mechanism for the sample transfer system. Thermal contact is further enhanced with a frame screwed on the surface of the copper holder. Samples presenting safety risks (\emph{e.g.} toxic, radioactive) can be sealed by inserting a thin Kapton foil in between the copper pieces.

The transfer system puts the sample holder in mechanical contact with the copper body of the sample receptacle, which is connected via a flexible copper braid to the cold head of a liquid He cryostat mounted directly on the spectrometer chamber on a CF63 flange. Thermal isolation from the goniometer is realized by a polyether ether ketone (PEEK) interface block supporting the sample receptacle. The required freedom in motion for the sample and necessary large optical access lead to thermal losses over the length of the copper braid limiting the working sample temperature to a minimum of approximately 22~K with the cold head staying at 12~K. The initial cool-down time for the cryostat is approximately three hours but after a sample transfer the base temperature is recovered within ten minutes.

\section{Commissioning results}\label{sec:results}

The commissioning of the spectrometer was performed at the beamline ID26 of the ESRF during multiple experimental sessions. The beamline was equipped with a cryogenically cooled double Si(111) monochromator, giving a resolving power $E/\Delta E \approx 7000$, that is, an energy bandwidth of 0.3~eV at 2~keV. Harmonic rejection was achieved by three Si mirrors working in total reflection. The beam was focused to 90 $\mu$m vertical by 350 $\mu$m horizontal. The incoming beam was monitored as previously described.

\subsection{Johansson crystal analyzers}\label{sec:comm_analyzers}
The performance of the spectrometer critically depends on the quality of the Johansson crystals. At the early stage of the design phase we tested commercially available crystals from Saint-Gobain (France), Rigaku (United States), AlpyX (France), and compared them with those fabricated at the ESRF. The performance of one commercial crystal was close to theoretical simulations (supplementary Figure~S9), demonstrating the feasibility of producing high quality crystals. This test validated the concept of the non-dispersive geometry, for which the energy bandwidth crucially depends on the optical quality of the crystals, in contrast to the dispersive geometry. A full set of high quality crystals could not be manufactured yet, therefore the results presented here do not represent the optimal performance of the instrument. We report in Section~\ref{sec:comm-eres} the results obtained with Si(111) and Si(110) crystals of 25\(\times\)80 mm\(^2\) surface area and 1~m bending radius produced at the ESRF.

We followed two manufacturing processes, a single machining and a double machining. A schematic view is reported as supplementary Figure~S10. The single machining process consists in bending the crystal wafer on a cylindrical substrate of radius 2R and machining the surface to a radius R. This approach is relatively simple and results in an analyzer thicker at the borders than at the centre, thus requiring to bend a thick crystal. The second approach consists in machining the two sides of a crystal to a thin meniscus (150--250~$\mu$m) of radius 2R, and bending it afterwards on a substrate of radius R. Although more time-consuming, this procedure provides uniformly thin crystals. The double machining process is the method of choice for reaching R~$\le$~500 mm, as plastic deformation methods would be needed to bend a single machined crystal.

\begin{figure}
  \centering
  \includegraphics[width=0.48\textwidth]{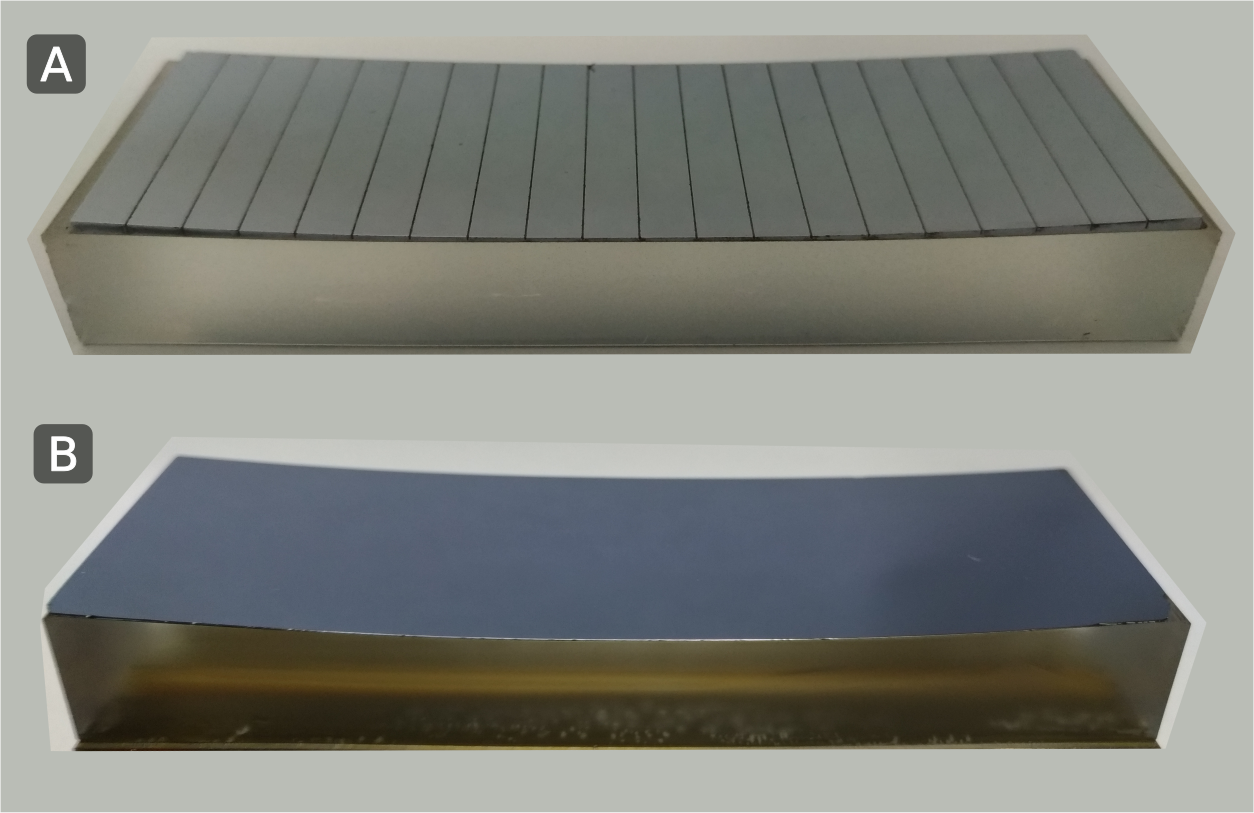}
  \caption{\label{fig8-analyzers}Johansson cylindrical crystal analyzers of 25\(\times\)80 mm\(^2\) size and 1~m bending radius as produced at the ESRF:\@ single-machined (A) and double-machined (B).}
\end{figure}

Figure~\ref{fig8-analyzers} shows the two types of 1~m bending radius analyzers produced at the ESRF. For the single machined one, the starting wafer has a thickness of 1.2~mm and is cut in stripes of 4~mm along the short side, with a groove of 1~mm height (incomplete cuts). Then, the Si wafer is bent on a borosilicate glass substrate with a tested anodic bonding procedure~\cite{Rovezzi:2017_RSI}. The grooves allow releasing part of the strain, resulting in a high quality bending on the substrate. In the absence of grooves, the underlying substrate cracks after anodic bonding. The machining removes 0.8~mm at the centre of the analyzer. This is performed by grinding the wafer with a silicon carbide powder from 17~\(\mu\)m down to 1~\(\mu\)m in size. The strain and damages introduced by the grinding are released via etching in a solution of nitric and hydrofluoric acid (20:1) for 40~minutes. At the end, the surface is polished with a 1~\(\mu\)m diamond powder. For the double machined crystals, the process starts with a thick 3~mm Si wafer. The convex face of radius 2R~=~1~m is prepared first with a grinding plus etching and polishing steps, as described previously. This face is temporary glued with bee wax on a glass substrate of the same radius and the concave face is prepared with grinding and polishing. A meniscus of 250~$\mu$m in thickness is obtained and afterwards bonded to the borosilicate glass substrate of radius R~=~0.5~m through anodic bonding. The final step is an etching plus polishing of the surface.

\subsection{Initial spectrometer alignment}\label{sec:comm-align}
The first step of the spectrometer alignment consists in aligning all mechanical parts with a precision of \(\le\)100~\(\mu\)m. This is performed once at the commissioning stage using a laser tracker, which gives an absolute precision in space (X, Y and Z) of 10~\(\mu\)m at best. The translation stages that move the entire spectrometer along Y and Z allow bringing the incoming X-ray beam into the source volume of the spectrometer. The alignment of the beam footprint on the sample along X is important to achieve the optimal performance, and is assured by carefully considering the dimensions of the sample mount and sample thickness. We plan in a future upgrade to align the sample surface along X by an in-line optical system in which the focus coincides with the spectrometer source volume.

The \(\theta\) rotation (TR) and all crystal analyzer mounts are calibrated at 90$^\circ$ with a precision below 5~\(\mu\)rad using an inclinometer. The centre of the detector window is positioned on the Rowland circle by introducing an offset in the control system representing the origin of the two inclined translation axes defining the trajectory space (DY, DZ). The $\chi$ angles of each crystal are aligned using a laser. This is performed only once, when mounting the analyzers modules on the pantograph.

The final alignment is performed with X-rays. Once a strong fluorescence line is excited, the spectrometer is driven to the tabulated emission energy and then the \(\theta\)-angle of each crystal is scanned with the other crystals moved out of diffraction, to find the maximum intensity of the line. At the end of this procedure, each crystal has its own \(\theta\) offset that is stored in a lookup table and kept constant during the emission scan of the spectrometer. We found that this is the only fine alignment that needs to be performed for each fluorescence line or group of lines. We did not find any benefit in fine tuning the vertical correction for each analyzer.

The bending radii of the crystals are not necessarily identical to those of the substrate. For a new set of analyzers, the actual bending radius is found by optimizing the diameter of the Rowland circle (2R). The procedure which we followed consists in measuring the emission line of an element against the 2R value of the Rowland circle for each analyzer. The width and height of the lines are extracted through peak fitting. The best 2R value is at the minimum of a quadratic fit. An example of this procedure applied to sulfur K$\alpha_{1,2}$ lines is shown in supplementary Figure~S13. For the same production series of analyzers, the optimized 2R is usually found with a variance below 5~mm, thus the average value is set in the control software for all analyzers and is kept fixed for all Bragg angles.

\subsection{Instrumental energy resolution}\label{sec:comm-eres}
The figure of merit of the spectrometer performance is the instrumental energy broadening as a function of the Bragg angle. It can be obtained with two Gaussian deconvolution methods, either from elastic peak scans (1) or from the emission lines (2). The first method is usually employed for high energy at synchrotron radiation facilities, where a scanning monochromatic beam is readily available, as we have adopted previously~\cite{Rovezzi:2017_RSI}. The second method was historically applied to laboratory-based instruments. Nowadays there is a re-gained interest in laboratory spectrometers, thus it is worth harmonising the procedure employed. In the tender X-ray range and for the chosen 90$^\circ$ scattering geometry, the elastic scattering is weak. Furthermore, the incoming beam double Si(111) monochromator has a relatively large bandwidth (\emph{e.g.} 0.3~eV at 2 keV). For these reasons we adopt the following procedure.

First, one measures an emission spectrum, composed of a single or multiple lines. The spectrum is fitted with Voigt profiles, that is, a convolution of Gaussian and Lorentzian profiles, accounting for the instrumental and intrinsic broadening, respectively. The instrumental/Gaussian width, $w_{G}$, is taken from the full-width-at-half-maximum (FWHM) of the Voigt profile, $w_{V}$, using the modified Whiting relation~\cite{Olivero:1977_JQSRT}: $w_{V} \approx 0.5346 w_{L} + \sqrt{0.2169 w_{L}^2 + w_{G}^2}$, where the intrinsic/Lorentzian width, $w_{L}$, is kept fixed at the tabulated value. If the measured emission spectrum contains two lines (\emph{e.g.} K\(\alpha_{1,2}\), L\(\alpha_{1,2}\)), the energy separation and peak ratio of the two lines are kept fixed at tabulated values, not fitted. The peak-fitting was performed with the \emph{Lmfit} code~\cite{scisoft:lmfit} and the signal of each wire was taken separately. The details of the procedure are reported in the supplementary section~S11. The assumption made in the peak fitting for the Lorentzian component results in negligible error bars on the fitted variables. Nevertheless, a systematic error is introduced assuming a Gaussian profile for the spectrometer response function. Furthermore, the overall energy alignment of the 11 analyzers may not be perfect. For these reasons we decided to report as error bars the standard deviation of the results obtained for each wire signal and for each analyzer crystal mounted on the 11 holders (supplementary Figure~S12C).

\begin{figure}
  \centering
  \includegraphics[width=0.48\textwidth]{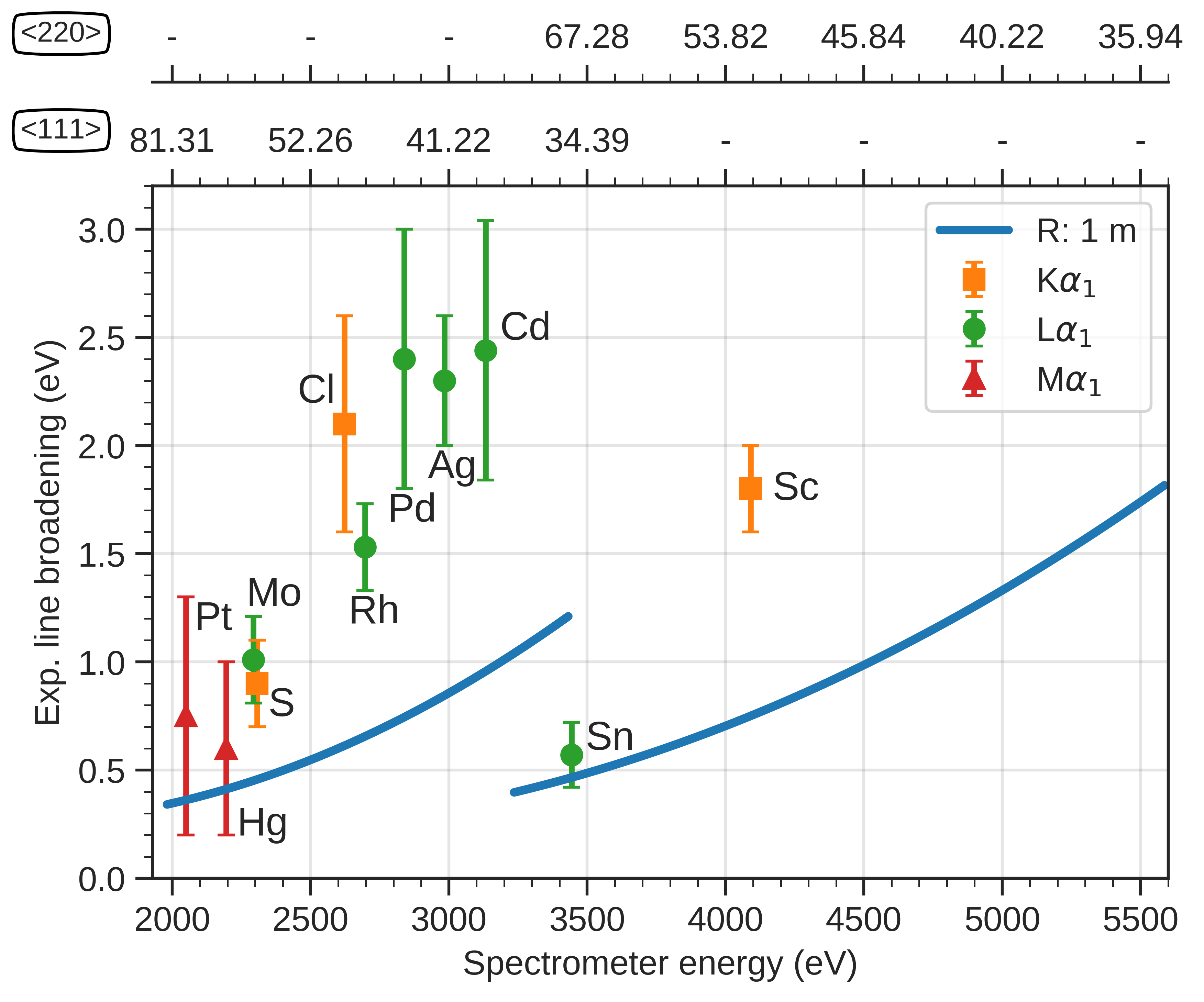}
  \caption{\label{fig9-eres}Spectrometer resolution function de-convoluted from the measured emission lines (symbols) and compared with the expected performances simulated with ray tracing (lines).}
\end{figure}

The results of the fits are shown in Figure~\ref{fig9-eres} and reported in Table~\ref{tab1}. A series of emission lines from K, L and M edges were measured with Si(111) and Si(110) crystals, sampling the entire angular range. The averaged data are shown in supplementary Figure~S11. Metallic foils were used as samples, when available. Otherwise, the emission lines from simple compounds (\emph{e.g.} oxides) were used. The broadening reported here may be larger than the real instrumental broadening, because line splittings (\emph{e.g.} multiplets) are not considered in the analysis.

The experimental line broadening increases from 0.5~eV to 2.5~eV as the Bragg angle decreases. Comparison of the experimental results with ray tracing simulations (Section~\ref{sec:concept}) shows that the angular dependence of the spectrometer resolution is reproduced in the simulations. The measured energy bandwidth is compatible with calculation down to a Bragg angle of 70\(^\circ\), but deviates from calculation at lower angles reaching approximately twice the simulated values at 35\(^\circ\). We attribute this discrepancy to the high residual strain in the single-machined crystals because of the thickness gradient. This was verified at high energy by measuring the Cu K$\alpha_{1,2}$ lines with two R\(=\)1~m Si(111) analyzers produced via double machining, one from a commercial provider (C) and another from the ESRF (E), using the same experimental set-up. The Cu K$\alpha_{1,2}$ lines were measured with the Si(444) and Si(333) reflections at Bragg angles of 79.309\(^\circ\) and 47.475\(^\circ\), respectively (Table~\ref{tab1}). The ESRF and commercial analyzers performed equally well at high Bragg angle (1.3(2)~eV \emph{vs} 0.96(40)~eV), whereas the ESRF one had a lower performance at low Bragg angle (3.3(3)~eV \emph{vs} 1.83(40)~eV). The reason for this discrepancy lies in the deviation between the crystal surface and substrate bending radius. The travel ranges of the spectrometer stages did not suffice to test the ESRF double-machined crystal at the optimal bending radius. However, the excellent performance of the commercial analyzer at high energy over the full Bragg angular range is not confirmed at low energy for sulfur (Si(111)\(_{\rm C}\) in Table~\ref{tab1}). The energy bandwidth is 0.7(1)~eV, which is more than 0.2~eV above the simulated value of 0.45~eV . We attribute this discrepancy to the optical quality of the crystal surface. We think that the energy bandwidth can be further reduced by improving the machining process.

\begin{table}
  \begin{center}
    \caption{Peak-fitting results for a series of emission lines covering the whole Bragg angular range (\(\theta_{\rm B}\)), partially shown in Figure~\ref{fig9-eres}.\\ {\footnotesize The analyzers without a label are single-machined, whereas those double-machined are labelled by the provider: commercial (C) or ESRF (E). The samples consist of powders (compound or metal) pressed into pellets or metallic foils. Last two columns report, respectively, the intrinsic broadening of the line (\(w_{L}\)) and the experimental contribution (\(w_{G}\)). The error bar on the last digit is reported in parentheses.}}\label{tab1}
    {\rowcolors{1}{gray!20}{white}
      \begin{tabular}{llllllll}
        \toprule
        \rowcolor{white} El. &  Line &  Energy &  Analyzer         & \(\theta_{\rm B}\) & Sample & \(w_{L}\) &  \(w_{G}\) \\
        \rowcolor{white}     &       &  (eV)   &  Mat(\emph{hkl}) &     (\(^{\circ}\)) &        &              (eV) &                (eV) \\
        \midrule
        Pt & M\(\alpha_1\) & 2050.0 &             Si(111) & 74.667 & Pt\(_{\rm foil}\) & 2.39 & 0.75(55) \\
        Hg & M\(\alpha_1\) & 2195.0 &             Si(111) & 64.250 &              HgSe & 2.59 &   0.6(4) \\
        Mo & L\(\alpha_1\) & 2293.2 &             Si(111) & 59.550 & Mo\(_{\rm foil}\) & 1.81 &   1.0(2) \\
         S & K\(\alpha_1\) & 2307.8 &             Si(111) & 58.945 &        ZnSO\(_4\) & 0.61 &   0.9(2) \\
         S & K\(\alpha_1\) & 2307.8 & Si(111)\(_{\rm C}\) & 58.945 &               ZnS & 0.61 &   0.7(1) \\
        Cl & K\(\alpha_1\) & 2622.4 &             Si(111) & 48.929 &               KCl & 0.68 &   2.1(5) \\
        Rh & L\(\alpha_1\) & 2696.8 &             Si(111) & 47.147 & Rh\(_{\rm met.}\) & 2.17 &   1.7(2) \\
        Pd & L\(\alpha_1\) & 2838.6 &             Si(111) & 44.145 & Pd\(_{\rm foil}\) & 2.31 &   2.4(6) \\
        Ag & L\(\alpha_1\) & 2984.4 &             Si(111) & 41.487 & Ag\(_{\rm foil}\) & 2.45 &   2.3(3) \\
        Cd & L\(\alpha_1\) & 3133.8 &             Si(111) & 39.114 & Cd\(_{\rm foil}\) & 2.58 &   2.4(6) \\
        Sn & L\(\alpha_1\) & 3444.0 &             Si(111) & 35.033 & Sn\(_{\rm met.}\) & 2.87 &   3.6(6) \\
        Sn & L\(\alpha_1\) & 3444.0 &             Si(220) & 69.622 & Sn\(_{\rm met.}\) & 2.87 &   0.6(2) \\
        Sn & L\(\beta_3\)  & 3750.3 &             Si(220) & 59.413 & Sn\(_{\rm met.}\) & 5.70 &   1.9(6) \\
        Sc & K\(\alpha_1\) & 4090.6 &             Si(220) & 52.115 &   Sc\(_2\)O\(_3\) & 1.06 &   1.8(2) \\
        Cu & K\(\alpha_1\) & 8047.8 & Si(444)\(_{\rm C}\) & 79.309 & Cu\(_{\rm foil}\) & 2.10 & 0.96(40) \\
        Cu & K\(\alpha_1\) & 8047.8 & Si(444)\(_{\rm E}\) & 79.309 & Cu\(_{\rm foil}\) & 2.10 &   1.3(2) \\
        Cu & K\(\alpha_1\) & 8047.8 & Si(333)\(_{\rm C}\) & 47.475 & Cu\(_{\rm foil}\) & 2.10 & 1.83(40) \\
        Cu & K\(\alpha_1\) & 8047.8 & Si(333)\(_{\rm E}\) & 47.475 & Cu\(_{\rm foil}\) & 2.10 &   3.3(3) \\
        \bottomrule
      \end{tabular}
    }
  \end{center} 
\end{table}

\subsection{First results for HERFD-XANES}
\begin{figure}
  \centering
  \includegraphics[width=0.48\textwidth]{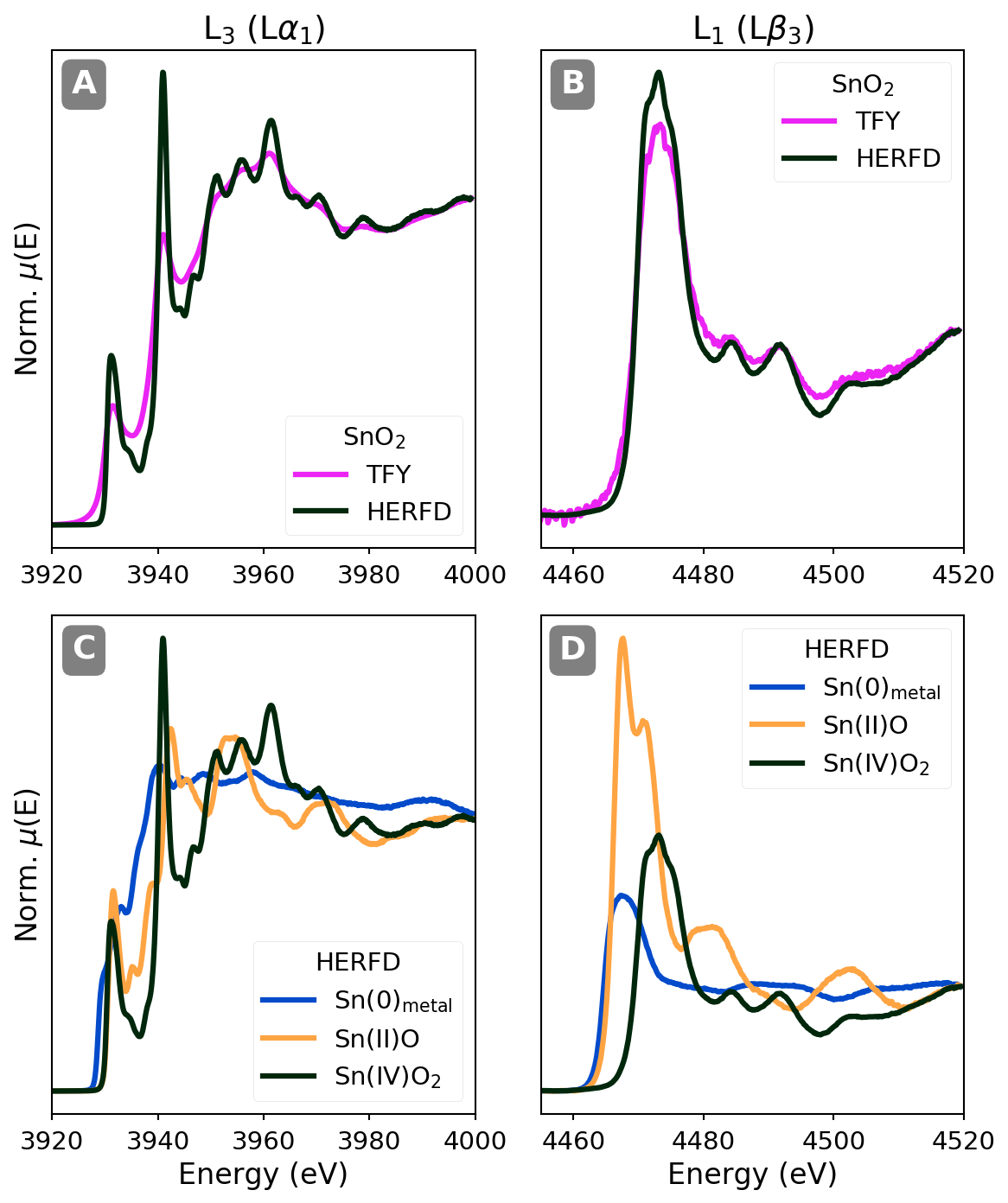}
  \caption{\label{fig10-sn}Example of collected data on Sn at L\(_3\) and L\(_1\) edges: (A, B) HERFD compared to TFY for SnO\(_2\); (C, D) HERFD spectra for three nominal oxidation states.}
\end{figure}

The non-dispersive geometry was chosen because an important application of the instrument is to perform HERFD-XANES spectroscopy. An example at the L\(_3\) and L\(_1\) edges of Tin (Sn) is shown in Figure~\ref{fig10-sn}. The experimental energy broadening obtained with the Si(110) crystals is 0.6(2)~eV for the L\(\alpha_1\) (L\(_3\)-M\(_5\)) line and 1.9(6)~eV for the L\(\beta_3\) (L\(_1\)-M\(_3\)) line. The natural bandwidths of these lines are 2.87 eV (L\(\alpha_1\)) and 5.7 eV (L\(\beta_3\)). This introduces a sharpening effect on the XANES features that is clearly visible when compared to the total fluorescence yield (TFY) measurements (Figure~\ref{fig10-sn}A,B). The additional spectral features that become visible provide important information when comparing the experimental data to quantum chemical calculations.

The gain in energy resolution provided by the spectrometer also allows one to suppress the parasitic fluorescence from elements in the sample having an absorption edge just below the edge of the target element, \emph{e.g.} As K-edge parasitic fluorescence when measuring at the Au L\(_3\) edge~\cite{Merkulova:2019_ESC}. At low energy, the line separation can be used to measure the L\(_1\) edge of an element without interference of the L\(_3\) and L\(_2\) signals. As an example, metallic Sn has the nominal electronic configuration of 4\(d^{10}\)5\(s^{2}\)5\(p^{2}\) and the highest oxidation state is 4\(^+\). This means that the 4\(d\) shell is formally always filled, thus the L\(_3\) edge has little sensitivity to the Sn valence shells. Therefore, correlating the edge position to the nominal oxidation state is difficult at the L\(_3\) edge (Figure~\ref{fig10-sn}C). In contrast, it is straightforward at the L\(_1\) edge (Figure~\ref{fig10-sn}D) because dipole transitions to the 5\(p\) orbitals directly probe their occupation and energy and thus the oxidation state. The L\(_3\) edge probes the unoccupied \(s\) and \(d\) orbitals that are strongly sensitive to the ligand environment as shown by the rich spectral features in Figure~\ref{fig10-sn}C with Sn and demonstrated previously for 5\(d^{10}\) in Hg~\cite{Manceau:2015_IC}. In general, probing several edges provides complementary information on the electronic structure and bonding environment, and is thus highly desirable. This is greatly facilitated using an emission spectrometer with high energy resolution covering a wide energy range in a single configuration. 

\section{Conclusions}
We have presented the design and performances of a tender X-ray emission spectrometer based on an array of 11 cylindrically bent Johansson crystals in non-dispersive scanning geo\-me\-try. This design ensures that the entire available solid angle (up to 87~millisterad for 0.5~m bending radius) is used within the operating energy of the instrument, thereby achieving high detection efficiency for background-free and HERFD XANES spectroscopy. An innovative mechanics for the table supporting the analyzer crystals allows optimizing the sagittal focusing dynamically with only two actuators. It can be adapted to a larger number of crystals and to different bending radii. A series of emission lines was measured to evaluate the performance of the instrument. Currently, the energy resolution is limited by the quality of the single-machined Johansson crystals produced at the ESRF. Higher resolution will be obtained in the near-future with their replacement with double-machined crystals. The performance of the instrument in terms of detection rate and background noise will then be presented for such optimized optics. We have reported an example of application of HERFD-XANES at the L$_3$ and L$_1$ edges of Sn. During the commissioning phase, the spectrometer was successfully employed for recording full RIXS planes, valence-to-core XES and HERFD-XANES on all the elements reported in Table~\ref{tab1}. Those measurements will be presented elsewhere.

\begin{acknowledgments}
We are indebted to the instrumentation services, administration, directorate and technical infrastructure divisions of the ESRF for help during the design, procurement and installation phases of the project. M.~Rovezzi is grateful to Jean-Louis Hazemann and the FAME group for their support during the commissioning of the instrument, data analysis and writing of the manuscript. A.~Svyazhin acknowledges the state assignment by the Ministry of Science and Higher Education of the Russian Federation (theme ``Electron'' No.~AAAA-A18-118020190098-5). This work was supported financially by the French National Research Agency (ANR) under Grant ANR-10-EQPX-27-01 (EcoX Equipex). We thank the five anonymous referees who helped us improving the quality of the manuscript with their constructive reviews.
\end{acknowledgments}
\bibliographystyle{apsrev4-1}
\input{refs_pap.bbl}

\part*{SUPPORTING INFORMATION}
\addcontentsline{toc}{chapter}{Supporting information}

\setcounter{equation}{0}
\setcounter{figure}{0}
\setcounter{table}{0}
\setcounter{page}{1}
\setcounter{section}{0}
\setcounter{subsection}{0}
\makeatletter
\renewcommand{\theequation}{S\arabic{equation}}
\renewcommand{\thefigure}{S\arabic{figure}}
\renewcommand{\thetable}{S\arabic{table}}
\renewcommand{\thepage}{S\arabic{page}}
\renewcommand{\thesection}{S\arabic{section}}
\renewcommand{\thesubsection}{S\arabic{subsection}}
\renewcommand{\bibnumfmt}[1]{\(^{\rm S#1}\)}
\renewcommand{\citenumfont}[1]{S#1}


\section{Detector alternatives to the proportional counter}
\label{sec:detector-alternatives}
We examined other solutions for the detector among exi\-sting commercial alternatives or promising projects in development. The most suited detector for such an application would be a scientific-quality large area charged-coupled device (CCD). CCDs are widely used in soft X-ray RIXS spectrometers for their low electronic noise and small pixel size. Furthermore, it is possible to run them in spectroscopy mode~\cite{Zhao:2017_RSI}, enhancing the possibility to reduce the background level of the measured signal. Commercial sensors have a standard size of approximately 27~\(\times\)~27~mm\(^2\), that is, 2048 square pixels of 13~\(\mu\)m side. The drawback of such sensors is  the read-out time, being as slow as 500~ms for unbind mode, well above the required 2~ms. Fast readout and large area CCDs exist, like the pnCCD~\cite{Struder:2000_NIMA} or the fastCCD~\cite{Doering:2011_RSI}. These extremely high performance detectors, initially developed for astro\-no\-mi\-cal applications in space telescopes, were out of our budget. An alternative to CCDs are silicon drift detectors (SDDs). Typical sensors sizes are of 50~mm\(^2\) area, thus to cover the required area of 2000~mm\(^2\) one could opt for a multi-element array~\cite{Hafizh:2019_JINST} or pixelated sensors~\cite{Bufon:2014_JINST}. Nevertheless, this solution is challenging and was not available at the time of the design. Hybrid pixel detectors, that is, a pixelated sensor (usually Si) bump-bonded to a large area CMOS have also demonstrated some applications in the low energy side~\cite{Donath:2013_conf,Klackova:2019_JINST}. Recently, direct reading of mass production CMOS cameras was shown as a possible alternative for tender X-rays~\cite{Holden:2018_RSI, Haro:2019_RPC}. The drawback of these CMOS-based detectors is their quantum efficiency, usually below 20\%. In fact, either the high noise level imposes an electronic threshold that barely allows the signal to be measured (the case of hybrid pixel detectors) or the active layer of the CMOS is too thin (the case of direct CMOS reading).

\section{Naming convention for wire signals and analyzer modules}
\label{sec:eres-naming}
\begin{figure*}[!htb]
  \centering
  \includegraphics[width=0.65\textwidth]{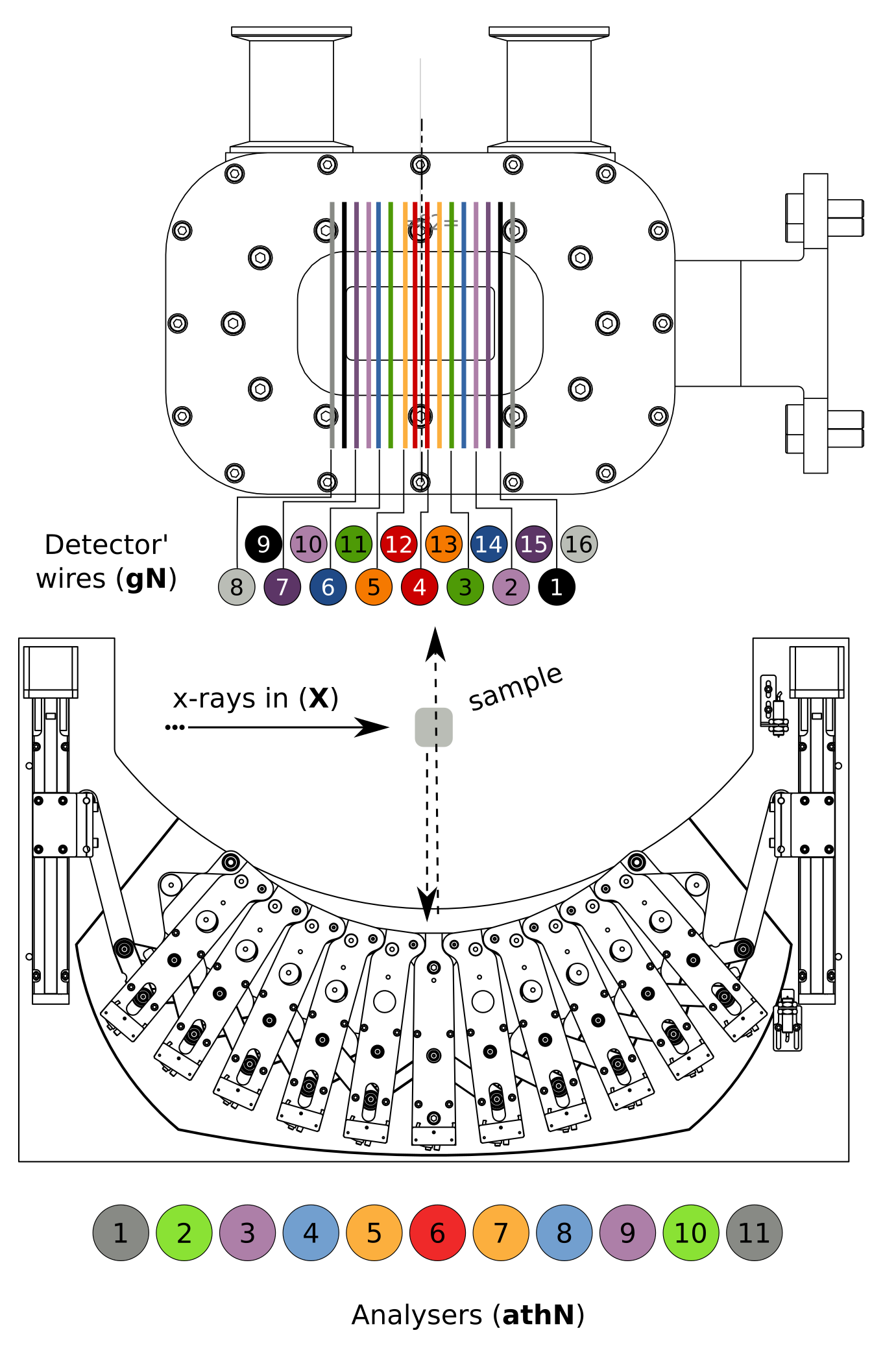}
  \caption{\label{figS1-wires-aths}Naming of the detector wires signals (gN) and analyzer modules (athN) used during the commissioning and data analysis. Detector and analyzer table are given in a projected view such as looking from the analyzer door of the spectrometer, that is, the incoming X-ray beam is coming from the left side. Dimensions are not to scale. An independent colour coding for the analyzer table and detector is used to show the central symmetry.}
\end{figure*}

The naming convention for the detector wires (gN) and analyzers modules (athN) is given in Figure~\ref{figS1-wires-aths}. The figure allows understanding the origin of the signal detected on each wire and the arrangement of the analyzers with respect to the incoming X-ray beam (X direction).

\section{Detector trajectories and hoses}
\label{sec:detector-suppl}
\begin{figure}[!htb]
  \centering
  \includegraphics[width=0.49\textwidth]{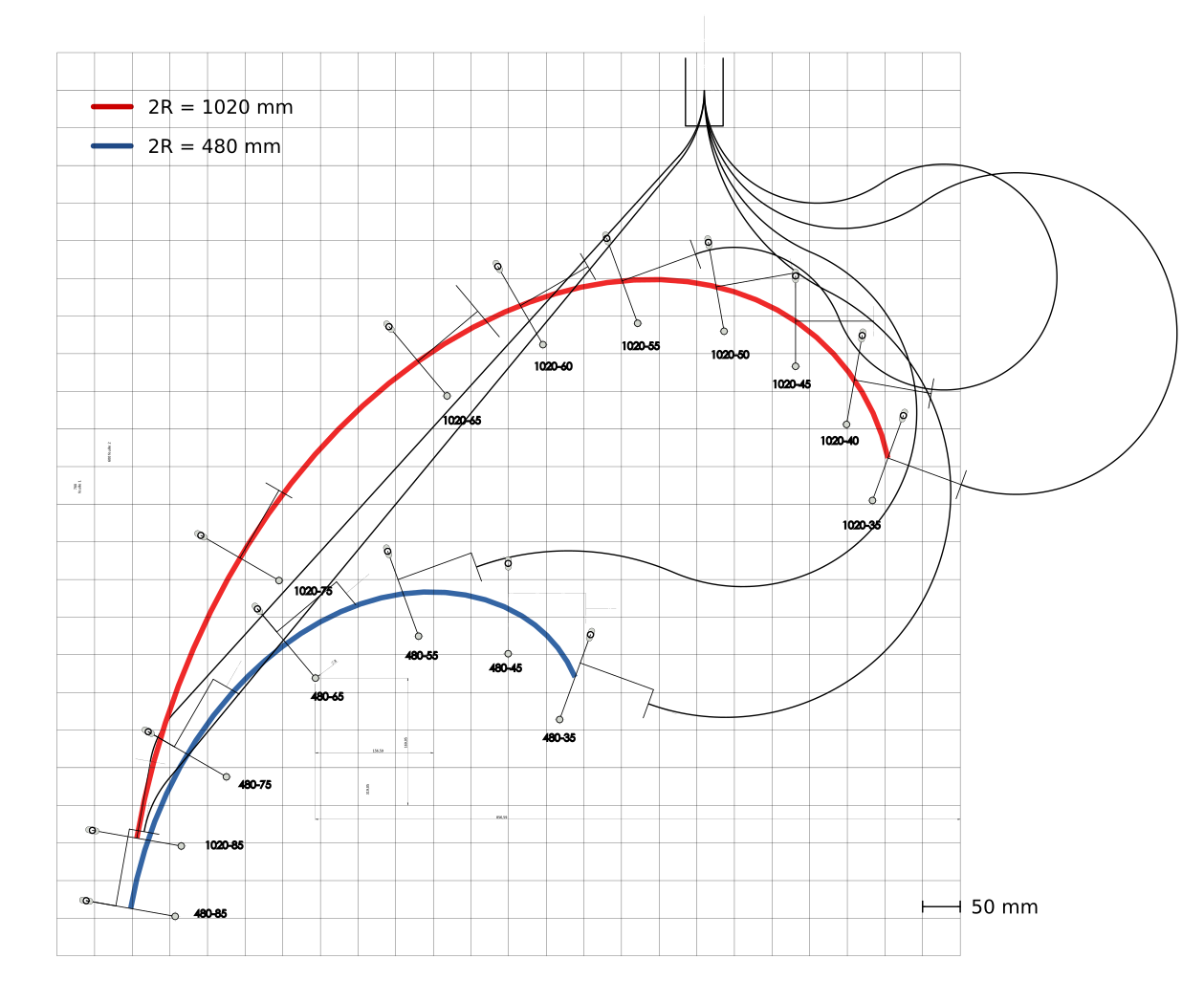}
  \caption{\label{figS2-det-trajs}Trajectories of the detector chamber with the simulation of the hoses movements for two extreme radii.}
\end{figure}

\begin{figure}[!htb]
  \centering
  \includegraphics[width=0.49\textwidth]{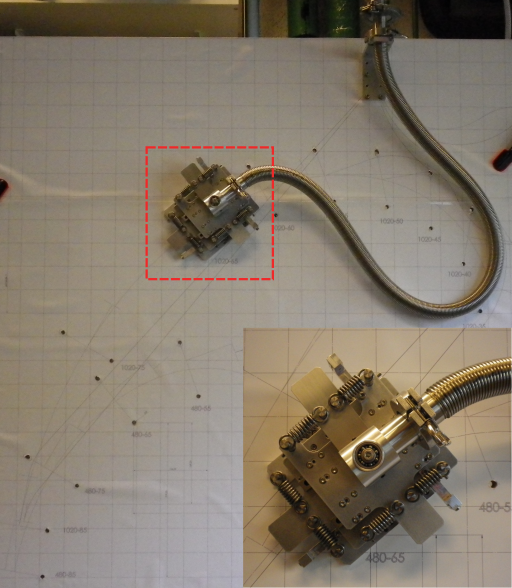}
  \caption{\label{figS3-det-hoses}Prototype mechanics for testing the metallic hoses in real conditions and measuring the forces acting on the detector arm.}
\end{figure}

The trajectories for 2R~=~1020~mm and 2R~=~480~mm, respectively, maximum and minimum allowed diameter of the Rowland circle are shown in Figure~\ref{figS2-det-trajs}. The simulation of the hoses position at given angular positions is also shown. Those simulations have been used for testing the behaviour of the metallic hoses and measuring the forces on the detector's arm, as shown in Figure~\ref{figS3-det-hoses}.

\section{Gas detector efficiency}
\label{sec:detector-efficiency}
\begin{figure}[!htb]
  \centering
  \includegraphics[width=0.49\textwidth]{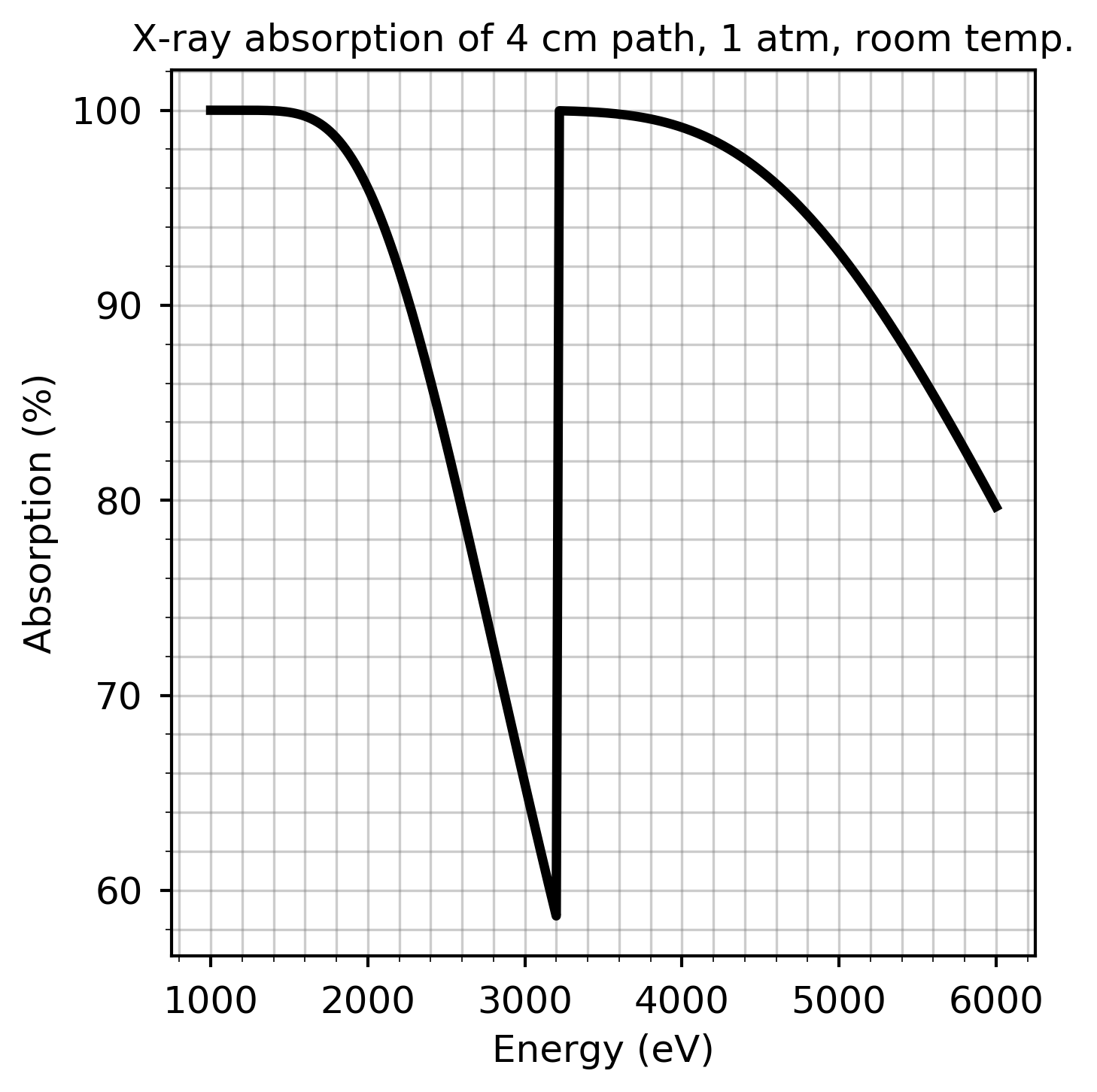}
  \caption{\label{figS4-det-abs}Gas detector efficiency in the energy range of interest.}
\end{figure}

The gas detector efficiency, that is, the X-ray absorption (in the energy range of interest) by the 15\% CO$_2$ in Ar mixture at 1~bar and at room temperature in the 4~cm path is shown in Figure~\ref{figS4-det-abs}.

\section{Gas detector linearity}
\label{sec:detector-linearity}
\begin{figure*}[!htb]
  \centering
  \includegraphics[width=0.65\textwidth]{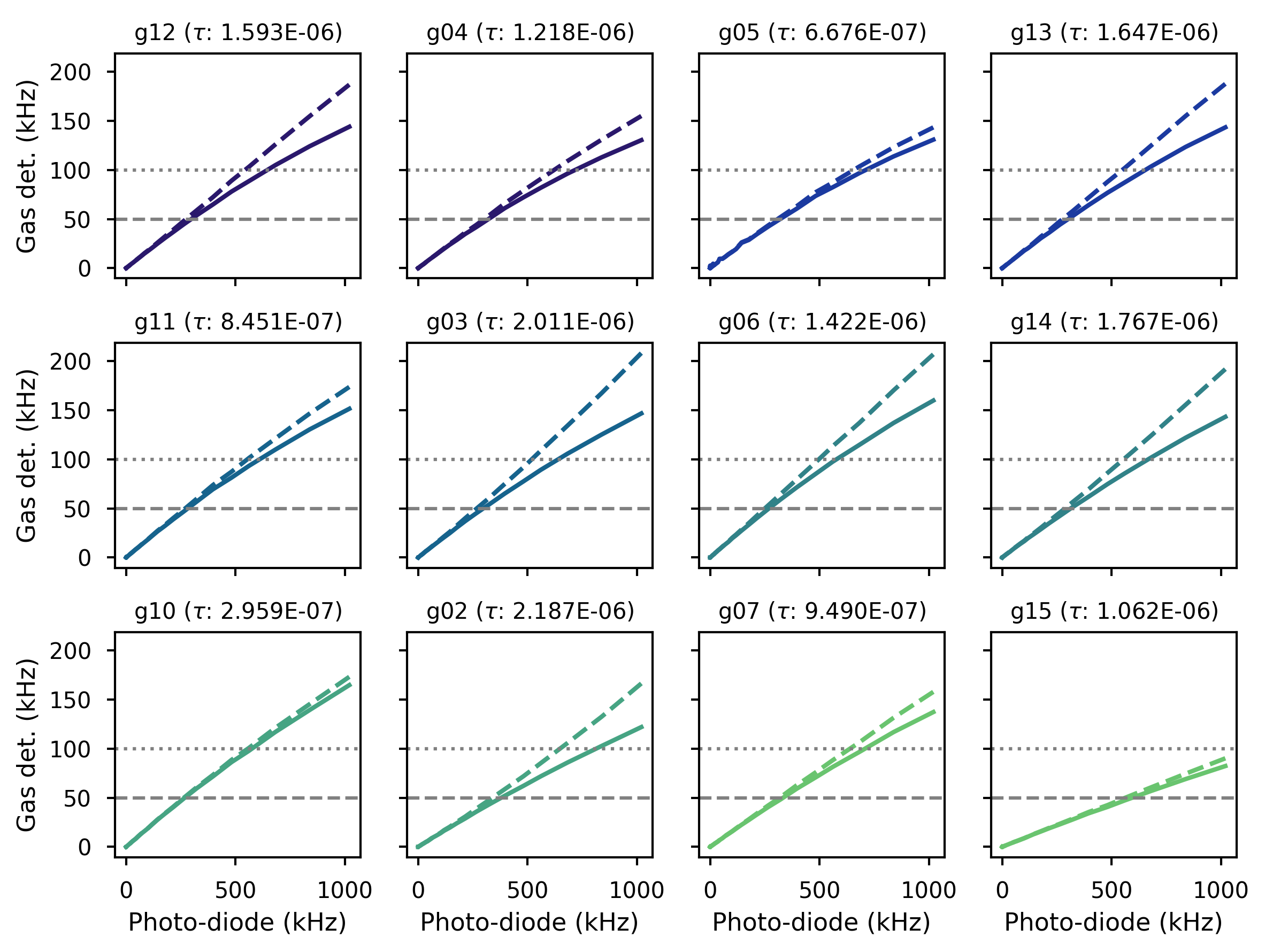}
  \caption{\label{figS:det-lin}Typical single-wire signal response of the gas detector as a function of the incoming beam flux (photo-diode signal) in kHz (1000 counts/s). The dashed lines are the dead-time corrected signal, using a dead-time constant ($\tau$) fitted in the linear response region below 25~kHz. The two horizontal dashed and dotted lines are a guide for the eye, respectively, at 50~kHz and 100~kHz.}
\end{figure*}

The gas detector linearity as a function of the incoming beam flux is measured for each wire at each experimental session, as shown in Figure~\ref{figS:det-lin}. The intensity of the incoming beam is varied by closing the incoming beam slits or scanning the undulators' gap. In fact, in the tender X-ray energy range the use of filters to attenuate the beam is not possible because of their strong absorption. A proportional signal is measured via the current of a Silicon photo-diode reading the back-scattering signal from a Kapton foil of 8~$\mu$m thickness (Figure~\ref{figS:samp-env}C.4). The dead time constant ($\tau$) for each wire of the gas detector is found by fitting the response in a linear region, that is, below 25~kHz (counts/s) per wire, using the standard formula of the nonparalyzable model (\onlinecite{Knoll:2000_book}, pag. 122): $m = kn/(1+kn\tau)$, where $m$ is the gas detector signal in the linear region, $n$ is the photo-diode signal, assumed as the true signal of the incoming beam, and $k$ a constant. In fact, the absolute number of counts of the photo-diode are not the absolute number of incoming photons/s.

\section{Sample environment}
\label{sec:sample-env}
\begin{figure*}[!htb]
  \centering
  \includegraphics[width=0.9\textwidth]{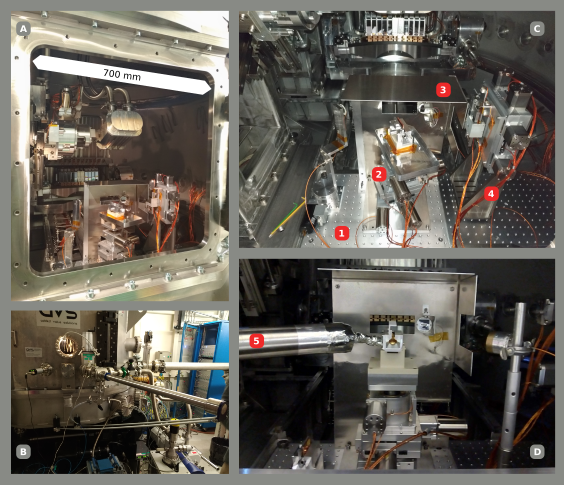}
  \caption{\label{figS:samp-env}Sample environment space. Front door (A). External view of the load-lock system (B). Internal views (C and D): base breadboard (1), sample tower (2), shielding for blocking background signals (3), incoming beam slits and intensity monitor (4), head of the liquid helium cryostat (5).}
\end{figure*}

\begin{figure*}[!htb]
  \centering
  \includegraphics[width=0.6\textwidth]{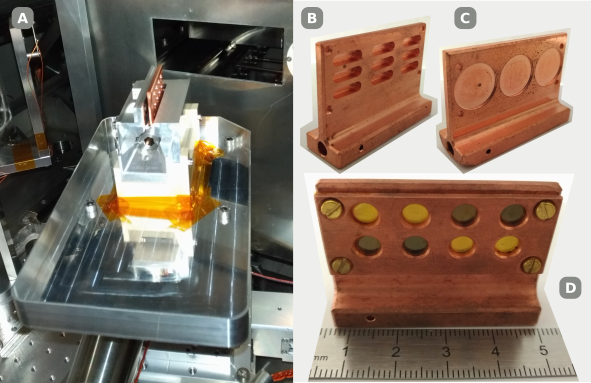}
  \caption{\label{figS6:samp-hold}Receptacle of the sample holder at the top of the sample goniometer tower (A). Example of sample holders copper blocks: Cells for frozen liquids (B), circular pellets of 13~mm diameter (C) and 5~mm diameter, showing eight mounted pellets sealed with a Kapton foil kept in place by a front mask (D).}
\end{figure*}

The sample environment space is shown in Figure~\ref{figS:samp-env}. Examples of the sample holders are shown in Figure~\ref{figS6:samp-hold}.

\section{\emph{Operando} cell for gas sensors}
\label{sec:david-cell}

\emph{In-situ} and \emph{operando} XAS are an essential method to understand synthesis-structure-function-relationships of metal oxide based gas sensing materials \cite{Barsan:2007_SAB, Degler:2018_Sensors}. The elements of interest are those from base materials such as SnO$_2$, In$_2$O$_3$ and WO$_3$, and additives like Rh, Pd, Pt and Au. In case of 5\emph{d}-elements there are various examples using HERFD-XAS to study the behavior of additives during gas sensing \cite{Degler:2016_ACSSens, Hubner:2011_Angewandte}, while in case of 4\emph{d}-elements K-edges were studied using conventional XAS \cite{Safonova:2005_ChemCom, Koziej:2009_PCCP, Staerz:2018_Nanomaterials}. Extending the spectral range to tender X-rays allows studying the L-edges of 4\emph{d}-elements and will provide unprecedented insights in the gas sensing process of pristine, doped and loaded metal oxide gas sensing materials.

\begin{figure*}[!htb]
  \centering
  \includegraphics[width=0.99\textwidth]{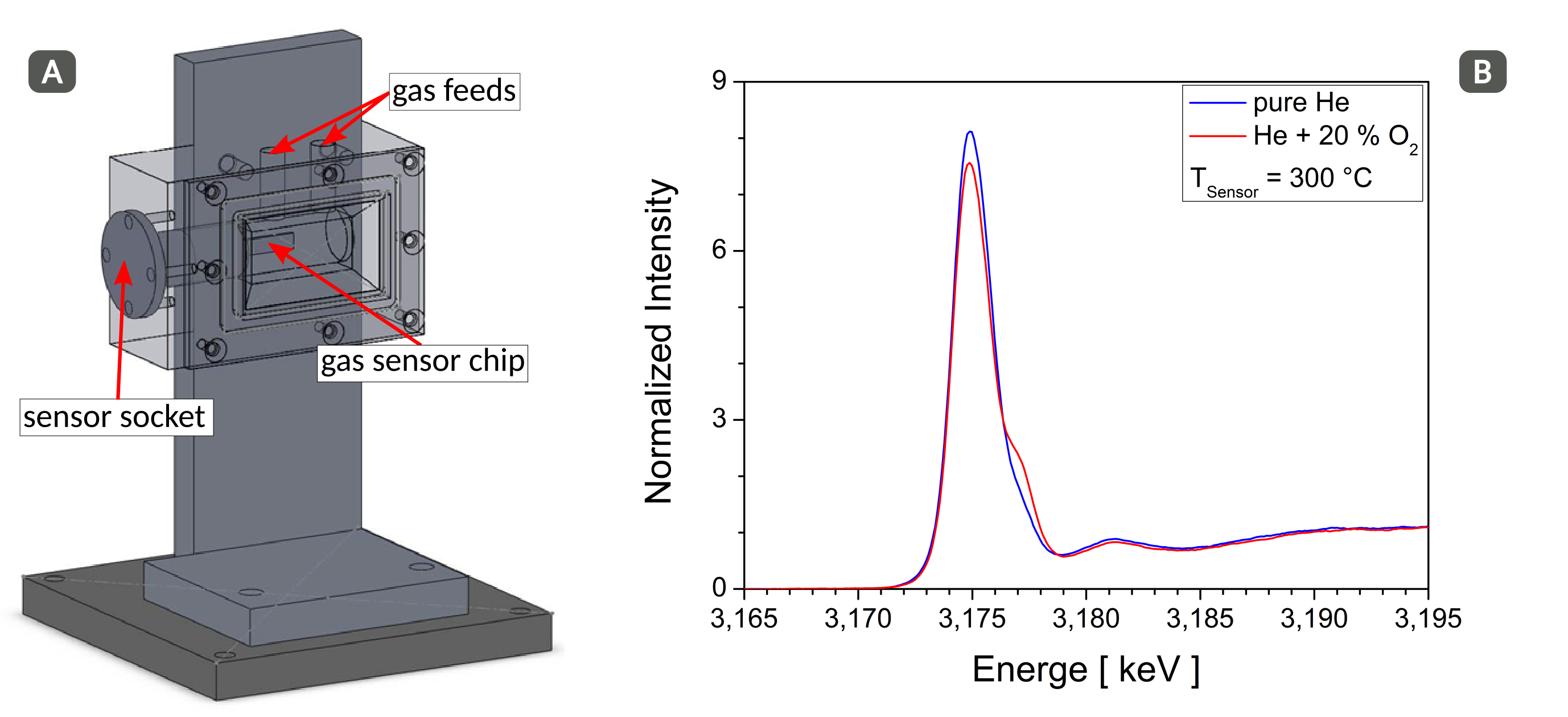}
  \caption{\label{figS:opcell}Schematic set-up of the operando XAS cell for gas sensors (A) and measurements at Pd L\(_3\)-edge.}
\end{figure*}

For TEXS a suitable operando gas sensor cell (Figure~\ref{figS:opcell}A) was designed, which allows dynamically dosing different gas flows. The sensor socket is designed to mount research-type gas sensor substrates~\cite{Barsan:2003_JPCM} and is equipped with electrical connections to supply a heating voltage and continuously read the sensor response during the experiment. Successful tests were done studying a 2 wt\% Pd-loaded SnO\(_2\) gas sensor at 300~\(^\circ\)C in different atmospheres (Figure~\ref{figS:opcell}B). The recorded HERFD-XANES spectra show a clear effect of the atmospheric composition on the Pd sites.

\section{Initial characterization of commercial analyzers}
\label{sec:jscyl-commercial}
\begin{figure*}[!htb]
  \centering
  \includegraphics[width=0.65\textwidth]{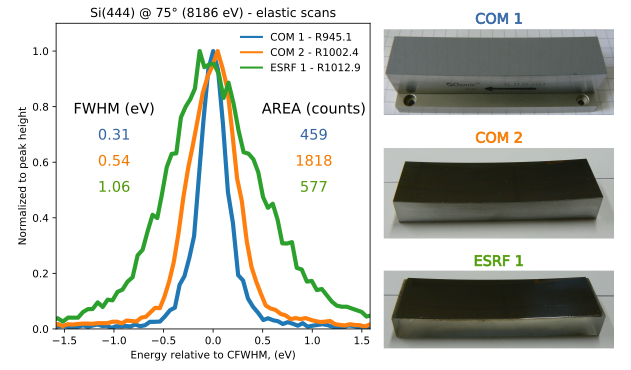}
  \caption{\label{figS:jscyl15} Elastic peak scans normalised to the peak height (left panel) for two commercially available Johansson analyzers (``COM 1'' and ``COM 2'') versus one at the ESRF (``ESRF 1'').}
\end{figure*}

A brief summary of the characterization of two selected commercial Johansson crystal analyzers (double machining) and one produced at the ESRF (single machining) is shown in Figure~\ref{figS:jscyl15}. The measurements were performed on the high energy spectrometer of ID26 at 8186~eV, using Si(444) reflection at a Bragg angle of 75\(^\circ\), following a well established procedure~\cite{Rovezzi:2017_RSI}. In summary, the elastic scattering peaks are collected at the best bending radius by scanning the energy of the incoming beam with a cryogenically cooled double Si(311) monochromator. During these measurements, a square avalanche photo diode detector of 10~mm side was used. This means that only a central area of 5~mm side along the flat direction was probed. The energy bandwidth measured with the ``COM 1'' analyzer corresponds to what is expected by a ray tracing simulation, that is, the experimental curve overlaps the simulated (normalized to peak maximum). The ``COM 2'' analyzer shows higher reflectivity, with an integrated area almost four times the area of ``COM 1''. The ESRF analyzer corresponds to a first production series of single-machined analyzers, that is, without grooves. This explains the lower performance with respect to the commercial analyzers. The grooves and the optimization of the surface polishing (as reported in the main text) have permitted to remove the side tails visible in the reflectivity curve.

\section{Machining cylindrical Johansson crystal analyzers}
\label{sec:jscyl-machining}
\begin{figure}[!htb]
  \centering
  \includegraphics[width=0.49\textwidth]{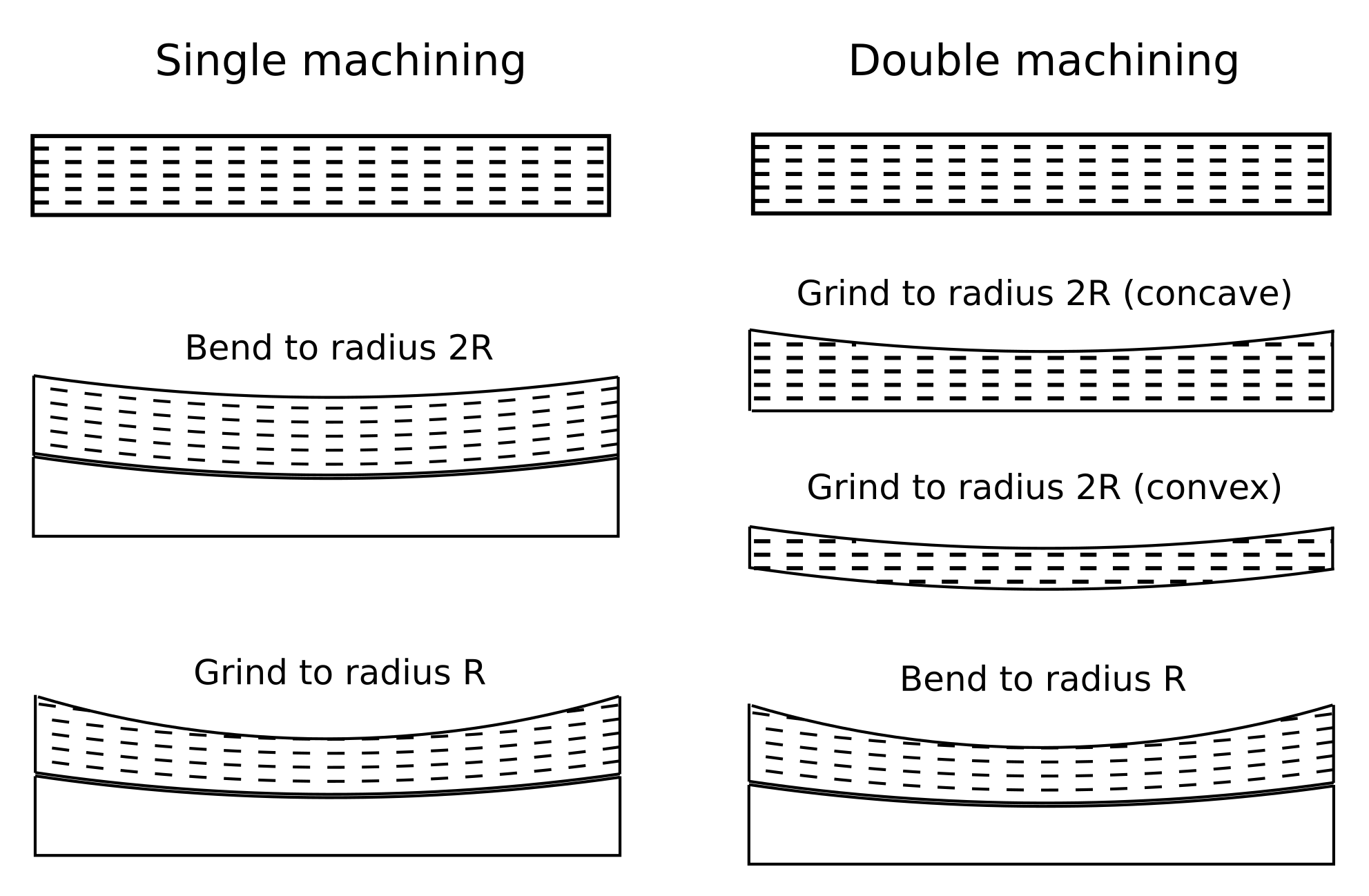}
  \caption{\label{figS:jscyl-machining}Machining approaches in Johansson crystal analysers production: single machining versus double machining. The dashed lines represent the atomic planes. Dimensions are not in scale.}
\end{figure}

The two standard machining approaches in the production of cylindrical Johansson crystal analyzers, namely, single and double machining, are schematized in Figure~\ref{figS:jscyl-machining}.

\section{Measured emission lines}
\label{sec:allXES}
\begin{figure*}[!htb]
  \centering
  \includegraphics[width=0.9\textwidth]{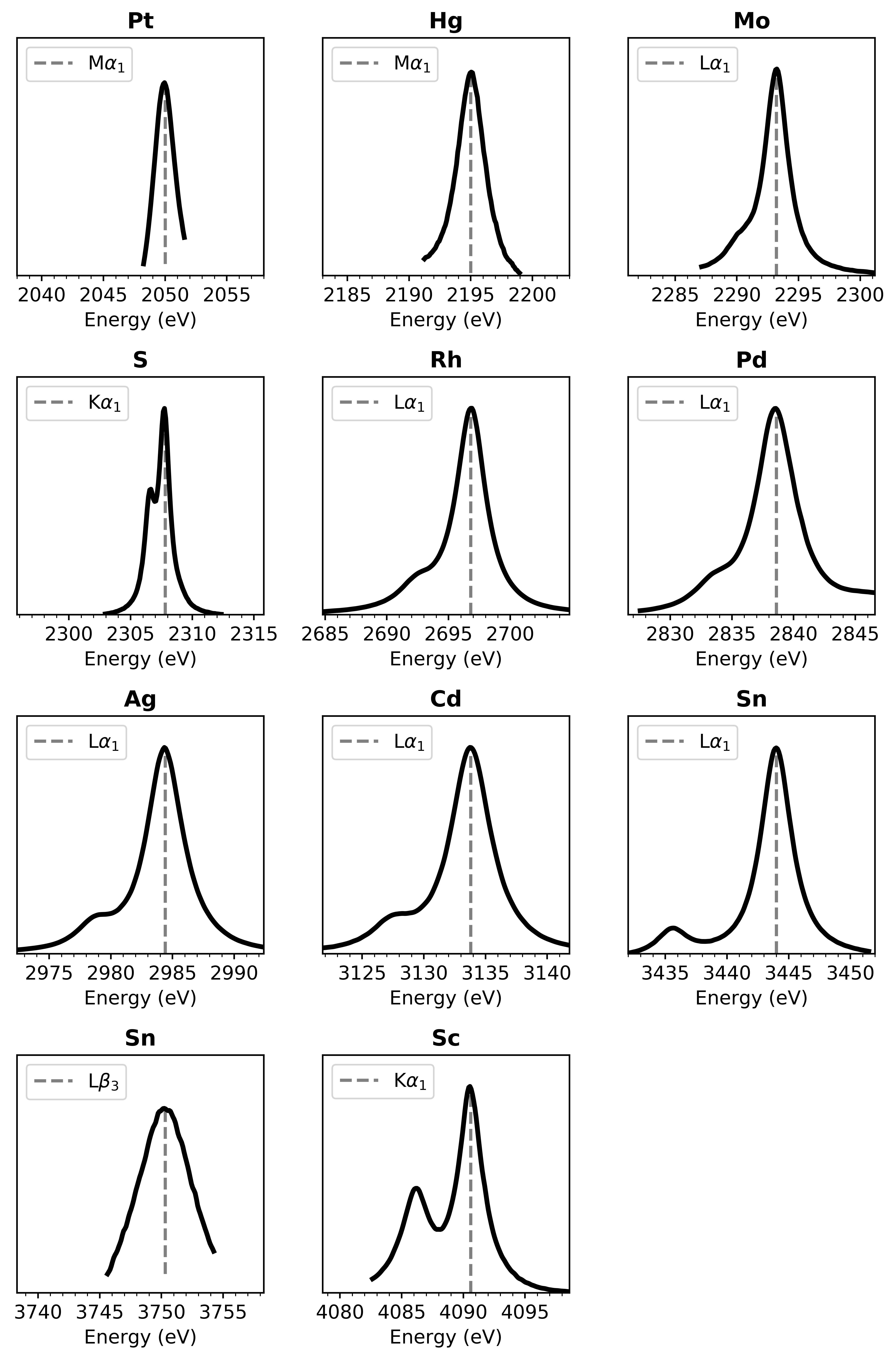}
  \caption{\label{figS:allXES}Measured emission lines. The energy range of all the plots is 20~eV.}
\end{figure*}

The measured emission lines reported in Table~1 of the main text are shown in Figure~\ref{figS:allXES}.

\section{Example of the data analysis workflow}
\label{sec:eres-fits}
\begin{figure*}[!htb]
  \centering
  \includegraphics[width=0.95\textwidth]{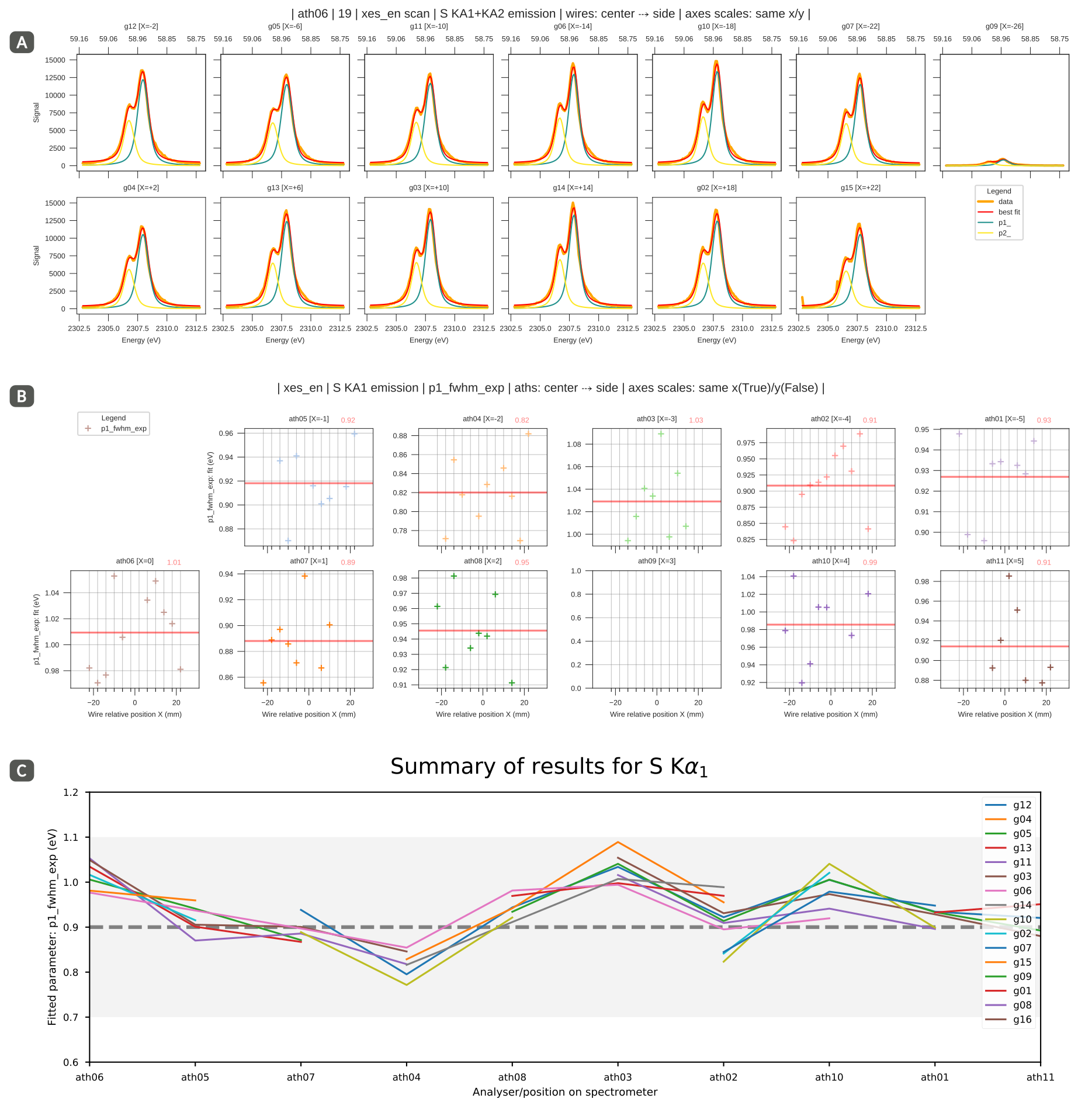}
  \caption{\label{figS:xesfit-S-KA12}Data analysis example for S K\(\alpha_{1,2}\) lines. Single-wire signal peak-fitting for a given analyzer, ath06 (A): the sub-plots have the same abscissa and coordinate scales. The distribution of the plots is from the central wires (g12, g04) to the side ones (g07, g15). The two rows of wires are symmetrically distributed. The signal from the side wires is partially cut by the detector entrance window. Evaluation of one fitted property, p1\_fwhm\_exp (B) and summary of the results (C). The horizontal dashed line is the average value and the gray area represents the experimental error bar.}
\end{figure*}

An example of the data analysis workflow, that is, the procedure we have employed to extract the instrumental energy resolution reported in Table 1 of the main text, is shown here for the S K$\alpha_{1,2}$ lines (Figure~\ref{figS:xesfit-S-KA12}). It is implemented in Python as a series of Jupyter notebooks and available on demand. The naming convention is given in Figure~\ref{figS1-wires-aths}. The first step of the analysis consists in a peak fitting (Figure~\ref{figS:xesfit-S-KA12}A): for each crystal analyzer mounted on the 11 holders of the spectrometer, the signal of each wire out of the sixteen available of the gas detector is analyzed independently. This corresponds to an average statistics of approximately 100 fits, out of the 176 possible, as not all wires read a good signal for each scan. In the second step (Figure~\ref{figS:xesfit-S-KA12}B), each property of the peak fitting, here the experimental full-width-at-half-maximum of the main peak (p1\_fwhm\_exp) is plotted per analyzer and following the spatial distribution of the wires. This step allows us to establish any issue related to mis-alignment of the spectrometer. It also allows to reject outliers. If no major problems are found, the results are taken all together (Figure~\ref{figS:xesfit-S-KA12}C) to establish the average reported value and the experimental standard deviation (error bar). We note that the error bar of the fit is negligible with respect to the experimental standard deviation.

\section{Best radius optimisation procedure}
\label{sec:eres-radius}
\begin{figure*}[!htb]
  \centering
  \includegraphics[width=0.95\textwidth]{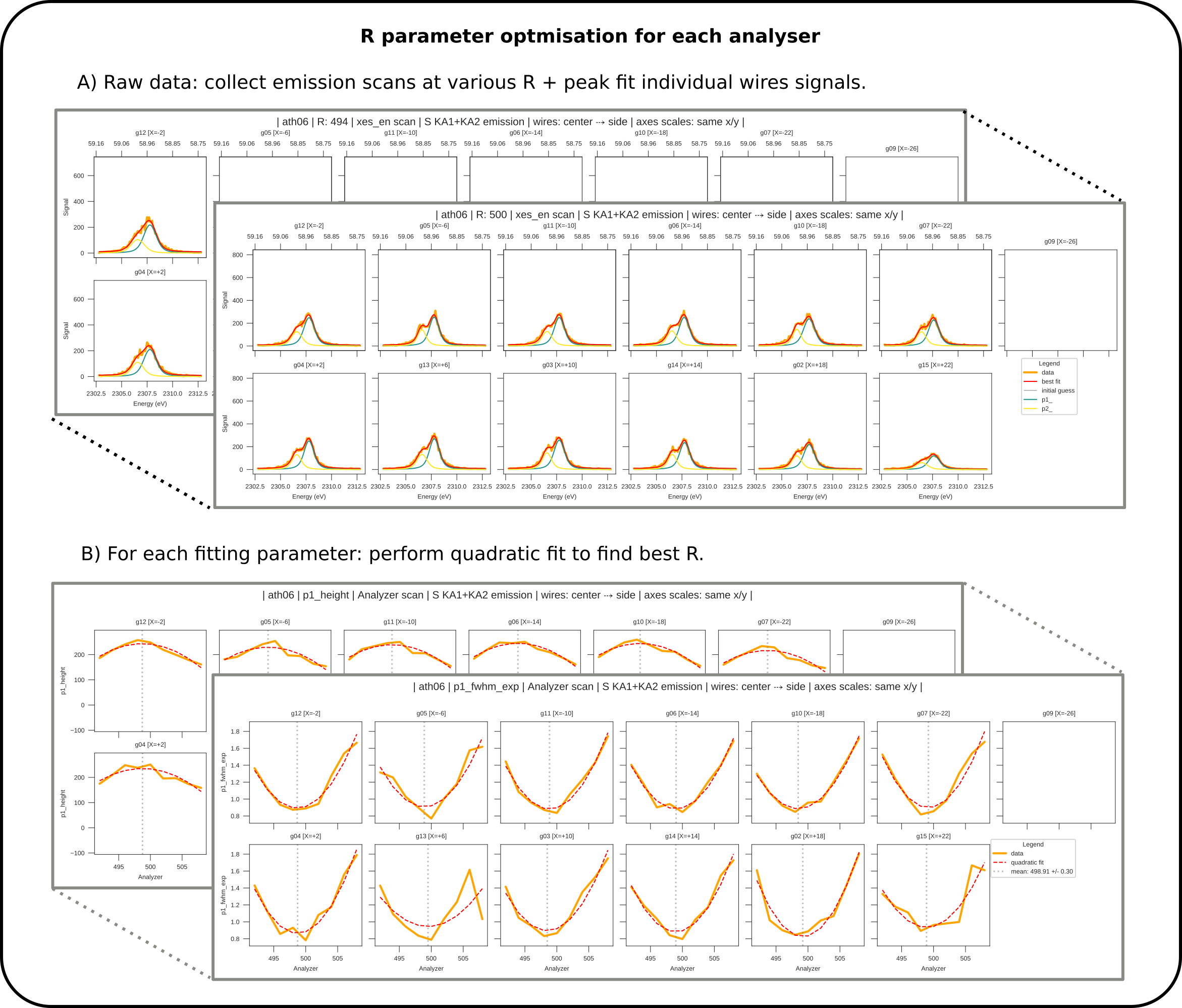}
  \caption{\label{figS:anaR}Procedure for finding the best R parameter for each crystal. Here the example of sulfur K$\alpha_{1,2}$ lines at around 59$^\circ$ Bragg angle with Si(111) single-machined Johansson analyzers is shown.}
\end{figure*}

The procedure employed for finding the best Rowland circle radius parameter (R) for each analyzer is summarized in Figure~\ref{figS:anaR}.

\bibliographystyle{apsrev4-1}
\input{refs_sup.bbl}

\end{document}

%% file: refs_pap.bbl
%

%% file: refs_sup.bbl
%

%% file: TEXSpap.bbl
\begin{thebibliography}{46}%
\makeatletter
\providecommand \@ifxundefined [1]{%
 \@ifx{#1\undefined}
}%
\providecommand \@ifnum [1]{%
 \ifnum #1\expandafter \@firstoftwo
 \else \expandafter \@secondoftwo
 \fi
}%
\providecommand \@ifx [1]{%
 \ifx #1\expandafter \@firstoftwo
 \else \expandafter \@secondoftwo
 \fi
}%
\providecommand \natexlab [1]{#1}%
\providecommand \enquote  [1]{``#1''}%
\providecommand \bibnamefont  [1]{#1}%
\providecommand \bibfnamefont [1]{#1}%
\providecommand \citenamefont [1]{#1}%
\providecommand \href@noop [0]{\@secondoftwo}%
\providecommand \href [0]{\begingroup \@sanitize@url \@href}%
\providecommand \@href[1]{\@@startlink{#1}\@@href}%
\providecommand \@@href[1]{\endgroup#1\@@endlink}%
\providecommand \@sanitize@url [0]{\catcode `\\12\catcode `\$12\catcode
  `\&12\catcode `\#12\catcode `\^12\catcode `\_12\catcode `\%12\relax}%
\providecommand \@@startlink[1]{}%
\providecommand \@@endlink[0]{}%
\providecommand \url  [0]{\begingroup\@sanitize@url \@url }%
\providecommand \@url [1]{\endgroup\@href {#1}{\urlprefix }}%
\providecommand \urlprefix  [0]{URL }%
\providecommand \Eprint [0]{\href }%
\providecommand \doibase [0]{http://dx.doi.org/}%
\providecommand \selectlanguage [0]{\@gobble}%
\providecommand \bibinfo  [0]{\@secondoftwo}%
\providecommand \bibfield  [0]{\@secondoftwo}%
\providecommand \translation [1]{[#1]}%
\providecommand \BibitemOpen [0]{}%
\providecommand \bibitemStop [0]{}%
\providecommand \bibitemNoStop [0]{.\EOS\space}%
\providecommand \EOS [0]{\spacefactor3000\relax}%
\providecommand \BibitemShut  [1]{\csname bibitem#1\endcsname}%
\let\auto@bib@innerbib\@empty
\bibitem [{\citenamefont {Meisel}\ \emph {et~al.}(1989)\citenamefont {Meisel},
  \citenamefont {Leonhardt},\ and\ \citenamefont {Szargan}}]{Meisel:1989_book}%
  \BibitemOpen
  \bibfield  {author} {\bibinfo {author} {\bibfnamefont {A.}~\bibnamefont
  {Meisel}}, \bibinfo {author} {\bibfnamefont {G.}~\bibnamefont {Leonhardt}}, \
  and\ \bibinfo {author} {\bibfnamefont {R.}~\bibnamefont {Szargan}},\ }\href
  {https://www.springer.com/gp/book/9783642822643} {\emph {\bibinfo {title}
  {{X-Ray Spectra and Chemical Binding}}}},\ edited by\ \bibinfo {editor}
  {\bibfnamefont {R.}~\bibnamefont {Gomer}}\ (\bibinfo  {publisher}
  {Springer-Verlag Berlin Heidelberg},\ \bibinfo {year} {1989})\ p.\ \bibinfo
  {pages} {458}\BibitemShut {NoStop}%
\bibitem [{\citenamefont {Glatzel}\ and\ \citenamefont
  {Bergmann}(2005)}]{Glatzel:2005_CCR}%
  \BibitemOpen
  \bibfield  {author} {\bibinfo {author} {\bibfnamefont {P.}~\bibnamefont
  {Glatzel}}\ and\ \bibinfo {author} {\bibfnamefont {U.}~\bibnamefont
  {Bergmann}},\ }\href {\doibase 10.1016/j.ccr.2004.04.011} {\bibfield
  {journal} {\bibinfo  {journal} {Coordination Chemistry Reviews}\ }\textbf
  {\bibinfo {volume} {249}},\ \bibinfo {pages} {65} (\bibinfo {year}
  {2005})}\BibitemShut {NoStop}%
\bibitem [{\citenamefont {Rueff}\ and\ \citenamefont
  {Shukla}(2013)}]{Rueff:2013_JESRP}%
  \BibitemOpen
  \bibfield  {author} {\bibinfo {author} {\bibfnamefont {J.}~\bibnamefont
  {Rueff}}\ and\ \bibinfo {author} {\bibfnamefont {A.}~\bibnamefont {Shukla}},\
  }\href {\doibase 10.1016/j.elspec.2013.04.014} {\bibfield  {journal}
  {\bibinfo  {journal} {Journal of Electron Spectroscopy and Related
  Phenomena}\ }\textbf {\bibinfo {volume} {188}},\ \bibinfo {pages} {10}
  (\bibinfo {year} {2013})}\BibitemShut {NoStop}%
\bibitem [{\citenamefont {Hayashi}(2000)}]{Hayashi:2013_book}%
  \BibitemOpen
  \bibfield  {author} {\bibinfo {author} {\bibfnamefont {H.}~\bibnamefont
  {Hayashi}},\ }in\ \href {\doibase 10.1002/9780470027318.a9389} {\emph
  {\bibinfo {booktitle} {Encyclopedia of Analytical Chemistry}}}\ (\bibinfo
  {publisher} {John Wiley {\&} Sons, Ltd},\ \bibinfo {address} {Chichester,
  UK},\ \bibinfo {year} {2000})\ pp.\ \bibinfo {pages} {1--31}\BibitemShut
  {NoStop}%
\bibitem [{\citenamefont {Rovezzi}\ and\ \citenamefont
  {Glatzel}(2014)}]{Rovezzi:2014_SST}%
  \BibitemOpen
  \bibfield  {author} {\bibinfo {author} {\bibfnamefont {M.}~\bibnamefont
  {Rovezzi}}\ and\ \bibinfo {author} {\bibfnamefont {P.}~\bibnamefont
  {Glatzel}},\ }\href {\doibase 10.1088/0268-1242/29/2/023002} {\bibfield
  {journal} {\bibinfo  {journal} {Semiconductor Science and Technology}\
  }\textbf {\bibinfo {volume} {29}},\ \bibinfo {pages} {023002} (\bibinfo
  {year} {2014})}\BibitemShut {NoStop}%
\bibitem [{\citenamefont {Bauer}(2014)}]{Bauer:2014_PCCP}%
  \BibitemOpen
  \bibfield  {author} {\bibinfo {author} {\bibfnamefont {M.}~\bibnamefont
  {Bauer}},\ }\href {\doibase 10.1039/C4CP00904E} {\bibfield  {journal}
  {\bibinfo  {journal} {Phys. Chem. Chem. Phys.}\ }\textbf {\bibinfo {volume}
  {16}},\ \bibinfo {pages} {13827} (\bibinfo {year} {2014})}\BibitemShut
  {NoStop}%
\bibitem [{\citenamefont {Sa}(2014)}]{Sa:2014_book}%
  \BibitemOpen
  \bibinfo {editor} {\bibfnamefont {J.}~\bibnamefont {Sa}},\ ed.,\ \href
  {\doibase 10.1201/b17184} {\emph {\bibinfo {title} {{High-Resolution
  XAS/XES}}}}\ (\bibinfo  {publisher} {CRC Press},\ \bibinfo {year}
  {2014})\BibitemShut {NoStop}%
\bibitem [{\citenamefont {DeBeer}\ and\ \citenamefont
  {Bergmann}(2016)}]{DeBeer:2016_book}%
  \BibitemOpen
  \bibfield  {author} {\bibinfo {author} {\bibfnamefont {S.}~\bibnamefont
  {DeBeer}}\ and\ \bibinfo {author} {\bibfnamefont {U.}~\bibnamefont
  {Bergmann}},\ }in\ \href {\doibase 10.1002/9781119951438.eibc2158} {\emph
  {\bibinfo {booktitle} {Encyclopedia of Inorganic and Bioinorganic
  Chemistry}}}\ (\bibinfo  {publisher} {John Wiley {\&} Sons, Ltd},\ \bibinfo
  {address} {Chichester, UK},\ \bibinfo {year} {2016})\ pp.\ \bibinfo {pages}
  {1--14}\BibitemShut {NoStop}%
\bibitem [{\citenamefont {Proux}\ \emph {et~al.}(2017)\citenamefont {Proux},
  \citenamefont {Lahera}, \citenamefont {Del~Net}, \citenamefont {Kieffer},
  \citenamefont {Rovezzi}, \citenamefont {Testemale}, \citenamefont {Irar},
  \citenamefont {Thomas}, \citenamefont {Aguilar-Tapia}, \citenamefont
  {Bazarkina}, \citenamefont {Prat}, \citenamefont {Tella}, \citenamefont
  {Auffan}, \citenamefont {Rose},\ and\ \citenamefont
  {Hazemann}}]{Proux:2017_JEQ}%
  \BibitemOpen
  \bibfield  {author} {\bibinfo {author} {\bibfnamefont {O.}~\bibnamefont
  {Proux}}, \bibinfo {author} {\bibfnamefont {E.}~\bibnamefont {Lahera}},
  \bibinfo {author} {\bibfnamefont {W.}~\bibnamefont {Del~Net}}, \bibinfo
  {author} {\bibfnamefont {I.}~\bibnamefont {Kieffer}}, \bibinfo {author}
  {\bibfnamefont {M.}~\bibnamefont {Rovezzi}}, \bibinfo {author} {\bibfnamefont
  {D.}~\bibnamefont {Testemale}}, \bibinfo {author} {\bibfnamefont
  {M.}~\bibnamefont {Irar}}, \bibinfo {author} {\bibfnamefont {S.}~\bibnamefont
  {Thomas}}, \bibinfo {author} {\bibfnamefont {A.}~\bibnamefont
  {Aguilar-Tapia}}, \bibinfo {author} {\bibfnamefont {E.~F.}\ \bibnamefont
  {Bazarkina}}, \bibinfo {author} {\bibfnamefont {A.}~\bibnamefont {Prat}},
  \bibinfo {author} {\bibfnamefont {M.}~\bibnamefont {Tella}}, \bibinfo
  {author} {\bibfnamefont {M.}~\bibnamefont {Auffan}}, \bibinfo {author}
  {\bibfnamefont {J.}~\bibnamefont {Rose}}, \ and\ \bibinfo {author}
  {\bibfnamefont {J.-l.}\ \bibnamefont {Hazemann}},\ }\href {\doibase
  10.2134/jeq2017.01.0023} {\bibfield  {journal} {\bibinfo  {journal} {Journal
  of Environment Quality}\ }\textbf {\bibinfo {volume} {46}},\ \bibinfo {pages}
  {1146} (\bibinfo {year} {2017})}\BibitemShut {NoStop}%
\bibitem [{\citenamefont {Alonso~Mori}\ \emph {et~al.}(2009)\citenamefont
  {Alonso~Mori}, \citenamefont {Paris}, \citenamefont {Giuli}, \citenamefont
  {Eeckhout}, \citenamefont {Kavcic}, \citenamefont {Zitnik},
  \citenamefont {Bucar}, \citenamefont {Pettersson},\ and\ \citenamefont
  {Glatzel}}]{AlonsoMori:2009_AC}%
  \BibitemOpen
  \bibfield  {author} {\bibinfo {author} {\bibfnamefont {R.}~\bibnamefont
  {Alonso~Mori}}, \bibinfo {author} {\bibfnamefont {E.}~\bibnamefont {Paris}},
  \bibinfo {author} {\bibfnamefont {G.}~\bibnamefont {Giuli}}, \bibinfo
  {author} {\bibfnamefont {S.~G.}\ \bibnamefont {Eeckhout}}, \bibinfo {author}
  {\bibfnamefont {M.}~\bibnamefont {Kavcic}}, \bibinfo {author}
  {\bibfnamefont {M.}~\bibnamefont {Zitnik}}, \bibinfo {author}
  {\bibfnamefont {K.}~\bibnamefont {Bucar}}, \bibinfo {author} {\bibfnamefont
  {L.~G.~M.}\ \bibnamefont {Pettersson}}, \ and\ \bibinfo {author}
  {\bibfnamefont {P.}~\bibnamefont {Glatzel}},\ }\href {\doibase
  10.1021/ac900970z} {\bibfield  {journal} {\bibinfo  {journal} {Analytical
  Chemistry}\ }\textbf {\bibinfo {volume} {81}},\ \bibinfo {pages} {6516}
  (\bibinfo {year} {2009})}\BibitemShut {NoStop}%
\bibitem [{\citenamefont {Mori}\ \emph {et~al.}(2010)\citenamefont {Mori},
  \citenamefont {Paris}, \citenamefont {Giuli}, \citenamefont {Eeckhout},
  \citenamefont {Kavcic}, \citenamefont {Zitnik}, \citenamefont
  {Bucar}, \citenamefont {Pettersson},\ and\ \citenamefont
  {Glatzel}}]{AlonsoMori:2010_IC}%
  \BibitemOpen
  \bibfield  {author} {\bibinfo {author} {\bibfnamefont {R.~A.}\ \bibnamefont
  {Mori}}, \bibinfo {author} {\bibfnamefont {E.}~\bibnamefont {Paris}},
  \bibinfo {author} {\bibfnamefont {G.}~\bibnamefont {Giuli}}, \bibinfo
  {author} {\bibfnamefont {S.~G.}\ \bibnamefont {Eeckhout}}, \bibinfo {author}
  {\bibfnamefont {M.}~\bibnamefont {Kavcic}}, \bibinfo {author}
  {\bibfnamefont {M.}~\bibnamefont {Zitnik}}, \bibinfo {author}
  {\bibfnamefont {K.}~\bibnamefont {Bucar}}, \bibinfo {author} {\bibfnamefont
  {L.~G.~M.}\ \bibnamefont {Pettersson}}, \ and\ \bibinfo {author}
  {\bibfnamefont {P.}~\bibnamefont {Glatzel}},\ }\href {\doibase
  10.1021/ic100304z} {\bibfield  {journal} {\bibinfo  {journal} {Inorganic
  Chemistry}\ }\textbf {\bibinfo {volume} {49}},\ \bibinfo {pages} {6468}
  (\bibinfo {year} {2010})}\BibitemShut {NoStop}%
\bibitem [{\citenamefont {Wilhelm}\ \emph {et~al.}(2016)\citenamefont
  {Wilhelm}, \citenamefont {Garbarino}, \citenamefont {Jacobs}, \citenamefont
  {Vitoux}, \citenamefont {Steinmann}, \citenamefont {Guillou}, \citenamefont
  {Snigirev}, \citenamefont {Snigireva}, \citenamefont {Voisin}, \citenamefont
  {Braithwaite}, \citenamefont {Aoki}, \citenamefont {Brison}, \citenamefont
  {Kantor}, \citenamefont {Lyatun},\ and\ \citenamefont
  {Rogalev}}]{Wilhelm:2016_HPR}%
  \BibitemOpen
  \bibfield  {author} {\bibinfo {author} {\bibfnamefont {F.}~\bibnamefont
  {Wilhelm}}, \bibinfo {author} {\bibfnamefont {G.}~\bibnamefont {Garbarino}},
  \bibinfo {author} {\bibfnamefont {J.}~\bibnamefont {Jacobs}}, \bibinfo
  {author} {\bibfnamefont {H.}~\bibnamefont {Vitoux}}, \bibinfo {author}
  {\bibfnamefont {R.}~\bibnamefont {Steinmann}}, \bibinfo {author}
  {\bibfnamefont {F.}~\bibnamefont {Guillou}}, \bibinfo {author} {\bibfnamefont
  {A.}~\bibnamefont {Snigirev}}, \bibinfo {author} {\bibfnamefont
  {I.}~\bibnamefont {Snigireva}}, \bibinfo {author} {\bibfnamefont
  {P.}~\bibnamefont {Voisin}}, \bibinfo {author} {\bibfnamefont
  {D.}~\bibnamefont {Braithwaite}}, \bibinfo {author} {\bibfnamefont
  {D.}~\bibnamefont {Aoki}}, \bibinfo {author} {\bibfnamefont {J.-P.}\
  \bibnamefont {Brison}}, \bibinfo {author} {\bibfnamefont {I.}~\bibnamefont
  {Kantor}}, \bibinfo {author} {\bibfnamefont {I.}~\bibnamefont {Lyatun}}, \
  and\ \bibinfo {author} {\bibfnamefont {A.}~\bibnamefont {Rogalev}},\ }\href
  {\doibase 10.1080/08957959.2016.1206092} {\bibfield  {journal} {\bibinfo
  {journal} {High Pressure Research}\ }\textbf {\bibinfo {volume} {36}},\
  \bibinfo {pages} {445} (\bibinfo {year} {2016})}\BibitemShut {NoStop}%
\bibitem [{\citenamefont {Thompson}\ \emph {et~al.}(2015)\citenamefont
  {Thompson}, \citenamefont {Nguyen}, \citenamefont {Nicholls}, \citenamefont
  {Bourne}, \citenamefont {Brazier}, \citenamefont {Lovelock}, \citenamefont
  {Brown}, \citenamefont {Wermeille}, \citenamefont {Bikondoa}, \citenamefont
  {Lucas}, \citenamefont {Hase},\ and\ \citenamefont
  {Newton}}]{Thompson:2015_JSR}%
  \BibitemOpen
  \bibfield  {author} {\bibinfo {author} {\bibfnamefont {P.~B.~J.}\
  \bibnamefont {Thompson}}, \bibinfo {author} {\bibfnamefont {B.~N.}\
  \bibnamefont {Nguyen}}, \bibinfo {author} {\bibfnamefont {R.}~\bibnamefont
  {Nicholls}}, \bibinfo {author} {\bibfnamefont {R.~A.}\ \bibnamefont
  {Bourne}}, \bibinfo {author} {\bibfnamefont {J.~B.}\ \bibnamefont {Brazier}},
  \bibinfo {author} {\bibfnamefont {K.~R.~J.}\ \bibnamefont {Lovelock}},
  \bibinfo {author} {\bibfnamefont {S.~D.}\ \bibnamefont {Brown}}, \bibinfo
  {author} {\bibfnamefont {D.}~\bibnamefont {Wermeille}}, \bibinfo {author}
  {\bibfnamefont {O.}~\bibnamefont {Bikondoa}}, \bibinfo {author}
  {\bibfnamefont {C.~A.}\ \bibnamefont {Lucas}}, \bibinfo {author}
  {\bibfnamefont {T.~P.~A.}\ \bibnamefont {Hase}}, \ and\ \bibinfo {author}
  {\bibfnamefont {M.~A.}\ \bibnamefont {Newton}},\ }\href {\doibase
  10.1107/S1600577515016148} {\bibfield  {journal} {\bibinfo  {journal}
  {Journal of Synchrotron Radiation}\ }\textbf {\bibinfo {volume} {22}},\
  \bibinfo {pages} {1426} (\bibinfo {year} {2015})}\BibitemShut {NoStop}%
\bibitem [{\citenamefont {Thomas}\ \emph {et~al.}(2015)\citenamefont {Thomas},
  \citenamefont {Kas}, \citenamefont {Glatzel}, \citenamefont {Al~Samarai},
  \citenamefont {de~Groot}, \citenamefont {Alonso~Mori}, \citenamefont
  {Kav{\v{c}}i{\v{c}}}, \citenamefont {Zitnik}, \citenamefont {Bucar},
  \citenamefont {Rehr},\ and\ \citenamefont {Tromp}}]{Thomas:2015_JPCC}%
  \BibitemOpen
  \bibfield  {author} {\bibinfo {author} {\bibfnamefont {R.}~\bibnamefont
  {Thomas}}, \bibinfo {author} {\bibfnamefont {J.}~\bibnamefont {Kas}},
  \bibinfo {author} {\bibfnamefont {P.}~\bibnamefont {Glatzel}}, \bibinfo
  {author} {\bibfnamefont {M.}~\bibnamefont {Al~Samarai}}, \bibinfo {author}
  {\bibfnamefont {F.~M.~F.}\ \bibnamefont {de~Groot}}, \bibinfo {author}
  {\bibfnamefont {R.}~\bibnamefont {Alonso~Mori}}, \bibinfo {author}
  {\bibfnamefont {M.}~\bibnamefont {Kav{\v{c}}i{\v{c}}}}, \bibinfo {author}
  {\bibfnamefont {M.}~\bibnamefont {Zitnik}}, \bibinfo {author} {\bibfnamefont
  {K.}~\bibnamefont {Bucar}}, \bibinfo {author} {\bibfnamefont {J.~J.}\
  \bibnamefont {Rehr}}, \ and\ \bibinfo {author} {\bibfnamefont
  {M.}~\bibnamefont {Tromp}},\ }\href {\doibase 10.1021/jp509376q} {\bibfield
  {journal} {\bibinfo  {journal} {The Journal of Physical Chemistry C}\ ,\
  \bibinfo {pages} {150126100226006}} (\bibinfo {year} {2015})}\BibitemShut
  {NoStop}%
\bibitem [{\citenamefont {Kav{\v{c}}i{\v{c}}}\ \emph
  {et~al.}(2016)\citenamefont {Kav{\v{c}}i{\v{c}}}, \citenamefont
  {Bu{\v{c}}ar}, \citenamefont {Petric}, \citenamefont {{\v{Z}}itnik},
  \citenamefont {Ar{\v{c}}on}, \citenamefont {Dominko},\ and\ \citenamefont
  {Vizintin}}]{Kavcic:2016_JPCC}%
  \BibitemOpen
  \bibfield  {author} {\bibinfo {author} {\bibfnamefont {M.}~\bibnamefont
  {Kav{\v{c}}i{\v{c}}}}, \bibinfo {author} {\bibfnamefont {K.}~\bibnamefont
  {Bu{\v{c}}ar}}, \bibinfo {author} {\bibfnamefont {M.}~\bibnamefont {Petric}},
  \bibinfo {author} {\bibfnamefont {M.}~\bibnamefont {{\v{Z}}itnik}}, \bibinfo
  {author} {\bibfnamefont {I.}~\bibnamefont {Ar{\v{c}}on}}, \bibinfo {author}
  {\bibfnamefont {R.}~\bibnamefont {Dominko}}, \ and\ \bibinfo {author}
  {\bibfnamefont {A.}~\bibnamefont {Vizintin}},\ }\href {\doibase
  10.1021/acs.jpcc.6b06705} {\bibfield  {journal} {\bibinfo  {journal} {The
  Journal of Physical Chemistry C}\ }\textbf {\bibinfo {volume} {120}},\
  \bibinfo {pages} {24568} (\bibinfo {year} {2016})}\BibitemShut {NoStop}%
\bibitem [{\citenamefont {Kvashnina}\ \emph {et~al.}(2013)\citenamefont
  {Kvashnina}, \citenamefont {Butorin}, \citenamefont {Martin},\ and\
  \citenamefont {Glatzel}}]{Kvashnina:2013_PRL}%
  \BibitemOpen
  \bibfield  {author} {\bibinfo {author} {\bibfnamefont {K.~O.}\ \bibnamefont
  {Kvashnina}}, \bibinfo {author} {\bibfnamefont {S.~M.}\ \bibnamefont
  {Butorin}}, \bibinfo {author} {\bibfnamefont {P.}~\bibnamefont {Martin}}, \
  and\ \bibinfo {author} {\bibfnamefont {P.}~\bibnamefont {Glatzel}},\ }\href
  {\doibase 10.1103/PhysRevLett.111.253002} {\bibfield  {journal} {\bibinfo
  {journal} {Physical Review Letters}\ }\textbf {\bibinfo {volume} {111}},\
  \bibinfo {pages} {253002} (\bibinfo {year} {2013})}\BibitemShut {NoStop}%
\bibitem [{\citenamefont {Kvashnina}\ \emph {et~al.}(2019)\citenamefont
  {Kvashnina}, \citenamefont {Romanchuk}, \citenamefont {Pidchenko},
  \citenamefont {Amidani}, \citenamefont {Gerber}, \citenamefont {Trigub},
  \citenamefont {Rossberg}, \citenamefont {Weiss}, \citenamefont {Popa},
  \citenamefont {Walter}, \citenamefont {Caciuffo}, \citenamefont {Scheinost},
  \citenamefont {Butorin},\ and\ \citenamefont
  {Kalmykov}}]{Kvashnina:2019_Ang}%
  \BibitemOpen
  \bibfield  {author} {\bibinfo {author} {\bibfnamefont {K.~O.}\ \bibnamefont
  {Kvashnina}}, \bibinfo {author} {\bibfnamefont {A.~Y.}\ \bibnamefont
  {Romanchuk}}, \bibinfo {author} {\bibfnamefont {I.}~\bibnamefont
  {Pidchenko}}, \bibinfo {author} {\bibfnamefont {L.}~\bibnamefont {Amidani}},
  \bibinfo {author} {\bibfnamefont {E.}~\bibnamefont {Gerber}}, \bibinfo
  {author} {\bibfnamefont {A.}~\bibnamefont {Trigub}}, \bibinfo {author}
  {\bibfnamefont {A.}~\bibnamefont {Rossberg}}, \bibinfo {author}
  {\bibfnamefont {S.}~\bibnamefont {Weiss}}, \bibinfo {author} {\bibfnamefont
  {K.}~\bibnamefont {Popa}}, \bibinfo {author} {\bibfnamefont {O.}~\bibnamefont
  {Walter}}, \bibinfo {author} {\bibfnamefont {R.}~\bibnamefont {Caciuffo}},
  \bibinfo {author} {\bibfnamefont {A.~C.}\ \bibnamefont {Scheinost}}, \bibinfo
  {author} {\bibfnamefont {S.~M.}\ \bibnamefont {Butorin}}, \ and\ \bibinfo
  {author} {\bibfnamefont {S.~N.}\ \bibnamefont {Kalmykov}},\ }\href {\doibase
  10.1002/anie.201911637} {\bibfield  {journal} {\bibinfo  {journal}
  {Angewandte Chemie International Edition}\ } (\bibinfo {year} {2019}),\
  10.1002/anie.201911637}\BibitemShut {NoStop}%
\bibitem [{\citenamefont {Doriese}\ \emph {et~al.}(2017)\citenamefont
  {Doriese}, \citenamefont {Abbamonte}, \citenamefont {Alpert}, \citenamefont
  {Bennett}, \citenamefont {Denison}, \citenamefont {Fang}, \citenamefont
  {Fischer}, \citenamefont {Fitzgerald}, \citenamefont {Fowler}, \citenamefont
  {Gard}, \citenamefont {Hays-Wehle}, \citenamefont {Hilton}, \citenamefont
  {Jaye}, \citenamefont {McChesney}, \citenamefont {Miaja-Avila}, \citenamefont
  {Morgan}, \citenamefont {Joe}, \citenamefont {O’Neil}, \citenamefont
  {Reintsema}, \citenamefont {Rodolakis}, \citenamefont {Schmidt},
  \citenamefont {Tatsuno}, \citenamefont {Uhlig}, \citenamefont {Vale},
  \citenamefont {Ullom},\ and\ \citenamefont {Swetz}}]{Doriese:2017_RSI}%
  \BibitemOpen
  \bibfield  {author} {\bibinfo {author} {\bibfnamefont {W.~B.}\ \bibnamefont
  {Doriese}}, \bibinfo {author} {\bibfnamefont {P.}~\bibnamefont {Abbamonte}},
  \bibinfo {author} {\bibfnamefont {B.~K.}\ \bibnamefont {Alpert}}, \bibinfo
  {author} {\bibfnamefont {D.~A.}\ \bibnamefont {Bennett}}, \bibinfo {author}
  {\bibfnamefont {E.~V.}\ \bibnamefont {Denison}}, \bibinfo {author}
  {\bibfnamefont {Y.}~\bibnamefont {Fang}}, \bibinfo {author} {\bibfnamefont
  {D.~A.}\ \bibnamefont {Fischer}}, \bibinfo {author} {\bibfnamefont {C.~P.}\
  \bibnamefont {Fitzgerald}}, \bibinfo {author} {\bibfnamefont {J.~W.}\
  \bibnamefont {Fowler}}, \bibinfo {author} {\bibfnamefont {J.~D.}\
  \bibnamefont {Gard}}, \bibinfo {author} {\bibfnamefont {J.~P.}\ \bibnamefont
  {Hays-Wehle}}, \bibinfo {author} {\bibfnamefont {G.~C.}\ \bibnamefont
  {Hilton}}, \bibinfo {author} {\bibfnamefont {C.}~\bibnamefont {Jaye}},
  \bibinfo {author} {\bibfnamefont {J.~L.}\ \bibnamefont {McChesney}}, \bibinfo
  {author} {\bibfnamefont {L.}~\bibnamefont {Miaja-Avila}}, \bibinfo {author}
  {\bibfnamefont {K.~M.}\ \bibnamefont {Morgan}}, \bibinfo {author}
  {\bibfnamefont {Y.~I.}\ \bibnamefont {Joe}}, \bibinfo {author} {\bibfnamefont
  {G.~C.}\ \bibnamefont {O’Neil}}, \bibinfo {author} {\bibfnamefont {C.~D.}\
  \bibnamefont {Reintsema}}, \bibinfo {author} {\bibfnamefont {F.}~\bibnamefont
  {Rodolakis}}, \bibinfo {author} {\bibfnamefont {D.~R.}\ \bibnamefont
  {Schmidt}}, \bibinfo {author} {\bibfnamefont {H.}~\bibnamefont {Tatsuno}},
  \bibinfo {author} {\bibfnamefont {J.}~\bibnamefont {Uhlig}}, \bibinfo
  {author} {\bibfnamefont {L.~R.}\ \bibnamefont {Vale}}, \bibinfo {author}
  {\bibfnamefont {J.~N.}\ \bibnamefont {Ullom}}, \ and\ \bibinfo {author}
  {\bibfnamefont {D.~S.}\ \bibnamefont {Swetz}},\ }\href {\doibase
  10.1063/1.4983316} {\bibfield  {journal} {\bibinfo  {journal} {Review of
  Scientific Instruments}\ }\textbf {\bibinfo {volume} {88}},\ \bibinfo {pages}
  {053108} (\bibinfo {year} {2017})}\BibitemShut {NoStop}%
\bibitem [{\citenamefont {Hoszowska}\ \emph {et~al.}(1996)\citenamefont
  {Hoszowska}, \citenamefont {Dousse}, \citenamefont {Kern},\ and\
  \citenamefont {Rh{\^{e}}me}}]{Hoszowska:1996_NIMA}%
  \BibitemOpen
  \bibfield  {author} {\bibinfo {author} {\bibfnamefont {J.}~\bibnamefont
  {Hoszowska}}, \bibinfo {author} {\bibfnamefont {J.-C.}\ \bibnamefont
  {Dousse}}, \bibinfo {author} {\bibfnamefont {J.}~\bibnamefont {Kern}}, \ and\
  \bibinfo {author} {\bibfnamefont {C.}~\bibnamefont {Rh{\^{e}}me}},\ }\href
  {\doibase 10.1016/0168-9002(96)00262-8} {\bibfield  {journal} {\bibinfo
  {journal} {Nuclear Instruments and Methods in Physics Research Section A:
  Accelerators, Spectrometers, Detectors and Associated Equipment}\ }\textbf
  {\bibinfo {volume} {376}},\ \bibinfo {pages} {129} (\bibinfo {year}
  {1996})}\BibitemShut {NoStop}%
\bibitem [{\citenamefont {Dousse}\ and\ \citenamefont
  {Hoszowska}(2014)}]{Dousse:2014_book}%
  \BibitemOpen
  \bibfield  {author} {\bibinfo {author} {\bibfnamefont {J.}\ \bibnamefont
  {Dousse}}\ and\ \bibinfo {author} {\bibfnamefont {J.}~\bibnamefont
  {Hoszowska}},\ }in\ \href {\doibase 10.1201/b17184-3} {\emph {\bibinfo
  {booktitle} {High-Resolution XAS/XES}}}\ (\bibinfo  {publisher} {CRC Press},\
  \bibinfo {year} {2014})\ Chap.~\bibinfo {chapter} {2}, pp.\ \bibinfo {pages}
  {27--58}\BibitemShut {NoStop}%
\bibitem [{\citenamefont {Welter}\ \emph {et~al.}(2005)\citenamefont {Welter},
  \citenamefont {Machek}, \citenamefont {Dr{\"{a}}ger}, \citenamefont
  {Br{\"{u}}ggmann},\ and\ \citenamefont {Fr{\"{o}}ba}}]{Welter:2005_JSR}%
  \BibitemOpen
  \bibfield  {author} {\bibinfo {author} {\bibfnamefont {E.}~\bibnamefont
  {Welter}}, \bibinfo {author} {\bibfnamefont {P.}~\bibnamefont {Machek}},
  \bibinfo {author} {\bibfnamefont {G.}~\bibnamefont {Dr{\"{a}}ger}}, \bibinfo
  {author} {\bibfnamefont {U.}~\bibnamefont {Br{\"{u}}ggmann}}, \ and\ \bibinfo
  {author} {\bibfnamefont {M.}~\bibnamefont {Fr{\"{o}}ba}},\ }\href {\doibase
  10.1107/S0909049505007843} {\bibfield  {journal} {\bibinfo  {journal}
  {Journal of synchrotron radiation}\ }\textbf {\bibinfo {volume} {12}},\
  \bibinfo {pages} {448} (\bibinfo {year} {2005})}\BibitemShut {NoStop}%
\bibitem [{\citenamefont {Huotari}\ \emph {et~al.}(2006)\citenamefont
  {Huotari}, \citenamefont {Albergamo}, \citenamefont {Vanko}, \citenamefont
  {Verbeni},\ and\ \citenamefont {Monaco}}]{Huotari:2006_RSI}%
  \BibitemOpen
  \bibfield  {author} {\bibinfo {author} {\bibfnamefont {S.}~\bibnamefont
  {Huotari}}, \bibinfo {author} {\bibfnamefont {F.}~\bibnamefont {Albergamo}},
  \bibinfo {author} {\bibfnamefont {G.}~\bibnamefont {Vanko}}, \bibinfo
  {author} {\bibfnamefont {R.}~\bibnamefont {Verbeni}}, \ and\ \bibinfo
  {author} {\bibfnamefont {G.}~\bibnamefont {Monaco}},\ }\href {\doibase
  10.1063/1.2198805} {\bibfield  {journal} {\bibinfo  {journal} {Review of
  Scientific Instruments}\ }\textbf {\bibinfo {volume} {77}},\ \bibinfo {pages}
  {053102} (\bibinfo {year} {2006})}\BibitemShut {NoStop}%
\bibitem [{\citenamefont {Holden}\ \emph {et~al.}(2017)\citenamefont {Holden},
  \citenamefont {Hoidn}, \citenamefont {Ditter}, \citenamefont {Seidler},
  \citenamefont {Kas}, \citenamefont {Stein}, \citenamefont {Cossairt},
  \citenamefont {Kozimor}, \citenamefont {Guo}, \citenamefont {Ye},
  \citenamefont {Marcus},\ and\ \citenamefont {Fakra}}]{Holden:2017_RSI}%
  \BibitemOpen
  \bibfield  {author} {\bibinfo {author} {\bibfnamefont {W.~M.}\ \bibnamefont
  {Holden}}, \bibinfo {author} {\bibfnamefont {O.~R.}\ \bibnamefont {Hoidn}},
  \bibinfo {author} {\bibfnamefont {A.~S.}\ \bibnamefont {Ditter}}, \bibinfo
  {author} {\bibfnamefont {G.~T.}\ \bibnamefont {Seidler}}, \bibinfo {author}
  {\bibfnamefont {J.}~\bibnamefont {Kas}}, \bibinfo {author} {\bibfnamefont
  {J.~L.}\ \bibnamefont {Stein}}, \bibinfo {author} {\bibfnamefont {B.~M.}\
  \bibnamefont {Cossairt}}, \bibinfo {author} {\bibfnamefont {S.~A.}\
  \bibnamefont {Kozimor}}, \bibinfo {author} {\bibfnamefont {J.}~\bibnamefont
  {Guo}}, \bibinfo {author} {\bibfnamefont {Y.}~\bibnamefont {Ye}}, \bibinfo
  {author} {\bibfnamefont {M.~A.}\ \bibnamefont {Marcus}}, \ and\ \bibinfo
  {author} {\bibfnamefont {S.}~\bibnamefont {Fakra}},\ }\href {\doibase
  10.1063/1.4994739} {\bibfield  {journal} {\bibinfo  {journal} {Review of
  Scientific Instruments}\ }\textbf {\bibinfo {volume} {88}},\ \bibinfo {pages}
  {073904} (\bibinfo {year} {2017})}\BibitemShut {NoStop}%
\bibitem [{\citenamefont {Wittry}\ and\ \citenamefont
  {Li}(1993)}]{Wittry:1993_RSI}%
  \BibitemOpen
  \bibfield  {author} {\bibinfo {author} {\bibfnamefont {D.~B.}\ \bibnamefont
  {Wittry}}\ and\ \bibinfo {author} {\bibfnamefont {R.~Y.}\ \bibnamefont
  {Li}},\ }\href {\doibase 10.1063/1.1143959} {\bibfield  {journal} {\bibinfo
  {journal} {Review of Scientific Instruments}\ }\textbf {\bibinfo {volume}
  {64}},\ \bibinfo {pages} {2195} (\bibinfo {year} {1993})}\BibitemShut
  {NoStop}%
\bibitem [{\citenamefont {Kav{\v{c}}i{\v{c}}}\ \emph
  {et~al.}(2012)\citenamefont {Kav{\v{c}}i{\v{c}}}, \citenamefont {Budnar},
  \citenamefont {M{\"{u}}hleisen}, \citenamefont {Gasser}, \citenamefont
  {{\v{Z}}itnik}, \citenamefont {Bu{\v{c}}ar},\ and\ \citenamefont
  {Bohinc}}]{Kavcic:2012_RSI}%
  \BibitemOpen
  \bibfield  {author} {\bibinfo {author} {\bibfnamefont {M.}~\bibnamefont
  {Kav{\v{c}}i{\v{c}}}}, \bibinfo {author} {\bibfnamefont {M.}~\bibnamefont
  {Budnar}}, \bibinfo {author} {\bibfnamefont {A.}~\bibnamefont
  {M{\"{u}}hleisen}}, \bibinfo {author} {\bibfnamefont {F.}~\bibnamefont
  {Gasser}}, \bibinfo {author} {\bibfnamefont {M.}~\bibnamefont
  {{\v{Z}}itnik}}, \bibinfo {author} {\bibfnamefont {K.}~\bibnamefont
  {Bu{\v{c}}ar}}, \ and\ \bibinfo {author} {\bibfnamefont {R.}~\bibnamefont
  {Bohinc}},\ }\href {\doibase 10.1063/1.3697862} {\bibfield  {journal}
  {\bibinfo  {journal} {The Review of scientific instruments}\ }\textbf
  {\bibinfo {volume} {83}},\ \bibinfo {pages} {033113} (\bibinfo {year}
  {2012})}\BibitemShut {NoStop}%
\bibitem [{\citenamefont {Abraham}\ \emph {et~al.}(2019)\citenamefont
  {Abraham}, \citenamefont {Nowak}, \citenamefont {Weninger}, \citenamefont
  {Armenta}, \citenamefont {Defever}, \citenamefont {Day}, \citenamefont
  {Carini}, \citenamefont {Nakahara}, \citenamefont {Gallo}, \citenamefont
  {Nelson}, \citenamefont {Nordlund}, \citenamefont {Kroll}, \citenamefont
  {Hunter}, \citenamefont {van Driel}, \citenamefont {Zhu}, \citenamefont
  {Weng}, \citenamefont {Alonso-Mori},\ and\ \citenamefont
  {Sokaras}}]{Abraham:2019_JSR}%
  \BibitemOpen
  \bibfield  {author} {\bibinfo {author} {\bibfnamefont {B.}~\bibnamefont
  {Abraham}}, \bibinfo {author} {\bibfnamefont {S.}~\bibnamefont {Nowak}},
  \bibinfo {author} {\bibfnamefont {C.}~\bibnamefont {Weninger}}, \bibinfo
  {author} {\bibfnamefont {R.}~\bibnamefont {Armenta}}, \bibinfo {author}
  {\bibfnamefont {J.}~\bibnamefont {Defever}}, \bibinfo {author} {\bibfnamefont
  {D.}~\bibnamefont {Day}}, \bibinfo {author} {\bibfnamefont {G.}~\bibnamefont
  {Carini}}, \bibinfo {author} {\bibfnamefont {K.}~\bibnamefont {Nakahara}},
  \bibinfo {author} {\bibfnamefont {A.}~\bibnamefont {Gallo}}, \bibinfo
  {author} {\bibfnamefont {S.}~\bibnamefont {Nelson}}, \bibinfo {author}
  {\bibfnamefont {D.}~\bibnamefont {Nordlund}}, \bibinfo {author}
  {\bibfnamefont {T.}~\bibnamefont {Kroll}}, \bibinfo {author} {\bibfnamefont
  {M.~S.}\ \bibnamefont {Hunter}}, \bibinfo {author} {\bibfnamefont
  {T.}~\bibnamefont {van Driel}}, \bibinfo {author} {\bibfnamefont
  {D.}~\bibnamefont {Zhu}}, \bibinfo {author} {\bibfnamefont {T.-C.}\
  \bibnamefont {Weng}}, \bibinfo {author} {\bibfnamefont {R.}~\bibnamefont
  {Alonso-Mori}}, \ and\ \bibinfo {author} {\bibfnamefont {D.}~\bibnamefont
  {Sokaras}},\ }\href {\doibase 10.1107/S1600577519002431} {\bibfield
  {journal} {\bibinfo  {journal} {Journal of Synchrotron Radiation}\ }\textbf
  {\bibinfo {volume} {26}},\ \bibinfo {pages} {629} (\bibinfo {year}
  {2019})}\BibitemShut {NoStop}%
\bibitem [{\citenamefont {Rowland}(1882)}]{Rowland:1882_PhilMag}%
  \BibitemOpen
  \bibfield  {author} {\bibinfo {author} {\bibfnamefont {H.~A.}\ \bibnamefont
  {Rowland}},\ }\href {http://zs.thulb.uni-jena.de/receive/
  jportal_jpvolume_00128126} {\bibfield  {journal} {\bibinfo  {journal}
  {Philosophical Magazine}\ }\textbf {\bibinfo {volume} {13}},\ \bibinfo
  {pages} {477} (\bibinfo {year} {1882})}\BibitemShut {NoStop}%
\bibitem [{\citenamefont {Wittry}\ and\ \citenamefont
  {Sun}(1990{\natexlab{a}})}]{Wittry:1990_JAP}%
  \BibitemOpen
  \bibfield  {author} {\bibinfo {author} {\bibfnamefont {D.~B.}\ \bibnamefont
  {Wittry}}\ and\ \bibinfo {author} {\bibfnamefont {S.}~\bibnamefont {Sun}},\
  }\href {\doibase 10.1063/1.345629} {\bibfield  {journal} {\bibinfo  {journal}
  {Journal of Applied Physics}\ }\textbf {\bibinfo {volume} {67}},\ \bibinfo
  {pages} {1633} (\bibinfo {year} {1990}{\natexlab{a}})}\BibitemShut {NoStop}%
\bibitem [{\citenamefont {Wittry}\ and\ \citenamefont
  {Sun}(1990{\natexlab{b}})}]{Wittry:1990_JAP_a}%
  \BibitemOpen
  \bibfield  {author} {\bibinfo {author} {\bibfnamefont {D.~B.}\ \bibnamefont
  {Wittry}}\ and\ \bibinfo {author} {\bibfnamefont {S.}~\bibnamefont {Sun}},\
  }\href {\doibase 10.1063/1.346834} {\bibfield  {journal} {\bibinfo  {journal}
  {Journal of Applied Physics}\ }\textbf {\bibinfo {volume} {68}},\ \bibinfo
  {pages} {387} (\bibinfo {year} {1990}{\natexlab{b}})}\BibitemShut {NoStop}%
\bibitem [{\citenamefont {Wittry}\ and\ \citenamefont
  {Sun}(1991)}]{Wittry:1991_JAP}%
  \BibitemOpen
  \bibfield  {author} {\bibinfo {author} {\bibfnamefont {D.~B.}\ \bibnamefont
  {Wittry}}\ and\ \bibinfo {author} {\bibfnamefont {S.}~\bibnamefont {Sun}},\
  }\href {\doibase 10.1063/1.348446} {\bibfield  {journal} {\bibinfo  {journal}
  {Journal of Applied Physics}\ }\textbf {\bibinfo {volume} {69}},\ \bibinfo
  {pages} {3886} (\bibinfo {year} {1991})}\BibitemShut {NoStop}%
\bibitem [{\citenamefont {Wittry}\ and\ \citenamefont
  {Sun}(1992)}]{Wittry:1992_JAP}%
  \BibitemOpen
  \bibfield  {author} {\bibinfo {author} {\bibfnamefont {D.~B.}\ \bibnamefont
  {Wittry}}\ and\ \bibinfo {author} {\bibfnamefont {S.}~\bibnamefont {Sun}},\
  }\href {\doibase 10.1063/1.350406} {\bibfield  {journal} {\bibinfo  {journal}
  {Journal of Applied Physics}\ }\textbf {\bibinfo {volume} {71}},\ \bibinfo
  {pages} {564} (\bibinfo {year} {1992})}\BibitemShut {NoStop}%
\bibitem [{\citenamefont {Schoonjans}\ \emph {et~al.}(2011)\citenamefont
  {Schoonjans}, \citenamefont {Brunetti}, \citenamefont {Golosio},
  \citenamefont {Sanchez~del Rio}, \citenamefont {Sol{\'{e}}}, \citenamefont
  {Ferrero},\ and\ \citenamefont {Vincze}}]{Schoonjans:2011_SAB}%
  \BibitemOpen
  \bibfield  {author} {\bibinfo {author} {\bibfnamefont {T.}~\bibnamefont
  {Schoonjans}}, \bibinfo {author} {\bibfnamefont {A.}~\bibnamefont
  {Brunetti}}, \bibinfo {author} {\bibfnamefont {B.}~\bibnamefont {Golosio}},
  \bibinfo {author} {\bibfnamefont {M.}~\bibnamefont {Sanchez~del Rio}},
  \bibinfo {author} {\bibfnamefont {V.~A.}\ \bibnamefont {Sol{\'{e}}}},
  \bibinfo {author} {\bibfnamefont {C.}~\bibnamefont {Ferrero}}, \ and\
  \bibinfo {author} {\bibfnamefont {L.}~\bibnamefont {Vincze}},\ }\href
  {\doibase 10.1016/j.sab.2011.09.011} {\bibfield  {journal} {\bibinfo
  {journal} {Spectrochimica Acta Part B: Atomic Spectroscopy}\ }\textbf
  {\bibinfo {volume} {66}},\ \bibinfo {pages} {776} (\bibinfo {year}
  {2011})}\BibitemShut {NoStop}%
\bibitem [{\citenamefont {Rovezzi}\ \emph {et~al.}(2017)\citenamefont
  {Rovezzi}, \citenamefont {Lapras}, \citenamefont {Manceau}, \citenamefont
  {Glatzel},\ and\ \citenamefont {Verbeni}}]{Rovezzi:2017_RSI}%
  \BibitemOpen
  \bibfield  {author} {\bibinfo {author} {\bibfnamefont {M.}~\bibnamefont
  {Rovezzi}}, \bibinfo {author} {\bibfnamefont {C.}~\bibnamefont {Lapras}},
  \bibinfo {author} {\bibfnamefont {A.}~\bibnamefont {Manceau}}, \bibinfo
  {author} {\bibfnamefont {P.}~\bibnamefont {Glatzel}}, \ and\ \bibinfo
  {author} {\bibfnamefont {R.}~\bibnamefont {Verbeni}},\ }\href {\doibase
  10.1063/1.4974100} {\bibfield  {journal} {\bibinfo  {journal} {Review of
  Scientific Instruments}\ }\textbf {\bibinfo {volume} {88}},\ \bibinfo {pages}
  {013108} (\bibinfo {year} {2017})}\BibitemShut {NoStop}%
\bibitem [{\citenamefont {Johansson}(1933)}]{Johansson:1933_ZP}%
  \BibitemOpen
  \bibfield  {author} {\bibinfo {author} {\bibfnamefont {T.}~\bibnamefont
  {Johansson}},\ }\href {\doibase 10.1007/BF01342254} {\bibfield  {journal}
  {\bibinfo  {journal} {Zeitschrift fuer Physik}\ }\textbf {\bibinfo {volume}
  {82}},\ \bibinfo {pages} {507} (\bibinfo {year} {1933})}\BibitemShut
  {NoStop}%
\bibitem [{\citenamefont {Sanchez~del Rio}\ \emph {et~al.}(2011)\citenamefont
  {Sanchez~del Rio}, \citenamefont {Canestrari}, \citenamefont {Jiang},\ and\
  \citenamefont {Cerrina}}]{SanchezdelRio:2011_JSR}%
  \BibitemOpen
  \bibfield  {author} {\bibinfo {author} {\bibfnamefont {M.}~\bibnamefont
  {Sanchez~del Rio}}, \bibinfo {author} {\bibfnamefont {N.}~\bibnamefont
  {Canestrari}}, \bibinfo {author} {\bibfnamefont {F.}~\bibnamefont {Jiang}}, \
  and\ \bibinfo {author} {\bibfnamefont {F.}~\bibnamefont {Cerrina}},\ }\href
  {\doibase 10.1107/S0909049511026306} {\bibfield  {journal} {\bibinfo
  {journal} {Journal of Synchrotron Radiation}\ }\textbf {\bibinfo {volume}
  {18}},\ \bibinfo {pages} {708} (\bibinfo {year} {2011})}\BibitemShut
  {NoStop}%
\bibitem [{\citenamefont {Sanchez~del Rio}\ \emph {et~al.}(2015)\citenamefont
  {Sanchez~del Rio}, \citenamefont {Perez-Bocanegra}, \citenamefont {Shi},
  \citenamefont {Honkim{\"{a}}ki},\ and\ \citenamefont
  {Zhang}}]{SanchezDelRio:2015_JAC}%
  \BibitemOpen
  \bibfield  {author} {\bibinfo {author} {\bibfnamefont {M.}~\bibnamefont
  {Sanchez~del Rio}}, \bibinfo {author} {\bibfnamefont {N.}~\bibnamefont
  {Perez-Bocanegra}}, \bibinfo {author} {\bibfnamefont {X.}~\bibnamefont
  {Shi}}, \bibinfo {author} {\bibfnamefont {V.}~\bibnamefont
  {Honkim{\"{a}}ki}}, \ and\ \bibinfo {author} {\bibfnamefont {L.}~\bibnamefont
  {Zhang}},\ }\href {\doibase 10.1107/S1600576715002782} {\bibfield  {journal}
  {\bibinfo  {journal} {Journal of Applied Crystallography}\ }\textbf {\bibinfo
  {volume} {48}},\ \bibinfo {pages} {477} (\bibinfo {year} {2015})}\BibitemShut
  {NoStop}%
\bibitem [{\citenamefont {Guijarro}\ and\ \citenamefont
  {{others}}(2018)}]{Guijarro:20180_conf}%
  \BibitemOpen
  \bibfield  {author} {\bibinfo {author} {\bibfnamefont {M.}~\bibnamefont
  {Guijarro}}\ and\ \bibinfo {author} {\bibnamefont {{others}}},\ }in\ \href
  {\doibase 10.18429/JACoW-ICALEPCS2017-WEBPL05} {\emph {\bibinfo {booktitle}
  {Proc. of International Conference on Accelerator and Large Experimental
  Control Systems (ICALEPCS'17), Barcelona, Spain, 8-13 October 2017}}},\
  \bibinfo {series and number} {\bibinfo {series} {International Conference on
  Accelerator and Large Experimental Control Systems}\ No.~\bibinfo {number}
  {16}}\ (\bibinfo  {publisher} {JACoW},\ \bibinfo {address} {Geneva,
  Switzerland},\ \bibinfo {year} {2018})\ pp.\ \bibinfo {pages}
  {1060--1066}\BibitemShut {NoStop}%
\bibitem [{\citenamefont {Rovezzi}(2017)}]{_fn:sloth}%
  \BibitemOpen
  \bibfield  {author} {\bibinfo {author} {\bibfnamefont {M.}~\bibnamefont
  {Rovezzi}},\ }\href {\doibase 10.5281/zenodo.167068} {\bibfield  {journal}
  {\bibinfo  {journal} {Zenodo}\ } (\bibinfo {year} {2017}),\
  10.5281/zenodo.167068}\BibitemShut {NoStop}%
\bibitem [{\citenamefont {Moretti~Sala}\ \emph {et~al.}(2018)\citenamefont
  {Moretti~Sala}, \citenamefont {Martel}, \citenamefont {Henriquet},
  \citenamefont {Al~Zein}, \citenamefont {Simonelli}, \citenamefont {Sahle},
  \citenamefont {Gonzalez}, \citenamefont {Lagier}, \citenamefont {Ponchut},
  \citenamefont {Huotari}, \citenamefont {Verbeni}, \citenamefont {Krisch},\
  and\ \citenamefont {Monaco}}]{MorettiSala:2018_JSR}%
  \BibitemOpen
  \bibfield  {author} {\bibinfo {author} {\bibfnamefont {M.}~\bibnamefont
  {Moretti~Sala}}, \bibinfo {author} {\bibfnamefont {K.}~\bibnamefont
  {Martel}}, \bibinfo {author} {\bibfnamefont {C.}~\bibnamefont {Henriquet}},
  \bibinfo {author} {\bibfnamefont {A.}~\bibnamefont {Al~Zein}}, \bibinfo
  {author} {\bibfnamefont {L.}~\bibnamefont {Simonelli}}, \bibinfo {author}
  {\bibfnamefont {C.~J.}\ \bibnamefont {Sahle}}, \bibinfo {author}
  {\bibfnamefont {H.}~\bibnamefont {Gonzalez}}, \bibinfo {author}
  {\bibfnamefont {M.-C.}\ \bibnamefont {Lagier}}, \bibinfo {author}
  {\bibfnamefont {C.}~\bibnamefont {Ponchut}}, \bibinfo {author} {\bibfnamefont
  {S.}~\bibnamefont {Huotari}}, \bibinfo {author} {\bibfnamefont
  {R.}~\bibnamefont {Verbeni}}, \bibinfo {author} {\bibfnamefont
  {M.}~\bibnamefont {Krisch}}, \ and\ \bibinfo {author} {\bibfnamefont
  {G.}~\bibnamefont {Monaco}},\ }\href {\doibase 10.1107/S1600577518001200}
  {\bibfield  {journal} {\bibinfo  {journal} {Journal of Synchrotron
  Radiation}\ }\textbf {\bibinfo {volume} {25}},\ \bibinfo {pages} {580}
  (\bibinfo {year} {2018})}\BibitemShut {NoStop}%
\bibitem [{\citenamefont {Knoll}(2000)}]{Knoll:2000_book}%
  \BibitemOpen
  \bibfield  {author} {\bibinfo {author} {\bibfnamefont {G.~F.}\ \bibnamefont
  {Knoll}},\ }\href@noop {} {\emph {\bibinfo {title} {{Radiation Detection and
  Measurement}}}},\ \bibinfo {edition} {3rd}\ ed.\ (\bibinfo {year}
  {2000})\BibitemShut {NoStop}%
\bibitem [{\citenamefont {Winkler}\ \emph {et~al.}(2015)\citenamefont
  {Winkler}, \citenamefont {Karadzhinova}, \citenamefont {Hild{\'{e}}n},
  \citenamefont {Garcia}, \citenamefont {Fedi}, \citenamefont {Devoto},\ and\
  \citenamefont {Br{\"{u}}cken}}]{Winkler:2015_AJP}%
  \BibitemOpen
  \bibfield  {author} {\bibinfo {author} {\bibfnamefont {A.}~\bibnamefont
  {Winkler}}, \bibinfo {author} {\bibfnamefont {A.}~\bibnamefont
  {Karadzhinova}}, \bibinfo {author} {\bibfnamefont {T.}~\bibnamefont
  {Hild{\'{e}}n}}, \bibinfo {author} {\bibfnamefont {F.}~\bibnamefont
  {Garcia}}, \bibinfo {author} {\bibfnamefont {G.}~\bibnamefont {Fedi}},
  \bibinfo {author} {\bibfnamefont {F.}~\bibnamefont {Devoto}}, \ and\ \bibinfo
  {author} {\bibfnamefont {E.~J.}\ \bibnamefont {Br{\"{u}}cken}},\ }\href
  {\doibase 10.1119/1.4923022} {\bibfield  {journal} {\bibinfo  {journal}
  {American Journal of Physics}\ }\textbf {\bibinfo {volume} {83}},\ \bibinfo
  {pages} {733} (\bibinfo {year} {2015})}\BibitemShut {NoStop}%
\bibitem [{\citenamefont {Wuhrer}\ and\ \citenamefont
  {Moran}(2018)}]{Wuhrer:2018_conf}%
  \BibitemOpen
  \bibfield  {author} {\bibinfo {author} {\bibfnamefont {R.}~\bibnamefont
  {Wuhrer}}\ and\ \bibinfo {author} {\bibfnamefont {K.}~\bibnamefont {Moran}},\
  }\href {\doibase 10.1088/1757-899X/304/1/012021} {\bibfield  {journal}
  {\bibinfo  {journal} {IOP Conference Series: Materials Science and
  Engineering}\ }\textbf {\bibinfo {volume} {304}},\ \bibinfo {pages} {012021}
  (\bibinfo {year} {2018})}\BibitemShut {NoStop}%
\bibitem [{\citenamefont {Olivero}\ and\ \citenamefont
  {Longbothum}(1977)}]{Olivero:1977_JQSRT}%
  \BibitemOpen
  \bibfield  {author} {\bibinfo {author} {\bibfnamefont {J.}~\bibnamefont
  {Olivero}}\ and\ \bibinfo {author} {\bibfnamefont {R.}~\bibnamefont
  {Longbothum}},\ }\href {\doibase 10.1016/0022-4073(77)90161-3} {\bibfield
  {journal} {\bibinfo  {journal} {Journal of Quantitative Spectroscopy and
  Radiative Transfer}\ }\textbf {\bibinfo {volume} {17}},\ \bibinfo {pages}
  {233} (\bibinfo {year} {1977})}\BibitemShut {NoStop}%
\bibitem [{\citenamefont {Newville}\ \emph {et~al.}(2019)\citenamefont
  {Newville}, \citenamefont {Otten}, \citenamefont {Nelson}, \citenamefont
  {Ingargiola}, \citenamefont {Stensitzki}, \citenamefont {Allan},
  \citenamefont {Fox}, \citenamefont {Carter}, \citenamefont {{Micha{\l}}},
  \citenamefont {Pustakhod}, \citenamefont {Ram}, \citenamefont {{Glenn}},
  \citenamefont {Deil}, \citenamefont {{Stuermer}}, \citenamefont {Beelen},
  \citenamefont {Frost}, \citenamefont {Zobrist}, \citenamefont {Pasquevich},
  \citenamefont {Hansen}, \citenamefont {Stark}, \citenamefont {Spillane},
  \citenamefont {Caldwell}, \citenamefont {Polloreno}, \citenamefont
  {{andrewhannum}}, \citenamefont {Borreguero}, \citenamefont {Fraine},
  \citenamefont {{deep-42-thought}}, \citenamefont {Maier}, \citenamefont
  {Gamari},\ and\ \citenamefont {Almarza}}]{scisoft:lmfit}%
  \BibitemOpen
  \bibfield  {author} {\bibinfo {author} {\bibfnamefont {M.}~\bibnamefont
  {Newville}}, \bibinfo {author} {\bibfnamefont {R.}~\bibnamefont {Otten}},
  \bibinfo {author} {\bibfnamefont {A.}~\bibnamefont {Nelson}}, \bibinfo
  {author} {\bibfnamefont {A.}~\bibnamefont {Ingargiola}}, \bibinfo {author}
  {\bibfnamefont {T.}~\bibnamefont {Stensitzki}}, \bibinfo {author}
  {\bibfnamefont {D.}~\bibnamefont {Allan}}, \bibinfo {author} {\bibfnamefont
  {A.}~\bibnamefont {Fox}}, \bibinfo {author} {\bibfnamefont {F.}~\bibnamefont
  {Carter}}, \bibinfo {author} {\bibnamefont {{Micha{\l}}}}, \bibinfo {author}
  {\bibfnamefont {D.}~\bibnamefont {Pustakhod}}, \bibinfo {author}
  {\bibfnamefont {Y.}~\bibnamefont {Ram}}, \bibinfo {author} {\bibnamefont
  {{Glenn}}}, \bibinfo {author} {\bibfnamefont {C.}~\bibnamefont {Deil}},
  \bibinfo {author} {\bibnamefont {{Stuermer}}}, \bibinfo {author}
  {\bibfnamefont {A.}~\bibnamefont {Beelen}}, \bibinfo {author} {\bibfnamefont
  {O.}~\bibnamefont {Frost}}, \bibinfo {author} {\bibfnamefont
  {N.}~\bibnamefont {Zobrist}}, \bibinfo {author} {\bibfnamefont
  {G.}~\bibnamefont {Pasquevich}}, \bibinfo {author} {\bibfnamefont {A.~L.~R.}\
  \bibnamefont {Hansen}}, \bibinfo {author} {\bibfnamefont {A.}~\bibnamefont
  {Stark}}, \bibinfo {author} {\bibfnamefont {T.}~\bibnamefont {Spillane}},
  \bibinfo {author} {\bibfnamefont {S.}~\bibnamefont {Caldwell}}, \bibinfo
  {author} {\bibfnamefont {A.}~\bibnamefont {Polloreno}}, \bibinfo {author}
  {\bibnamefont {{andrewhannum}}}, \bibinfo {author} {\bibfnamefont
  {J.}~\bibnamefont {Borreguero}}, \bibinfo {author} {\bibfnamefont
  {J.}~\bibnamefont {Fraine}}, \bibinfo {author} {\bibnamefont
  {{deep-42-thought}}}, \bibinfo {author} {\bibfnamefont {B.~F.}\ \bibnamefont
  {Maier}}, \bibinfo {author} {\bibfnamefont {B.}~\bibnamefont {Gamari}}, \
  and\ \bibinfo {author} {\bibfnamefont {A.}~\bibnamefont {Almarza}},\ }\href
  {\doibase 10.5281/ZENODO.3381550} {\enquote {\bibinfo {title}
  {{lmfit/lmfit-py 0.9.14}},}\ } (\bibinfo {year} {2019})\BibitemShut {NoStop}%
\bibitem [{\citenamefont {Merkulova}\ \emph {et~al.}(2019)\citenamefont
  {Merkulova}, \citenamefont {Mathon}, \citenamefont {Glatzel}, \citenamefont
  {Rovezzi}, \citenamefont {Batanova}, \citenamefont {Marion}, \citenamefont
  {Boiron},\ and\ \citenamefont {Manceau}}]{Merkulova:2019_ESC}%
  \BibitemOpen
  \bibfield  {author} {\bibinfo {author} {\bibfnamefont {M.}~\bibnamefont
  {Merkulova}}, \bibinfo {author} {\bibfnamefont {O.}~\bibnamefont {Mathon}},
  \bibinfo {author} {\bibfnamefont {P.}~\bibnamefont {Glatzel}}, \bibinfo
  {author} {\bibfnamefont {M.}~\bibnamefont {Rovezzi}}, \bibinfo {author}
  {\bibfnamefont {V.}~\bibnamefont {Batanova}}, \bibinfo {author}
  {\bibfnamefont {P.}~\bibnamefont {Marion}}, \bibinfo {author} {\bibfnamefont
  {M.-C.}\ \bibnamefont {Boiron}}, \ and\ \bibinfo {author} {\bibfnamefont
  {A.}~\bibnamefont {Manceau}},\ }\href {\doibase
  10.1021/acsearthspacechem.9b00099} {\bibfield  {journal} {\bibinfo  {journal}
  {ACS Earth and Space Chemistry}\ }\textbf {\bibinfo {volume} {3}},\ \bibinfo
  {pages} {1905} (\bibinfo {year} {2019})}\BibitemShut {NoStop}%
\bibitem [{\citenamefont {Manceau}\ \emph {et~al.}(2015)\citenamefont
  {Manceau}, \citenamefont {Lemouchi}, \citenamefont {Rovezzi}, \citenamefont
  {Lanson}, \citenamefont {Glatzel}, \citenamefont {Nagy}, \citenamefont
  {Gautier-Luneau}, \citenamefont {Joly},\ and\ \citenamefont
  {Enescu}}]{Manceau:2015_IC}%
  \BibitemOpen
  \bibfield  {author} {\bibinfo {author} {\bibfnamefont {A.}~\bibnamefont
  {Manceau}}, \bibinfo {author} {\bibfnamefont {C.}~\bibnamefont {Lemouchi}},
  \bibinfo {author} {\bibfnamefont {M.}~\bibnamefont {Rovezzi}}, \bibinfo
  {author} {\bibfnamefont {M.}~\bibnamefont {Lanson}}, \bibinfo {author}
  {\bibfnamefont {P.}~\bibnamefont {Glatzel}}, \bibinfo {author} {\bibfnamefont
  {K.~L.}\ \bibnamefont {Nagy}}, \bibinfo {author} {\bibfnamefont
  {I.}~\bibnamefont {Gautier-Luneau}}, \bibinfo {author} {\bibfnamefont
  {Y.}~\bibnamefont {Joly}}, \ and\ \bibinfo {author} {\bibfnamefont
  {M.}~\bibnamefont {Enescu}},\ }\href {\doibase 10.1021/acs.inorgchem.5b01932}
  {\bibfield  {journal} {\bibinfo  {journal} {Inorganic Chemistry}\ }\textbf
  {\bibinfo {volume} {54}},\ \bibinfo {pages} {11776} (\bibinfo {year}
  {2015})}\BibitemShut {NoStop}%
\end{thebibliography}

\begin{thebibliography}{18}%
\makeatletter
\providecommand \@ifxundefined [1]{%
 \@ifx{#1\undefined}
}%
\providecommand \@ifnum [1]{%
 \ifnum #1\expandafter \@firstoftwo
 \else \expandafter \@secondoftwo
 \fi
}%
\providecommand \@ifx [1]{%
 \ifx #1\expandafter \@firstoftwo
 \else \expandafter \@secondoftwo
 \fi
}%
\providecommand \natexlab [1]{#1}%
\providecommand \enquote  [1]{``#1''}%
\providecommand \bibnamefont  [1]{#1}%
\providecommand \bibfnamefont [1]{#1}%
\providecommand \citenamefont [1]{#1}%
\providecommand \href@noop [0]{\@secondoftwo}%
\providecommand \href [0]{\begingroup \@sanitize@url \@href}%
\providecommand \@href[1]{\@@startlink{#1}\@@href}%
\providecommand \@@href[1]{\endgroup#1\@@endlink}%
\providecommand \@sanitize@url [0]{\catcode `\\12\catcode `\$12\catcode
  `\&12\catcode `\#12\catcode `\^12\catcode `\_12\catcode `\%12\relax}%
\providecommand \@@startlink[1]{}%
\providecommand \@@endlink[0]{}%
\providecommand \url  [0]{\begingroup\@sanitize@url \@url }%
\providecommand \@url [1]{\endgroup\@href {#1}{\urlprefix }}%
\providecommand \urlprefix  [0]{URL }%
\providecommand \Eprint [0]{\href }%
\providecommand \doibase [0]{http://dx.doi.org/}%
\providecommand \selectlanguage [0]{\@gobble}%
\providecommand \bibinfo  [0]{\@secondoftwo}%
\providecommand \bibfield  [0]{\@secondoftwo}%
\providecommand \translation [1]{[#1]}%
\providecommand \BibitemOpen [0]{}%
\providecommand \bibitemStop [0]{}%
\providecommand \bibitemNoStop [0]{.\EOS\space}%
\providecommand \EOS [0]{\spacefactor3000\relax}%
\providecommand \BibitemShut  [1]{\csname bibitem#1\endcsname}%
\let\auto@bib@innerbib\@empty
\bibitem [{\citenamefont {Zhao}\ and\ \citenamefont
  {Sakurai}(2017)}]{Zhao:2017_RSI}%
  \BibitemOpen
  \bibfield  {author} {\bibinfo {author} {\bibfnamefont {W.}~\bibnamefont
  {Zhao}}\ and\ \bibinfo {author} {\bibfnamefont {K.}~\bibnamefont {Sakurai}},\
  }\href {\doibase 10.1063/1.4985149} {\bibfield  {journal} {\bibinfo
  {journal} {Review of Scientific Instruments}\ }\textbf {\bibinfo {volume}
  {88}},\ \bibinfo {pages} {063703} (\bibinfo {year} {2017})}\BibitemShut
  {NoStop}%
\bibitem [{\citenamefont {Str{\"{u}}der}(2000)}]{Struder:2000_NIMA}%
  \BibitemOpen
  \bibfield  {author} {\bibinfo {author} {\bibfnamefont {L.}~\bibnamefont
  {Str{\"{u}}der}},\ }\href {\doibase 10.1016/S0168-9002(00)00811-1} {\bibfield
   {journal} {\bibinfo  {journal} {Nuclear Instruments and Methods in Physics
  Research Section A: Accelerators, Spectrometers, Detectors and Associated
  Equipment}\ }\textbf {\bibinfo {volume} {454}},\ \bibinfo {pages} {73}
  (\bibinfo {year} {2000})}\BibitemShut {NoStop}%
\bibitem [{\citenamefont {Doering}\ \emph {et~al.}(2011)\citenamefont
  {Doering}, \citenamefont {Chuang}, \citenamefont {Andresen}, \citenamefont
  {Chow}, \citenamefont {Contarato}, \citenamefont {Cummings}, \citenamefont
  {Domning}, \citenamefont {Joseph}, \citenamefont {Pepper}, \citenamefont
  {Smith}, \citenamefont {Zizka}, \citenamefont {Ford}, \citenamefont {Lee},
  \citenamefont {Weaver}, \citenamefont {Patthey}, \citenamefont {Weizeorick},
  \citenamefont {Hussain},\ and\ \citenamefont {Denes}}]{Doering:2011_RSI}%
  \BibitemOpen
  \bibfield  {author} {\bibinfo {author} {\bibfnamefont {D.}~\bibnamefont
  {Doering}}, \bibinfo {author} {\bibfnamefont {Y.-D.}\ \bibnamefont {Chuang}},
  \bibinfo {author} {\bibfnamefont {N.}~\bibnamefont {Andresen}}, \bibinfo
  {author} {\bibfnamefont {K.}~\bibnamefont {Chow}}, \bibinfo {author}
  {\bibfnamefont {D.}~\bibnamefont {Contarato}}, \bibinfo {author}
  {\bibfnamefont {C.}~\bibnamefont {Cummings}}, \bibinfo {author}
  {\bibfnamefont {E.}~\bibnamefont {Domning}}, \bibinfo {author} {\bibfnamefont
  {J.}~\bibnamefont {Joseph}}, \bibinfo {author} {\bibfnamefont {J.~S.}\
  \bibnamefont {Pepper}}, \bibinfo {author} {\bibfnamefont {B.}~\bibnamefont
  {Smith}}, \bibinfo {author} {\bibfnamefont {G.}~\bibnamefont {Zizka}},
  \bibinfo {author} {\bibfnamefont {C.}~\bibnamefont {Ford}}, \bibinfo {author}
  {\bibfnamefont {W.~S.}\ \bibnamefont {Lee}}, \bibinfo {author} {\bibfnamefont
  {M.}~\bibnamefont {Weaver}}, \bibinfo {author} {\bibfnamefont
  {L.}~\bibnamefont {Patthey}}, \bibinfo {author} {\bibfnamefont
  {J.}~\bibnamefont {Weizeorick}}, \bibinfo {author} {\bibfnamefont
  {Z.}~\bibnamefont {Hussain}}, \ and\ \bibinfo {author} {\bibfnamefont
  {P.}~\bibnamefont {Denes}},\ }\href {\doibase 10.1063/1.3609862} {\bibfield
  {journal} {\bibinfo  {journal} {Review of Scientific Instruments}\ }\textbf
  {\bibinfo {volume} {82}},\ \bibinfo {pages} {073303} (\bibinfo {year}
  {2011})}\BibitemShut {NoStop}%
\bibitem [{\citenamefont {Hafizh}\ \emph {et~al.}(2019)\citenamefont {Hafizh},
  \citenamefont {Bellotti}, \citenamefont {Carminati}, \citenamefont {Utica},
  \citenamefont {Gugiatti}, \citenamefont {Balerna}, \citenamefont {Tullio},
  \citenamefont {Lepore}, \citenamefont {Borghi}, \citenamefont {Ficorella},
  \citenamefont {Picciotto}, \citenamefont {Zorzi}, \citenamefont {Capsoni},
  \citenamefont {Coelli}, \citenamefont {Bombelli},\ and\ \citenamefont
  {Fiorini}}]{Hafizh:2019_JINST}%
  \BibitemOpen
  \bibfield  {author} {\bibinfo {author} {\bibfnamefont {I.}~\bibnamefont
  {Hafizh}}, \bibinfo {author} {\bibfnamefont {G.}~\bibnamefont {Bellotti}},
  \bibinfo {author} {\bibfnamefont {M.}~\bibnamefont {Carminati}}, \bibinfo
  {author} {\bibfnamefont {G.}~\bibnamefont {Utica}}, \bibinfo {author}
  {\bibfnamefont {M.}~\bibnamefont {Gugiatti}}, \bibinfo {author}
  {\bibfnamefont {A.}~\bibnamefont {Balerna}}, \bibinfo {author} {\bibfnamefont
  {V.}~\bibnamefont {Tullio}}, \bibinfo {author} {\bibfnamefont
  {G.}~\bibnamefont {Lepore}}, \bibinfo {author} {\bibfnamefont
  {G.}~\bibnamefont {Borghi}}, \bibinfo {author} {\bibfnamefont
  {F.}~\bibnamefont {Ficorella}}, \bibinfo {author} {\bibfnamefont
  {A.}~\bibnamefont {Picciotto}}, \bibinfo {author} {\bibfnamefont
  {N.}~\bibnamefont {Zorzi}}, \bibinfo {author} {\bibfnamefont
  {A.}~\bibnamefont {Capsoni}}, \bibinfo {author} {\bibfnamefont
  {S.}~\bibnamefont {Coelli}}, \bibinfo {author} {\bibfnamefont
  {L.}~\bibnamefont {Bombelli}}, \ and\ \bibinfo {author} {\bibfnamefont
  {C.}~\bibnamefont {Fiorini}},\ }\href {\doibase
  10.1088/1748-0221/14/06/P06027} {\bibfield  {journal} {\bibinfo  {journal}
  {Journal of Instrumentation}\ }\textbf {\bibinfo {volume} {14}},\ \bibinfo
  {pages} {P06027} (\bibinfo {year} {2019})}\BibitemShut {NoStop}%
\bibitem [{\citenamefont {Bufon}\ \emph {et~al.}(2014)\citenamefont {Bufon},
  \citenamefont {Ahangarianabhari}, \citenamefont {Bellutti}, \citenamefont
  {Bertuccio}, \citenamefont {Carrato}, \citenamefont {Cautero}, \citenamefont
  {Fabiani}, \citenamefont {Giacomini}, \citenamefont {Gianoncelli},
  \citenamefont {Giuressi}, \citenamefont {Grassi}, \citenamefont {Malcovati},
  \citenamefont {Menk}, \citenamefont {Picciotto}, \citenamefont {Piemonte},
  \citenamefont {Rashevskaya}, \citenamefont {Rachevski}, \citenamefont
  {Stolfa}, \citenamefont {Vacchi}, \citenamefont {Zampa},\ and\ \citenamefont
  {Zampa}}]{Bufon:2014_JINST}%
  \BibitemOpen
  \bibfield  {author} {\bibinfo {author} {\bibfnamefont {J.}~\bibnamefont
  {Bufon}}, \bibinfo {author} {\bibfnamefont {M.}~\bibnamefont
  {Ahangarianabhari}}, \bibinfo {author} {\bibfnamefont {P.}~\bibnamefont
  {Bellutti}}, \bibinfo {author} {\bibfnamefont {G.}~\bibnamefont {Bertuccio}},
  \bibinfo {author} {\bibfnamefont {S.}~\bibnamefont {Carrato}}, \bibinfo
  {author} {\bibfnamefont {G.}~\bibnamefont {Cautero}}, \bibinfo {author}
  {\bibfnamefont {S.}~\bibnamefont {Fabiani}}, \bibinfo {author} {\bibfnamefont
  {G.}~\bibnamefont {Giacomini}}, \bibinfo {author} {\bibfnamefont
  {A.}~\bibnamefont {Gianoncelli}}, \bibinfo {author} {\bibfnamefont
  {D.}~\bibnamefont {Giuressi}}, \bibinfo {author} {\bibfnamefont
  {M.}~\bibnamefont {Grassi}}, \bibinfo {author} {\bibfnamefont
  {P.}~\bibnamefont {Malcovati}}, \bibinfo {author} {\bibfnamefont
  {R.}~\bibnamefont {Menk}}, \bibinfo {author} {\bibfnamefont {A.}~\bibnamefont
  {Picciotto}}, \bibinfo {author} {\bibfnamefont {C.}~\bibnamefont {Piemonte}},
  \bibinfo {author} {\bibfnamefont {I.}~\bibnamefont {Rashevskaya}}, \bibinfo
  {author} {\bibfnamefont {A.}~\bibnamefont {Rachevski}}, \bibinfo {author}
  {\bibfnamefont {A.}~\bibnamefont {Stolfa}}, \bibinfo {author} {\bibfnamefont
  {A.}~\bibnamefont {Vacchi}}, \bibinfo {author} {\bibfnamefont
  {G.}~\bibnamefont {Zampa}}, \ and\ \bibinfo {author} {\bibfnamefont
  {N.}~\bibnamefont {Zampa}},\ }\href {\doibase 10.1088/1748-0221/9/12/C12017}
  {\bibfield  {journal} {\bibinfo  {journal} {Journal of Instrumentation}\
  }\textbf {\bibinfo {volume} {9}},\ \bibinfo {pages} {C12017} (\bibinfo {year}
  {2014})}\BibitemShut {NoStop}%
\bibitem [{\citenamefont {Donath}\ \emph {et~al.}(2013)\citenamefont {Donath},
  \citenamefont {Brandstetter}, \citenamefont {Cibik}, \citenamefont
  {Commichau}, \citenamefont {Hofer}, \citenamefont {Krumrey}, \citenamefont
  {L{\"{u}}thi}, \citenamefont {Marggraf}, \citenamefont {M{\"{u}}ller},
  \citenamefont {Schneebeli}, \citenamefont {Schulze-Briese},\ and\
  \citenamefont {Wernecke}}]{Donath:2013_conf}%
  \BibitemOpen
  \bibfield  {author} {\bibinfo {author} {\bibfnamefont {T.}~\bibnamefont
  {Donath}}, \bibinfo {author} {\bibfnamefont {S.}~\bibnamefont
  {Brandstetter}}, \bibinfo {author} {\bibfnamefont {L.}~\bibnamefont {Cibik}},
  \bibinfo {author} {\bibfnamefont {S.}~\bibnamefont {Commichau}}, \bibinfo
  {author} {\bibfnamefont {P.}~\bibnamefont {Hofer}}, \bibinfo {author}
  {\bibfnamefont {M.}~\bibnamefont {Krumrey}}, \bibinfo {author} {\bibfnamefont
  {B.}~\bibnamefont {L{\"{u}}thi}}, \bibinfo {author} {\bibfnamefont
  {S.}~\bibnamefont {Marggraf}}, \bibinfo {author} {\bibfnamefont
  {P.}~\bibnamefont {M{\"{u}}ller}}, \bibinfo {author} {\bibfnamefont
  {M.}~\bibnamefont {Schneebeli}}, \bibinfo {author} {\bibfnamefont
  {C.}~\bibnamefont {Schulze-Briese}}, \ and\ \bibinfo {author} {\bibfnamefont
  {J.}~\bibnamefont {Wernecke}},\ }\href {\doibase
  10.1088/1742-6596/425/6/062001} {\bibfield  {journal} {\bibinfo  {journal}
  {Journal of Physics: Conference Series}\ }\textbf {\bibinfo {volume} {425}},\
  \bibinfo {pages} {062001} (\bibinfo {year} {2013})}\BibitemShut {NoStop}%
\bibitem [{\citenamefont {Kla{\v{c}}kov{\'{a}}}\ \emph
  {et~al.}(2019)\citenamefont {Kla{\v{c}}kov{\'{a}}}, \citenamefont {Blaj},
  \citenamefont {Denes}, \citenamefont {Dragone}, \citenamefont {G{\"{o}}de},
  \citenamefont {Hauf}, \citenamefont {Januschek}, \citenamefont {Joseph},\
  and\ \citenamefont {Kuster}}]{Klackova:2019_JINST}%
  \BibitemOpen
  \bibfield  {author} {\bibinfo {author} {\bibfnamefont {I.}~\bibnamefont
  {Kla{\v{c}}kov{\'{a}}}}, \bibinfo {author} {\bibfnamefont {G.}~\bibnamefont
  {Blaj}}, \bibinfo {author} {\bibfnamefont {P.}~\bibnamefont {Denes}},
  \bibinfo {author} {\bibfnamefont {A.}~\bibnamefont {Dragone}}, \bibinfo
  {author} {\bibfnamefont {S.}~\bibnamefont {G{\"{o}}de}}, \bibinfo {author}
  {\bibfnamefont {S.}~\bibnamefont {Hauf}}, \bibinfo {author} {\bibfnamefont
  {F.}~\bibnamefont {Januschek}}, \bibinfo {author} {\bibfnamefont
  {J.}~\bibnamefont {Joseph}}, \ and\ \bibinfo {author} {\bibfnamefont
  {M.}~\bibnamefont {Kuster}},\ }\href {\doibase
  10.1088/1748-0221/14/01/C01008} {\bibfield  {journal} {\bibinfo  {journal}
  {Journal of Instrumentation}\ }\textbf {\bibinfo {volume} {14}},\ \bibinfo
  {pages} {C01008} (\bibinfo {year} {2019})}\BibitemShut {NoStop}%
\bibitem [{\citenamefont {Holden}\ \emph {et~al.}(2018)\citenamefont {Holden},
  \citenamefont {Hoidn}, \citenamefont {Seidler},\ and\ \citenamefont
  {DiChiara}}]{Holden:2018_RSI}%
  \BibitemOpen
  \bibfield  {author} {\bibinfo {author} {\bibfnamefont {W.~M.}\ \bibnamefont
  {Holden}}, \bibinfo {author} {\bibfnamefont {O.~R.}\ \bibnamefont {Hoidn}},
  \bibinfo {author} {\bibfnamefont {G.~T.}\ \bibnamefont {Seidler}}, \ and\
  \bibinfo {author} {\bibfnamefont {A.~D.}\ \bibnamefont {DiChiara}},\ }\href
  {\doibase 10.1063/1.5047934} {\bibfield  {journal} {\bibinfo  {journal}
  {Review of Scientific Instruments}\ }\textbf {\bibinfo {volume} {89}},\
  \bibinfo {pages} {093111} (\bibinfo {year} {2018})}\BibitemShut {NoStop}%
\bibitem [{\citenamefont {Haro}\ \emph {et~al.}(2019)\citenamefont {Haro},
  \citenamefont {Bessia}, \citenamefont {P{\'{e}}rez}, \citenamefont
  {Blostein}, \citenamefont {Balmaceda}, \citenamefont {Berisso},\ and\
  \citenamefont {Lipovetzky}}]{Haro:2019_RPC}%
  \BibitemOpen
  \bibfield  {author} {\bibinfo {author} {\bibfnamefont {M.~S.}\ \bibnamefont
  {Haro}}, \bibinfo {author} {\bibfnamefont {F.~A.}\ \bibnamefont {Bessia}},
  \bibinfo {author} {\bibfnamefont {M.}~\bibnamefont {P{\'{e}}rez}}, \bibinfo
  {author} {\bibfnamefont {J.~J.}\ \bibnamefont {Blostein}}, \bibinfo {author}
  {\bibfnamefont {D.~F.}\ \bibnamefont {Balmaceda}}, \bibinfo {author}
  {\bibfnamefont {M.~G.}\ \bibnamefont {Berisso}}, \ and\ \bibinfo {author}
  {\bibfnamefont {J.}~\bibnamefont {Lipovetzky}},\ }\href {\doibase
  10.1016/j.radphyschem.2019.108354} {\bibfield  {journal} {\bibinfo  {journal}
  {Radiation Physics and Chemistry}\ ,\ \bibinfo {pages} {108354}} (\bibinfo
  {year} {2019})}\BibitemShut {NoStop}%
\bibitem [{\citenamefont {BARSAN}\ \emph {et~al.}(2007)\citenamefont {BARSAN},
  \citenamefont {KOZIEJ},\ and\ \citenamefont {WEIMAR}}]{Barsan:2007_SAB}%
  \BibitemOpen
  \bibfield  {author} {\bibinfo {author} {\bibfnamefont {N.}~\bibnamefont
  {BARSAN}}, \bibinfo {author} {\bibfnamefont {D.}~\bibnamefont {KOZIEJ}}, \
  and\ \bibinfo {author} {\bibfnamefont {U.}~\bibnamefont {WEIMAR}},\ }\href
  {\doibase 10.1016/j.snb.2006.09.047} {\bibfield  {journal} {\bibinfo
  {journal} {Sensors and Actuators B: Chemical}\ }\textbf {\bibinfo {volume}
  {121}},\ \bibinfo {pages} {18} (\bibinfo {year} {2007})}\BibitemShut
  {NoStop}%
\bibitem [{\citenamefont {Degler}(2018)}]{Degler:2018_Sensors}%
  \BibitemOpen
  \bibfield  {author} {\bibinfo {author} {\bibfnamefont {D.}~\bibnamefont
  {Degler}},\ }\href {\doibase 10.3390/s18103544} {\bibfield  {journal}
  {\bibinfo  {journal} {Sensors}\ }\textbf {\bibinfo {volume} {18}},\ \bibinfo
  {pages} {3544} (\bibinfo {year} {2018})}\BibitemShut {NoStop}%
\bibitem [{\citenamefont {Degler}\ \emph {et~al.}(2016)\citenamefont {Degler},
  \citenamefont {Rank}, \citenamefont {M{\"{u}}ller}, \citenamefont {Pereira~de
  Carvalho}, \citenamefont {Grunwaldt}, \citenamefont {Weimar},\ and\
  \citenamefont {Barsan}}]{Degler:2016_ACSSens}%
  \BibitemOpen
  \bibfield  {author} {\bibinfo {author} {\bibfnamefont {D.}~\bibnamefont
  {Degler}}, \bibinfo {author} {\bibfnamefont {S.}~\bibnamefont {Rank}},
  \bibinfo {author} {\bibfnamefont {S.}~\bibnamefont {M{\"{u}}ller}}, \bibinfo
  {author} {\bibfnamefont {H.~W.}\ \bibnamefont {Pereira~de Carvalho}},
  \bibinfo {author} {\bibfnamefont {J.-D.}\ \bibnamefont {Grunwaldt}}, \bibinfo
  {author} {\bibfnamefont {U.}~\bibnamefont {Weimar}}, \ and\ \bibinfo {author}
  {\bibfnamefont {N.}~\bibnamefont {Barsan}},\ }\href {\doibase
  10.1021/acssensors.6b00477} {\bibfield  {journal} {\bibinfo  {journal} {ACS
  Sensors}\ }\textbf {\bibinfo {volume} {1}},\ \bibinfo {pages} {1322}
  (\bibinfo {year} {2016})}\BibitemShut {NoStop}%
\bibitem [{\citenamefont {H{\"{u}}bner}\ \emph {et~al.}(2011)\citenamefont
  {H{\"{u}}bner}, \citenamefont {Koziej}, \citenamefont {Bauer}, \citenamefont
  {Barsan}, \citenamefont {Kvashnina}, \citenamefont {Rossell}, \citenamefont
  {Weimar},\ and\ \citenamefont {Grunwaldt}}]{Hubner:2011_Angewandte}%
  \BibitemOpen
  \bibfield  {author} {\bibinfo {author} {\bibfnamefont {M.}~\bibnamefont
  {H{\"{u}}bner}}, \bibinfo {author} {\bibfnamefont {D.}~\bibnamefont
  {Koziej}}, \bibinfo {author} {\bibfnamefont {M.}~\bibnamefont {Bauer}},
  \bibinfo {author} {\bibfnamefont {N.}~\bibnamefont {Barsan}}, \bibinfo
  {author} {\bibfnamefont {K.}~\bibnamefont {Kvashnina}}, \bibinfo {author}
  {\bibfnamefont {M.~D.}\ \bibnamefont {Rossell}}, \bibinfo {author}
  {\bibfnamefont {U.}~\bibnamefont {Weimar}}, \ and\ \bibinfo {author}
  {\bibfnamefont {J.-D.}\ \bibnamefont {Grunwaldt}},\ }\href {\doibase
  10.1002/anie.201004499} {\bibfield  {journal} {\bibinfo  {journal}
  {Angewandte Chemie International Edition}\ }\textbf {\bibinfo {volume}
  {50}},\ \bibinfo {pages} {2841} (\bibinfo {year} {2011})}\BibitemShut
  {NoStop}%
\bibitem [{\citenamefont {Safonova}\ \emph {et~al.}(2005)\citenamefont
  {Safonova}, \citenamefont {Neisius}, \citenamefont {Ryzhikov}, \citenamefont
  {Chenevier}, \citenamefont {Gaskov},\ and\ \citenamefont
  {Labeau}}]{Safonova:2005_ChemCom}%
  \BibitemOpen
  \bibfield  {author} {\bibinfo {author} {\bibfnamefont {O.~V.}\ \bibnamefont
  {Safonova}}, \bibinfo {author} {\bibfnamefont {T.}~\bibnamefont {Neisius}},
  \bibinfo {author} {\bibfnamefont {A.}~\bibnamefont {Ryzhikov}}, \bibinfo
  {author} {\bibfnamefont {B.}~\bibnamefont {Chenevier}}, \bibinfo {author}
  {\bibfnamefont {A.~M.}\ \bibnamefont {Gaskov}}, \ and\ \bibinfo {author}
  {\bibfnamefont {M.}~\bibnamefont {Labeau}},\ }\href {\doibase
  10.1039/b509826b} {\bibfield  {journal} {\bibinfo  {journal} {Chemical
  Communications}\ ,\ \bibinfo {pages} {5202}} (\bibinfo {year}
  {2005})}\BibitemShut {NoStop}%
\bibitem [{\citenamefont {Koziej}\ \emph {et~al.}(2009)\citenamefont {Koziej},
  \citenamefont {H{\"{u}}bner}, \citenamefont {Barsan}, \citenamefont {Weimar},
  \citenamefont {Sikora},\ and\ \citenamefont {Grunwaldt}}]{Koziej:2009_PCCP}%
  \BibitemOpen
  \bibfield  {author} {\bibinfo {author} {\bibfnamefont {D.}~\bibnamefont
  {Koziej}}, \bibinfo {author} {\bibfnamefont {M.}~\bibnamefont
  {H{\"{u}}bner}}, \bibinfo {author} {\bibfnamefont {N.}~\bibnamefont
  {Barsan}}, \bibinfo {author} {\bibfnamefont {U.}~\bibnamefont {Weimar}},
  \bibinfo {author} {\bibfnamefont {M.}~\bibnamefont {Sikora}}, \ and\ \bibinfo
  {author} {\bibfnamefont {J.-D.}\ \bibnamefont {Grunwaldt}},\ }\href {\doibase
  10.1039/b906829e} {\bibfield  {journal} {\bibinfo  {journal} {Physical
  Chemistry Chemical Physics}\ }\textbf {\bibinfo {volume} {11}},\ \bibinfo
  {pages} {8620} (\bibinfo {year} {2009})}\BibitemShut {NoStop}%
\bibitem [{\citenamefont {Staerz}\ \emph {et~al.}(2018)\citenamefont {Staerz},
  \citenamefont {Boehme}, \citenamefont {Degler}, \citenamefont {Bahri},
  \citenamefont {Doronkin}, \citenamefont {Zimina}, \citenamefont {Brinkmann},
  \citenamefont {Herrmann}, \citenamefont {Junker}, \citenamefont {Ersen},
  \citenamefont {Grunwaldt}, \citenamefont {Weimar},\ and\ \citenamefont
  {Barsan}}]{Staerz:2018_Nanomaterials}%
  \BibitemOpen
  \bibfield  {author} {\bibinfo {author} {\bibfnamefont {A.}~\bibnamefont
  {Staerz}}, \bibinfo {author} {\bibfnamefont {I.}~\bibnamefont {Boehme}},
  \bibinfo {author} {\bibfnamefont {D.}~\bibnamefont {Degler}}, \bibinfo
  {author} {\bibfnamefont {M.}~\bibnamefont {Bahri}}, \bibinfo {author}
  {\bibfnamefont {D.}~\bibnamefont {Doronkin}}, \bibinfo {author}
  {\bibfnamefont {A.}~\bibnamefont {Zimina}}, \bibinfo {author} {\bibfnamefont
  {H.}~\bibnamefont {Brinkmann}}, \bibinfo {author} {\bibfnamefont
  {S.}~\bibnamefont {Herrmann}}, \bibinfo {author} {\bibfnamefont
  {B.}~\bibnamefont {Junker}}, \bibinfo {author} {\bibfnamefont
  {O.}~\bibnamefont {Ersen}}, \bibinfo {author} {\bibfnamefont {J.-D.}\
  \bibnamefont {Grunwaldt}}, \bibinfo {author} {\bibfnamefont {U.}~\bibnamefont
  {Weimar}}, \ and\ \bibinfo {author} {\bibfnamefont {N.}~\bibnamefont
  {Barsan}},\ }\href {\doibase 10.3390/nano8110892} {\bibfield  {journal}
  {\bibinfo  {journal} {Nanomaterials}\ }\textbf {\bibinfo {volume} {8}},\
  \bibinfo {pages} {892} (\bibinfo {year} {2018})}\BibitemShut {NoStop}%
\bibitem [{\citenamefont {Barsan}\ and\ \citenamefont
  {Weimar}(2003)}]{Barsan:2003_JPCM}%
  \BibitemOpen
  \bibfield  {author} {\bibinfo {author} {\bibfnamefont {N.}~\bibnamefont
  {Barsan}}\ and\ \bibinfo {author} {\bibfnamefont {U.}~\bibnamefont
  {Weimar}},\ }\href {\doibase 10.1088/0953-8984/15/20/201} {\bibfield
  {journal} {\bibinfo  {journal} {Journal of Physics: Condensed Matter}\
  }\textbf {\bibinfo {volume} {15}},\ \bibinfo {pages} {R813} (\bibinfo {year}
  {2003})}\BibitemShut {NoStop}%
\bibitem [{\citenamefont {Rovezzi}\ \emph {et~al.}(2017)\citenamefont
  {Rovezzi}, \citenamefont {Lapras}, \citenamefont {Manceau}, \citenamefont
  {Glatzel},\ and\ \citenamefont {Verbeni}}]{Rovezzi:2017_RSI}%
  \BibitemOpen
  \bibfield  {author} {\bibinfo {author} {\bibfnamefont {M.}~\bibnamefont
  {Rovezzi}}, \bibinfo {author} {\bibfnamefont {C.}~\bibnamefont {Lapras}},
  \bibinfo {author} {\bibfnamefont {A.}~\bibnamefont {Manceau}}, \bibinfo
  {author} {\bibfnamefont {P.}~\bibnamefont {Glatzel}}, \ and\ \bibinfo
  {author} {\bibfnamefont {R.}~\bibnamefont {Verbeni}},\ }\href {\doibase
  10.1063/1.4974100} {\bibfield  {journal} {\bibinfo  {journal} {Review of
  Scientific Instruments}\ }\textbf {\bibinfo {volume} {88}},\ \bibinfo {pages}
  {013108} (\bibinfo {year} {2017})}\BibitemShut {NoStop}%
\end{thebibliography}
